\documentclass{lmcs} %%% last changed 2014-08-20
\pdfoutput=1

\usepackage{lastpage}
\lmcsdoi{22}{1}{24}
\lmcsheading{}{\pageref{LastPage}}{}{}%
{Nov.~13,~2024}{Mar.~12,~2026}{}

\usepackage{hyperref}
\theoremstyle{plain} %\crefname{satz}{Satz}{S\"atze}
\usepackage{mathtools}
\usepackage{stmaryrd}
\usepackage{todonotes}
\usepackage{cleveref}
\usepackage{algorithm}
\usepackage{algpseudocode}
\usepackage{tcolorbox}
\usepackage{xifthen}
\usepackage{amssymb}
\usepackage{subcaption}
\usepackage{thmtools}
\usepackage{thm-restate}
\usepackage[utf8]{inputenc}
\usepackage{booktabs}
\crefname{thm}{Theorem}{Theorems}
\crefname{lem}{Lemma}{Lemmas}
\crefname{cor}{Corollary}{Corollaries}
\crefname{prop}{Proposition}{Propositions}
\crefname{exa}{Example}{Examples}

\RenewDocumentCommand{\P}{}{\ensuremath{\mathsf{P}}}
\NewDocumentCommand{\NP}{}{\ensuremath{\mathsf{NP}}}

\DeclareMathOperator{\poly}{poly}
\DeclareMathOperator{\polylog}{polylog}
\NewDocumentCommand{\SAT}{}{\ensuremath{\mathsf{SAT}}}

\NewDocumentCommand{\dec}{}{\ensuremath{\mathsf{dec}}}

\NewDocumentCommand{\jcircuit}{}{$\{\bowtie, \dec\}$-circuit}

\NewDocumentCommand{\dcircuit}{}{$\{\times, \dec\}$-circuit}

\NewDocumentCommand{\ocircuit}{}{ordered $\{\times, \dec\}$-circuit}

\NewDocumentCommand{\var}{}{\ensuremath{\mathsf{var}}}
\NewDocumentCommand{\decvar}{}{\ensuremath{\mathsf{decvar}}}
\NewDocumentCommand{\out}{}{\ensuremath{\mathsf{out}}}

\NewDocumentCommand{\nrel}{O{C}}{\ensuremath{\mathsf{nrel}_{#1}}}
\NewDocumentCommand{\xrel}{O{C}}{\ensuremath{\mathsf{Xrel}_{#1}}}
\NewDocumentCommand{\sink}{}{\ensuremath{\mathsf{sink}}}
\NewDocumentCommand{\rel}{}{\ensuremath{\mathsf{rel}}}

\NewDocumentCommand{\size}{ m }{\ensuremath{\vert #1 \vert}}

\NewDocumentCommand{\atoms}{}{\ensuremath{\mathsf{atoms}}}
\NewDocumentCommand{\atomp}{O{Q}}{\ensuremath{\atoms^+({#1})}}
\NewDocumentCommand{\atomn}{O{Q}}{\ensuremath{\atoms^-({#1})}}

\NewDocumentCommand{\free}{}{\ensuremath{\mathsf{free}}}
\NewDocumentCommand{\ecup}{}{{\overline{\cup}}}
\NewDocumentCommand{\db}{}{\ensuremath{\mathbf{D}}}

\NewDocumentCommand{\rank}{}{\ensuremath{\mathsf{rank}}}
\NewDocumentCommand{\tup}{O{x}}{\ensuremath{\mathbf{#1}}}
\NewDocumentCommand{\ans}{O{Q} O{} O{\db}}{\ensuremath{{{\llbracket #1 \rrbracket}^{#3}_{#2}}}}
\NewDocumentCommand{\kso}{}{$k$-th solution}
\NewDocumentCommand{\kt}{}{$k^{\mathsf{th}}$ tuple}
\NewDocumentCommand{\kth}{}{\ensuremath{k^{\mathsf{th}}}}
\NewDocumentCommand{\igraph}{O{\tau} O{Q}}{\ensuremath{\mathcal{I}_{#1}^{#2}}}

\NewDocumentCommand{\simpd}{O{\tau} O{Q} O{\db}}{\ensuremath{#2 \Downarrow \langle #1, #3 \rangle}}
\NewDocumentCommand{\simp}{O{\tau} O{Q} O{\db}}{\ensuremath{#2 \Downarrow #1}}

\NewDocumentCommand{\qabs}{O{Q}}{\ensuremath{\overline{#1}}}

\NewDocumentCommand{\emptytup}{}{\ensuremath{\langle\rangle}}

\NewDocumentCommand{\bin}{O{R} O{b}}{\ensuremath{\widetilde{#1}^{#2}}}
\NewDocumentCommand{\debin}{O{R} O{b}}{\ensuremath{\bar{#1}^{#2}}}

\NewDocumentCommand{\DPLL}{}{\ensuremath{\mathsf{DPLL}}}
\NewDocumentCommand{\cache}{}{\ensuremath{\mathsf{cache}}}
\NewDocumentCommand{\New}{}{\ensuremath{\mathbf{new}}}
\NewDocumentCommand{\rec}{O{Q} O{\db}}{\ensuremath{\mathbf{R}_{#1}^{#2}}}

\NewDocumentCommand{\neigh}{O{v} O{H}}{\ensuremath{N_{#2}(#1)}}
\NewDocumentCommand{\oneigh}{O{v} O{H}}{\ensuremath{N^*_{#2}(#1)}}
\NewDocumentCommand{\bigo}{}{\ensuremath{\mathcal{O}}}
\NewDocumentCommand{\bigot}{}{\ensuremath{\tilde{\mathcal{O}}}}
\NewDocumentCommand{\da}{}{direct access}

\NewDocumentCommand{\qnn}{ O{N_1} O{N_2} }{\ensuremath{Q_{#1,#2}}}

\NewDocumentCommand{\width}{O{Q} O{\prec}}{\ensuremath{{\mathsf{width}_{#1}(#2)}}}
\NewDocumentCommand{\hwidth}{m m m}{%
  \ifthenelse{\isempty{#2}}%
  {\ensuremath{\mathsf{#3}(#1)}}%
  {\ensuremath{\mathsf{#3}(#1, #2)}}%
}
\NewDocumentCommand{\how}{m O{}}{\hwidth{#1}{#2}{how}}
\NewDocumentCommand{\fhow}{m O{}}{\hwidth{#1}{#2}{fhow}}
\NewDocumentCommand{\bhow}{m O{}}{\beta\text{-}\hwidth{#1}{#2}{how}}
\NewDocumentCommand{\bfhow}{m O{}}{\beta\text{-}\hwidth{#1}{#2}{fhow}}
\NewDocumentCommand{\sfhow}{m O{}}{\hwidth{#1}{#2}{sfhow}}
\RenewDocumentCommand{\show}{m O{}}{\hwidth{#1}{#2}{show}}
\NewDocumentCommand{\htw}{m O{}}{\hwidth{#1}{#2}{htw}}
\NewDocumentCommand{\fhtw}{m O{}}{\hwidth{#1}{#2}{fhtw}}
\NewDocumentCommand{\bfhtw}{m O{}}{\beta\text{-}\hwidth{#1}{#2}{fhtw}}
\NewDocumentCommand{\bhtw}{m O{}}{\beta\text{-}\hwidth{#1}{#2}{htw}}
\NewDocumentCommand{\bpfhtw}{m O{}}{\beta\text{-}\hwidth{#1}{#2}{fhtw'}}
\NewDocumentCommand{\bphtw}{m O{}}{\beta\text{-}\hwidth{#1}{#2}{htw'}}
\NewDocumentCommand{\nsw}{m O{}}{\hwidth{#1}{#2}{nsw}}
\NewDocumentCommand{\tw}{m O{}}{\hwidth{#1}{#2}{tw}}

\NewDocumentCommand{\hcv}{O{v} O{H} O{\prec}}{\ensuremath{\mathcal{C}^{#3}_{#1}(#2)}}

\RenewDocumentCommand{\leq}{}{\leqslant}
\RenewDocumentCommand{\geq}{}{\geqslant}
\RenewDocumentCommand{\log}{}{\mathsf{log}}

\NewDocumentCommand{\N}{}{\ensuremath{\mathbb{N}}}
\NewDocumentCommand{\Rp}{}{\ensuremath{\mathbb{R_+}}}
\NewDocumentCommand{\lexprec}{}{\ensuremath{\prec_{\mathsf{lex}}}}
\NewDocumentCommand{\lexpreceq}{}{\ensuremath{\preceq_{\mathsf{lex}}}}

\NewDocumentCommand{\from}{}{\ensuremath{\colon}}
\RenewDocumentCommand{\to}{}{\ensuremath{\rightarrow}}

\NewDocumentCommand{\eIf}{m m}{\State {\bf if} #1 {\bf then} #2}

\NewDocumentCommand{\flo}{ O{} m }{%
  \todo[linecolor=blue,
        backgroundcolor=blue!25,
        bordercolor=blue,
        tickmarkheight=0.15cm,
        #1]{
          \textbf{Florent : }#2}
}

\NewDocumentCommand{\nof}{ O{} m }{%
  \todo[linecolor=purple,
        backgroundcolor=purple!25,
        bordercolor=purple,
        tickmarkheight=0.15cm,
        #1]{
          \textbf{Nofar : }#2}
}

\NewDocumentCommand{\oli}{ O{} m }{%
  \todo[linecolor=green,
  backgroundcolor=green!25,
  bordercolor=green,
  tickmarkheight=0.15cm,
  #1]{
    \textbf{Oliver : }#2}
}
\NewDocumentCommand{\syl}{ O{} m }{%
  \todo[linecolor=orange,
        backgroundcolor=orange!25,
        bordercolor=purple,
        tickmarkheight=0.15cm,
        #1]{
          \textbf{Sylvain : }#2}
}

\newtheorem{openquestion}[thm]{Open Question}

\NewDocumentCommand{\clone}{O{H} O{u}}{\ensuremath{\mathsf{clone}(#1,#2)}}

\begin{document}

% If the title is longer than 55 characters, then specify a shorter running title as the optional argument to \title. The running title should be roughyl at most 55 characters:
\title[DA for CQ with Negations]{Direct Access for Conjunctive Queries with Negations}% {. }How to prepare papers
  % for LMCS using \texorpdfstring{\MakeLowercase{\texttt{lmcs.cls}}}{lmcs.cls}\rsuper*\\Version of
  % 2022-04-01}
\titlecomment{This paper is a longer version of~\cite{confversion} which appeared at ICDT 2024.}
\thanks{We would like to thank the anonymous reviewers of this paper for their numerous comments which helped improve the quality of this paper. This work was supported by project ANR KCODA, ANR-20-CE48-0004.}	%optional

% affiliations are numbered automatically with a, b, c (see below)
% use the optional argument to indicate the affiliation(s) of each author
% omit the argument if there is only one author, or only one affiliation
\author[F.~Capelli]{Florent Capelli\lmcsorcid{0000-0002-2842-8223}}[a]
\author[N.~Carmeli]{Nofar Carmeli\lmcsorcid{0000-0003-0673-5510}}[b]
\author[O.~Irwin]{Oliver Irwin\lmcsorcid{0000-0002-8986-1506}}[c]
\author[S.~Salvati]{Sylvain Salvati\lmcsorcid{0000-0002-6230-0098}}[c]

% affiliation 1 (automatically numbered a)
\address{Université d'Artois, CNRS, UMR 8188, Centre de Recherche en Informatique de Lens (CRIL), F-62300 Lens, France}	%optional
% write emails for all authors having that affiliation
\email{capelli@cril.fr}  %optional

% affiliation 2 (automatically numbered b)
\address{Inria, LIRMM, Université de Montpellier, CNRS, UMR 5506, F-34000 Montpellier, France}	%optional
\email{nofar.carmeli@inria.fr}  %optional

\address{Université de Lille, CNRS, Inria, UMR 9189 - CRIStAL, F-59000 Lille, France}	%optional
\email{oliver.irwin@univ-lille.fr, sylvain.salvati@univ-lille.fr}  %optional

%% etc.

%% required for running head on odd and even pages, use suitable
%% abbreviations in case of long titles and many authors:

%%%%%%%%%%%%%%%%%%%%%%%%%%%%%%%%%%%%%%%%%%%%%%%%%%%%%%%%%%%%%%%%%%%%%%%%%%%

%% the abstract has to PRECEDE the command \maketitle:
%% be sure not to issue the \maketitle command twice!

%abstract
\begin{abstract}
Given a conjunctive query $Q$ and a database $\db$, a direct access to the answers of $Q$ over $\db$ is the operation of returning, given an index $k$, the $k^{\mathsf{th}}$ answer for some order on its answers. While this problem is \#\P-hard in general with respect to combined complexity, many conjunctive queries have an underlying structure that allows for a direct access to their answers for some lexicographical ordering that takes polylogarithmic time in the size of the database after polynomial preprocessing time. Previous work has precisely characterised the tractable classes and given fine-grained lower bounds on the preprocessing time needed depending on the structure of the query.
In this paper, we generalise these tractability results to the case of signed conjunctive queries, that is, conjunctive queries that may contain negative atoms. Our technique is based on a class of circuits that can represent relational data. We first show that this class supports tractable direct access after a polynomial time preprocessing. We then give bounds on the size of the circuit needed to represent the answer set of signed conjunctive queries depending on their structure. Both results combined together allow us to prove the tractability of direct access for a large class of conjunctive queries. On the one hand, we recover the known tractable classes from the literature in the case of positive conjunctive queries. On the other hand, we generalise and unify known tractability results about negative conjunctive queries -- that is, queries having only negated atoms. In particular, we show that the class of $\beta$-acyclic negative conjunctive queries and the class of bounded nest set width negative conjunctive queries admit tractable direct access.
\end{abstract}

%%% Local Variables:
%%% mode: latex
%%% TeX-master: "main"
%%% End:

\maketitle

\allowdisplaybreaks

\section{Introduction}
\label{sec:introduction}

The \emph{direct access task}, given a database query $Q$ and a database $\db$, is the problem of outputing on input $k$, the $k$-th answer of $Q$ over $\db$ or an error when $k$ is greater than the number of answers of $Q$, where some order on $\ans$, the answers of $Q$ over $\db$, is assumed. This task was introduced by Bagan, Durand, Grandjean and Olive in~\cite{BDGO08} and is very natural in the context of databases. It can be used as a building block for many other interesting tasks such as counting, enumerating~\cite{BDGO08} or sampling without repetition~\cite{carmeli2020answering, KeppelerPhD} the answers of $Q$. Of course, if one has access to an ordered array containing $\ans$, answering direct access tasks simply consists in reading the right entry of the array. However, building such an array is often expensive, especially when the number of answers of $Q$ is large. Hence, a natural approach for solving this problem is to simulate this method by using a data structure to represent $\ans$ that still allows for efficient direct access tasks to be solved but that is cheaper to compute than the complete answer set. This approach is thus separated in two phases: a \emph{preprocessing phase} where the data structure is constructed followed by a phase where direct access tasks are solved. To measure the quality of an algorithm for solving direct access tasks, we hence separate the \emph{preprocessing time} -- that is the time needed for the preprocessing phase -- and the \emph{access time}, that is, the time needed to answer one direct access query after the preprocessing. For example, the approach consisting in building an indexed array for $\ans$ has a preprocessing time in at least the size of $\ans$ (and possibly higher) and constant access time (in the RAM model which will be made precise in \cref{sec:preliminaries}. While the access time is optimal in this case, the cost of preprocessing is often too high to pay in practice.

Previous work has consequently focused on devising methods with better preprocessing time while offering reasonable access time. In their seminal work~\cite{BDGO08}, Bagan, Durand, Grandjean and Olive give an algorithm for solving direct access tasks with linear preprocessing time and constant access time on a class of first order logic formulas and bounded degree databases. Bagan~\cite{BaganPhD} later studied the problem for monadic second order formulas over bounded treewidth databases. Another line of research has been to study classes of conjunctive queries that support efficient direct access over general databases. In~\cite{carmeli2020answering}, Carmeli, Zeevi, Berkholz, Kimelfeld, and Schweikardt prove that direct access tasks can be solved on  acyclic conjunctive queries  with linear preprocessing time and polylogarithmic access time for a well-chosen lexicographical order. The results are also generalised to the case of bounded fractional hypertree width queries, a number measuring how far a conjunctive query is from being acyclic. It generalises many results from the seminal paper by Yannakakis establishing the tractability of model checking on acyclic conjunctive queries~\cite{Yannakakis} to the tractability of counting the number of answers of conjunctive queries~\cite{pichler2013} having bounded hypertree width. This result was later improved by precisely characterising the lexicographical ordering allowing for this kind of complexity guarantees. Fine-grained characterisation of the complexity of answering direct access tasks on conjunctive queries, whose answers are assumed to be ordered using some lexographical order, has been given by Carmeli, Tziavelis, Gatterbauer, Kimelfeld and Riedewald in~\cite{Carmeli2023} for the special case of acyclic queries and by Bringmann, Carmeli and Mengel in~\cite{bringmann2022tight} for any join queries. More recently, Eldar, Carmeli and Kimelfeld~\cite{eldar2023direct} have studied the complexity of solving direct access tasks for conjunctive queries with aggregation.
% Related work; do not forget MSO stuff in Bagan's thesis, see paper with Stefan

In this paper, we devise new methods for solving direct access tasks on the answer set of \emph{signed conjunctive queries}, that is, conjunctive queries that may contain negated atoms. This is particularly challenging because only a few tractability results are known on signed conjunctive queries.
The model checking problem for signed conjunctive queries being $\NP$-hard on acyclic conjunctive queries with respect to combined complexity, % Citation?
it is not possible to directly build on the work cited in the last paragraph. Two classes of negative conjunctive queries (that is, conjunctive queries where every atom is negated) have been shown so far to support efficient model checking: the class of $\beta$-acyclic queries~\cite{OrdyniakPS13,brault2013pertinence} and the class of bounded nested-set width queries~\cite{lanzinger2023tractability}. The former has been shown to also support efficient (weighted) counting~\cite{BraultCM15, Capelli17}. Our main contribution is a generalisation of these results to direct access tasks. More precisely, we give an algorithm that efficiently solves direct access tasks on a large class of signed conjunctive queries, which contains in particular $\beta$-acyclic negative conjunctive queries, bounded nest-width negative conjunctive queries and bounded fractional hypertree width positive conjunctive query. For the latter case, the complexity we obtain is similar to the one presented in~\cite{bringmann2022tight} and we also get complexity guarantees depending on a lexicographical ordering that can be specified by the user. Hence our result both improves the understanding of the tractability of signed conjunctive queries and unifies the existing results with the positive case. In a nutshell, we prove that the complexity of solving direct access tasks for a lexicographical order of a signed conjunctive query $Q$ roughly matches the complexity proven in~\cite{bringmann2022tight} for the worst positive query we could construct  by removing some negative atoms of $Q$ and turning the others into positive atoms. 

As a byproduct, we introduce a new notion of hypergraph width based on elimination order, called the $\beta$-hyperorder width. It is a hereditary width notion, meaning that the width of every subhypergraph does not exceed the width of the original hypergraph. It makes it particularly well tailored for the study of the tractability of negative conjunctive queries. We show that this notion sits between nest-set width and $\beta$-hypertree width~\cite{gottlob2004hypergraphs}, but does not suffer from the main drawback of working with $\beta$-hypertree width: our width notion is based on a decomposition that works for every subhypergraph.

We give two different algorithms for solving direct access on signed queries: the first method uses the results on positive join queries from~\cite{bringmann2022tight} as an oracle. We show that it has optimal data complexity but has an exponential dependency on the size of the query. The second algorithm is based on a two-step preprocessing. Given a signed conjunctive query $Q$, a database $\db$ and an order $\prec$ on its variables, we start by constructing a circuit which represents $\ans$ in a factorised way, enjoying interesting syntactical properties. The size of this circuit depends on the order $\prec$ chosen on the variables of $Q$, some orders being harder than other and we are able to measure their complexity. We then show that with a second light preprocessing step on the circuit itself, we can answer direct access tasks on the circuit in time $\poly(n)\polylog(D)$ where $n$ is the number of variables of $Q$ and $D$ is the domain of $\db$. This approach is akin to the approach used in \emph{factorised databases}, introduced by Olteanu and Závodný~\cite{olteanu2012factorised}, a fruitful approach allowing efficient management of the answer sets of a query by working directly on a factorised representation of the answer set instead of working on the query itself~\cite{olteanu2016factorized,schleich2016learning,bakibayev2013aggregation,olteanu2015}.  However, the restrictions that we are considering in this paper are different from the ones used in previous work since we need to somehow account for the variable ordering in the circuit itself. The syntactic restrictions we use have already been considered in~\cite{Capelli17} where they are useful to deal with $\beta$-acyclic CNF formulas.

% - Tractability of aggregation on JQ/CQ
% - Direct access as an interesting framework [counting, random enumeration]. 
% - Well understood complexity for this kind of problem for *positive* CQ.
% - Far less is known for JQ/CQ having negated atoms: quick tour of the literature, CAPELLI15, link to \#SAT -- boolean domain -- and SZEIDER, BOVA15, CAPELLI17, LANZINGER
% - Our work aims at getting a better understanding on the tractability landscape of signed CQs. CONTRIBUTION: <showoff>
%   1. We leverage many known results for counting to direct access,
%   2. generalise structural parameter to handle negated atoms,
%   3. give a unifying framework
% </showoff>
% - Our technique relies on exploiting the structure of JQ to construct a factorized representation of Q(D) as relational circuits. This representation enjoys interesting syntaxical properties making the DA of the underlying circuit tractable. DPLL style procedure to construct it, only depends on an ordering of the variables of $Q$, which make it simpler than tree decompositions without compromising the tractability results. 

\paragraph{Difference with conference version.} This paper is a longer and improved version of~\cite{confversion}. The longer version contains every proof missing from~\cite{confversion} but also several new results. First, we improve the complexity of the preprocessing step by a $|\db|$ factor. This is achieved using a simple algorithmic trick during the computation of the circuit to avoid exploring every possible domain value when branching on a variable. This improvement allows us to match the complexity from~\cite{bringmann2022tight} when running our algorithm on a join query without negative atoms. Second, we match this improved upper bound with a lower bound, leveraging results from~\cite{bringmann2022tight}, which establishes the optimality of our approach. Indeed, we show that for every signed join query $Q$ and order $\prec$ on the variables of $Q$, there exists $k$ (depending on $Q$ and $\prec$) such that direct access on signed join queries is possible with preprocessing time $\bigot(|\db|^kf(|Q|))$ and direct access $\bigo(f(|Q|)\polylog(|\db|))$ but  if one is able to solve direct access for $Q$ with preprocessing time $\bigo(|\db|^{k-\varepsilon} g(|Q|) \polylog(|\db|))$ for some $g$ and $\varepsilon>0$, and access time $\bigo(\polylog(|\db|))$, then the Zero-Clique conjecture, a widely believed conjecture from fine-grained complexity, is false. 

\paragraph{Organisation of the paper.} The paper is organised as follows: \cref{sec:preliminaries} introduces the notations and concepts necessary to understand the paper. We give in \cref{sec:lowerbound} a reduction of the problem of solving direct access on signed join query to the problem of solving direct access on positive queries. It directly gives us a full characterisation of the tractable cases of signed join queries but with exponential query complexity. We then present the family of circuits we use to represent database relations and the direct access algorithm in \cref{sec:ocircuits}. \cref{sec:cqtocircuits} presents the algorithm used to construct a circuit representing $\ans$ from a join query $Q$ (that is a conjunctive query without existential quantifiers) and a database $\db$. Upper bounds on the size of the circuits produced are given in \cref{sec:compl-exha-dpll} using hypergraph decompositions defined in \cref{sec:hyperorder-width}. Finally \cref{sec:ra-scq} explicitly states the results we obtain by combining both techniques together and improves the combined complexity of the algorithm obtained in \cref{sec:lowerbound} by reduction. Moreover, it explains how one can go from join queries to conjunctive queries by existentially projecting variables directly in the circuit. Finally, \cref{sec:sat} studies the special cases of negative join queries and of SAT and compares our result to the existing literature on these topics.

%%% Local Variables:
%%% mode: latex
%%% TeX-master: "main"
%%% TeX-engine: luatex
%%% End:

\section{Preliminaries}
\label{sec:preliminaries}

\paragraph*{General mathematical notations.}
Given $n \in \N$, we denote by $[n]$ the set $\{1,\dots,n\}$. %
When writing down complexity, we use the notation $\poly(n)$ to denote that
the complexity is polynomial in $n$, $\poly_k(n)$ to denote that the
complexity is polynomial in $n$ when $k$ is considered a constant (in other
words, the coefficients and the degree of the polynomial may depend on $k$)
and $\polylog(n)$ to denote that the complexity is polynomial in
$\log(n)$. %
Moreover, we use $\bigot(n)$ to indicate that polylogarithmic factors are
ignored, that is, the complexity is $\bigo(n\polylog(n))$. %

\paragraph*{Tuples and relations.}

Let $D$ and $X$ be finite sets. %
A (named) \emph{tuple} on domain $D$ and variables set $X$ is a mapping from $X$ to
$D$. %
We denote by $D^X$ the set of all tuples on domain $D$ and variables set $X$. %
A \emph{relation $R$ on domain $D$ and variables set $X$} is a subset of tuples,
that is, $R \subseteq D^X$. We will denote the variables set $X$ of $R$ by $\var(R)$. %
The number of tuples in a relation \(R\), or the \emph{size of \(R\)} will
often be denoted by \(\#R\). %
Given a tuple $\tau \in D^X$ and $Y \subseteq X$, we denote by $\tau|_Y$ the tuple on domain
$D$ and variables set $Y$ such that $\tau|_Y(y) = \tau(y)$ for every $y \in
Y$. %
Given a variable $x \in X$ and $d \in D$, we denote by $[x \gets d]$ the tuple
on variables set $\{x\}$ that assigns the value $d \in D$ to $x$. %
We denote by $\emptytup$ the empty tuple, that is, the only element of
$D^\emptyset$. %
Given two tuples $\tau_1 \in D^{X_1}$ and $\tau_2 \in D^{X_2}$, we say that $\tau_1$ and
$\tau_2$ are \emph{compatible}, denoted by $\tau_1 \simeq \tau_2$, if
$\tau_1|_{X_1 \cap X_2} = \tau_2|_{X_1 \cap X_2}$. %
In this case, we denote by $\tau_1 \bowtie \tau_2$ the tuple on domain $D$ and variables set
$X_1 \cup X_2$ defined as
\[
  (\tau_1 \bowtie \tau_2)(x) =
  \begin{dcases*}
    \tau_1(x) \text{ if } x \in X_1 \\
    \tau_2(x) \text{ if } x \in X_2
  \end{dcases*}
\]
If $X_1 \cap X_2 = \emptyset$, we write $\tau_1 \times \tau_2$ instead of $\tau_1 \bowtie \tau_2$. %
The \emph{join $R_1 \bowtie R_2$ of $R_1$ and $R_2$}, for two relations $R_1$ and $R_2$ on
domain $D$ and variables set $X_1$ and $X_2$ respectively, is defined as
$\{ \tau_1 \bowtie \tau_2 \mid \tau_1 \in R_1, \tau_2 \in R_2, \tau_1 \simeq
\tau_2\}$. %
Observe that if $X_1 \cap X_2 = \emptyset$, $R_1 \bowtie R_2$ is simply the \emph{cartesian product
  of $R_1$ and $R_2$}. %
In this case, we denote it by $R_1 \times R_2$. %
The \emph{extended union of $R_1$ and $R_2$}, denoted by $R_1~\ecup~R_2$, is the
relation on domain $D$ and variables set $X_1 \cup X_2$ defined as
$(R_1 \times D^{X_2 \setminus X_1}) \cup (R_2 \times D^{X_1 \setminus
X_2})$. %
When $X_1 = X_2$, the extended union of $R_1$ and $R_2$ is simply
$R_1 \cup R_2$, that is, the set of tuples over $X_1$ that are either in
$R_1$ or in $R_2$. %

%\paragraph*{Extension of relations.}

Let $R \subseteq D^X$ be a relation from a variables set $X$ to a domain
$D$. %
We denote $\sigma_F(R)$ as the subset of $R$ where the formula $F$ is
true, where $F$ is a conjunction of atomic formulas of the form $x=d, x < d, x \leq d, x > d, x \geq d$ for a variable $x \in X$ and domain value $d \in D$. For example, $\sigma_{x \leq d}(R)$ contains every tuple $\tau \in R$ such that $\tau(x) \leq d$. 
Throughout the paper, we will assume that both the domain $D$ and the
variable set $X$ are totally ordered. %
The order on $D$ will be denoted as $<$ and the order on $X$ as $\prec$ and we
will often write $D = \{d_1, \dots, d_p\}$ with $d_1 < \dots < d_p$ and
$X = \{x_1, \dots, x_n\}$ with $x_1 \prec \dots \prec x_n$. %
For $Y \subseteq X$, we denote by $\min_{\prec}(Y)$ (or $\min(Y)$ when $\prec$ is clear from context), the minimal element of $Y$.%
Given $d \in D$, we denote by $\rank(d)$ the number of elements of $D$ that
are smaller or equal to $d$. %
Both $<$ and $\prec$ induce a lexicographical order $\lexprec$ on $D^X$ defined
as $\tau \lexprec \tau'$ if there exists $x \in X$ such that for every
$y \prec x$, $\tau(y)=\tau'(y)$ and $\tau(x) < \tau'(x)$. %
Given a integer $k \leqslant \#R$, we denote by $R[k]$ the
$k^{\mathsf{th}}$ tuple in $R$ for the $\lexprec$-order. %

% Add example

\begin{figure}[htp]
  \centering
  \includegraphics[scale=1.5]{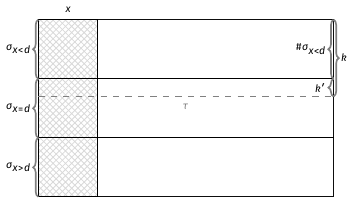}
  \caption[$k$ to $k'$]{Visual representation of
    \cref{lem:assignment-of-tau-x}. The rows represent tuples in increasing order and the
    columns represent variables. The value on $x$ of the $k^{\mathsf{th}}$ tuple in $R$, depicted by a dashed line, is the smallest value $d \in D$ such that $R$ contains more than $k$ tuples with a value smaller or equal to $d$. For a smaller $d$, we are below the dashed line because there are not enough tuple.}
  \label{fig:k_prime}
\end{figure}

We will often use the following observation:

\begin{restatable}{lem}{assignmentTauX}
  \label{lem:assignment-of-tau-x}
  Let $\tau = R[k]$ and $x = \mathsf{min}(\var(R))$. Then $\tau(x) =
  \mathsf{min}\{d \mid \#\sigma_{x \leqslant d}(R) \geqslant k\}$.
  Moreover, $\tau = R'[k']$, where $R'=\sigma_{x=d}(R)$ is the subset of $R$ where
  $x$ is equal to $d$ and $k' = k - \#\sigma_{x < d}(R)$.
\end{restatable}
\begin{proof}
  A visual representation of this proof and of the meaning of the statement
  is given in \cref{fig:k_prime}. %
  A complete proof can be found in \cref{app:assignment}. %
\end{proof}

\paragraph*{Queries.}
A \emph{(signed) join query $Q$} is an expression of the form \[
  Q \coloneq R_1(\tup[x_1]), \dots, R_\ell(\tup[x_\ell]), \neg S_{\ell+1}(\tup[x_{\ell+1}]), \dots, \neg S_{m}(\tup[x_{m}])
\]
where each $R_i$ and $S_j$ are relation symbols and $\tup[x_i]$ are tuples of
variables in $X$. %
In this paper, we consider \emph{self-join free} queries, that is, we assume
that any relation symbol appears at most once in each query. %
Elements of the form $R_i(\tup[x_i])$ are called positive atoms and elements
of the form $S_j(\tup[x_j])$ are called negative atoms. %
The set of variables of $Q$ is denoted by $\var(Q)$, the set of positive
(resp. negative) atoms of $Q$ is denoted by $\atomp$ (resp. $\atomn$). %
A \emph{positive join query} is a signed join query without negative
atoms. %
A \emph{negative join query} is a join query without positive atoms. %
The size $|Q|$ of $Q$ is defined as $\sum_{i=1}^m |\tup[x_i]|$, where $|\tup|$
denotes the number of variables in $\tup$. %
A \emph{database} $\db$ for $Q$ is an ordered finite set $D$ called the
\emph{domain} together with a set of relations $R_i^\db \subseteq D^{a_i}$,
$S_j^\db \subseteq D^{a_j}$ such that $a_i = |\tup[x_i]|$ and \(a_j =
\size{\tup[x_j]}\). %
The \emph{answers of $Q$ over $\db$} is the relation $\ans \subseteq D^{\var(Q)}$
defined as the set of $\sigma \in D^{\var(Q)}$ such that for every
$i \leqslant \ell$, $\sigma(\tup[x_i]) \in R_i^\db$ and for every \(\ell <
i \leq m\), $\sigma(\tup[x_i]) \notin S_i^\db$ where $\sigma(\tup[x_i])$ is defined as $(\sigma(x_i^1), \dots, \sigma(x_i^{a_i}))$ for $\tup[x_i] = (x_i^1, \dots, x_i^{a_i})$. %
The size $|\db|$ of the database $\db$ is defined to be the total number of
tuples in it plus the size of its domain\footnote{We follow the definition
  of~\cite{lanzinger2023tractability} concerning the size of the
  database. %
  Adding the size of the domain here is essential since we are dealing with
  negative atoms. %
  Hence a query may have answers even when the database is empty, for
  example the query $Q=\neg R(x)$ with $R^\db=\emptyset$ has $|D|$
  answers.}, that is,
$|D|+\sum_{i=1}^\ell|R_i^\db|+\sum_{i=\ell+1}^m|S_j^\db|$. %

A \emph{signed conjunctive query $Q(Y)$} is a signed join query $Q$ together
with $Y \subseteq \var(Q)$, called the \emph{free variables of $Q$} and
denoted by $\free(Q)$. %
The answers $\ans[Q(Y)]$ of a conjunctive query $Q$ over a database $D$ are
defined as $\ans|_Y$, that is, they are the projection over $Y$ of answers
of $Q$. %

\paragraph*{Direct access tasks.}

Given a query $Q$, a database instance $\db$ on ordered domain $D$ and a
total order $\prec$ on the variables of $Q$, a \emph{\da{}
  task}~\cite{Carmeli2023} is the problem of returning, on input $k$, the
$k$-th tuple $\ans{}[k]$ for the order $\lexprec$ if $k \leq \#\ans$ and
failing otherwise. %
We are interested in answering \da{} tasks using the same setting
as~\cite{Carmeli2023}: we allow a \emph{preprocessing} phase during which a
data structure is constructed, followed by an \emph{access} phase. %
Our goal is to obtain -- with a preprocessing time that is polynomial in
the size of $\db$ -- a data structure that can be used to answer any access
query in polylogarithmic time in the size of $\db$. %

\paragraph*{Hypergraphs.} A \emph{hypergraph} $H=(V, E)$ is defined as a set of \emph{vertices} $V$ and \emph{hyperedges} $E  \subseteq 2^V$, that is, a hyperedge $e \in E$ is a subset of $V$.

Let $H=(V,E)$ be a hypergraph. A \emph{subhypergraph} $H'$ of $H$, denoted by $H' \subseteq H$ is a hypergraph of the form $(V,E')$ with $E' \subseteq E$. In other words, a subhypergraph of $H$ is a hypergraph obtained by removing edges in $H$. For $S \subseteq V$, we denote by $H \setminus S$ the hypergraph $(V \setminus S, E')$ where $E' = \{e \setminus S \mid e \in E\}$. Given $v \in V$, we denote by $E(v) = \{e \in E \mid v \in e\}$ the set of edges containing $v$, by $\neigh[v][H] = \bigcup_{e \in E(v)} e$ the \emph{neighbourhood of $v$ in $H$} and by $\oneigh[v][H] = \neigh \setminus \{v\}$ the \emph{open neighbourhood of $v$}. We will be interested in the following vertex removal operation on $H$: given a vertex $v$ of $H$, we denote by $H/v = (V\setminus \{v\}, E/v)$ where $E/v$ is defined as $\{e \setminus \{v\} \mid e \in E\} \setminus \{\emptyset\} \cup \{\oneigh\}$, that is, $H/v$ is obtained from $H$ by removing $v$ from every edge of $H$ and by adding a new edge that contains the open neighbourhood of $v$.

Given sets $W$, $S \subseteq W$ and $K \subseteq 2^W$, a \emph{covering of $S$ with
  $K$} is a subset $F \subseteq K$ such that $S \subseteq \bigcup_{e \in F} e$. %
The \emph{cover number $\rho(S,K)$ of $S$ with respect to $K$} is defined as the minimal size
of a covering of $S$ with $K$, that is,
$\rho(S,K) = \min \{|F| \mid F \text{ is a covering of }S\text{ with }K\}$. %
A \emph{fractional covering of $S$ with $K$} is a function $c : K \rightarrow \Rp$ such
that for every $s \in S$, $\sum_{e \in K, s \in e} c(e) \geqslant 1$. %
Observe that a covering is a fractional covering where $c$ has values in
$\{0,1\}$. %
The \emph{fractional cover number $\rho^*(S,K)$ of $S$ wrt $K$} is defined as the
minimal size of a fractional covering of $S$ with $K$, that is,
$\rho^*(S,K) = \min \{ \sum_{e \in E} c(e) \mid c \text{ is a fractional covering of } S \text{ with
}K\}$. %

We will mostly use (fractional) covering for covering vertices of hypergraphs using a subset of its edges.  Fractional covers are particularly interesting because of the following theorem by Grohe and Marx: %

\begin{thmC}[\cite{grohe2014constraint}]
  \label{thm:agm} Let $Q$ be a join query and $\lambda$ be the fractional cover number of $\var(Q)$. Then for every database $\db$, $\ans$ has size at most $|\db|^\lambda$.
\end{thmC}
%Changed to thmC for Double Brackets

\paragraph{Signed hypergraphs.} A \emph{signed hypergraph} $H = (V, E_+, E_-)$ is defined as a set of \emph{vertices} $V$, \emph{positive edges} $E_+ \subseteq 2^V$ and \emph{negative edges} $E_- \subseteq 2^V$.  The \emph{signed hypergraph $H(Q)=(\var(Q), E_+, E_-)$ of a signed conjunctive query $Q(Y)$} is defined as the signed hypergraph whose vertex set is the variables of $Q$ such that $E_+ = \{\var(a) \mid a \text{ is a positive atom of } Q\}$ and $E_- = \{\var(a) \mid a \text{ is a negative atom of } Q\}$. We observe that when $Q$ is a positive query, $H(Q)$ corresponds to the usual definition of the hypergraph of a conjunctive query since $E_- = \emptyset$.

\paragraph*{Model of computation.} In this paper, we will always work in the word-RAM model of computation, with $\bigo(\log(n))$-bit words (where $n$ is the size of the input, which, in this paper, is often the size of the database) and unit-cost operations. The main consequences of this choice are the following: we can perform arithmetic operations on integers of size at most $n^k$ (hence encoded over $\bigo(k \log(n))$ bits) in time polynomial in $k$ only. In particular, it is known that addition can be performed in time $\bigo(k)$ (using the usual addition algorithm on numbers represented in base $\log(n)$) and both multiplication and division can be performed in time $\bigo(k \log(k))$ using Harvey and van der Hoeven algorithm~\cite{harvey2021integer}. We will heavily use this fact in the complexity analysis of our algorithm since we will need to manipulate integers representing the sizes of relations on domain $D$ and $k$ variables, hence, of size at most $|D|^k$. We will hence assume that every arithmetic operation here can be performed in time $\bigo(k\log(k))$. 

Moreover, we know that using radix sort, we can sort $m$ values whose total size is bounded by $n$ in time $O(m+n)$~\cite[Section 6.3]{cormen2022introduction}. We will mostly use this fact to sort relations of a database $\db$ in time $O(|\db|)$.

\section{An optimal yet inefficient algorithm}
\label{sec:lowerbound}

In this section, we reduce between the problem of answering direct access tasks
on join queries with negated atoms and the problem of answering direct access
tasks on join queries \emph{without negated atoms}. %
This will allow us to directly use results from~\cite{bringmann2022tight} and get,
for any given $Q$, an algorithm to answer direct access tasks with optimal data
complexity. %
Both the algorithm and its optimality will directly follow from the algorithm
and the lower bound of~\cite{bringmann2022tight}. %
The algorithm however has exponential complexity in the size of the query. %
We present in the next sections a direct and optimal algorithm with better
combined complexity. %

\subsection{From join queries to positive join queries}
\label{sec:reduction}

Here, we show that, in terms of data complexity, and if we ignore polylog
factors, \da{} for a signed query has the same complexity as \da{} for the
worst positive query obtained by considering some of the negated relations
to be positive. %

A signed join query \(Q\) is a join between two subqueries: the one
consisting of the positive atoms and the one consisting of the negated
atoms. %
With a slight abuse of notation, we denote this by \(Q = P \wedge N\), where
\(P\) is the join of the positive atoms and \(N\) the join of the negated
atoms. %

Let \(Q\) be a signed join query. For a set of relations \(N \subseteq \atomn[Q]\),
we denote by \(\overline{N}\) the same relations but considered as positive
atoms. Given a disjoint pair \((N_1, N_2)\) of subsets of \(\atomn[Q]\), we
denote by \qnn{} the query computed by considering all the relations in
\(N_1\) as if they were positive atoms, keeping the negated atoms of
\(N_2\) and removing all other negated atoms. That is, \(\qnn = P \wedge
\overline{N_1} \wedge N_2\). %

\begin{exa}
For $Q =  U(a,b) \wedge \neg R(x,y) \wedge \neg S(y,z) \wedge \neg T(z,x)$, $N_1 = \{\neg R(x,y)\}$ and $N_2 = \{\neg T(z,x)\}$, we have $Q_{N_1,N_2} = U(a,b) \wedge R(x,y) \wedge \neg T(z,x)$.
\end{exa}

\begin{figure}[htp]
  \centering
  \includegraphics[scale=1.5]{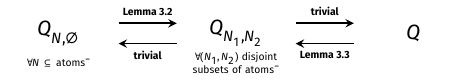}
  \caption{Relations between \da{} for signed queries and \da{} for
    positive queries}
  \label{fig:signed_to_pos}
\end{figure}

Our lower bound relies on the following strategy: we show that having direct access to $Q$ is equivalent to having direct access to $Q_{N,\emptyset}$ for every $N \subseteq \atomn$, which is a positive query and for which we already know lower bounds. To do so, we actually prove that this is equivalent to having a direct access to $Q_{N_1,N_2}$ for every disjoint $N_1,N_2 \subseteq \atomn$ by induction. The proof structure is summarized by \cref{fig:signed_to_pos}.

\begin{lem}[Subtraction Lemma]
  \label{lem:substraction}
  Let \(S\) be a set ordered by \(\prec\) and two subsets \(S_1, S_2\) of \(S\) such that \(S_1 \subseteq
  S_2\). %
  Assume that, for any value \(k\), we can output the \kth element of
  \(S_1\) (respectively of \(S_2\)) for the order \(\prec\) in time \(t_1\)
  (respectively in time \(t_2\)). %
  Then, given any \(k\), we can output the \kth element of \(S_2 \setminus S_1\) for
  the order \(\prec\) in time \(C \cdot (t_2 + t_1 \cdot \log\size{S_1}) \cdot
  \log\size{S_2}\) for some constant $C$. %
\end{lem}
\begin{proof}
  We define \emph{ranking} as the reverse problem to \da{}. That is, given a
  tuple \(\tau\) from an ordered set, finding its rank. %
  If \(\tau\) is not in the set, we return the rank of the largest smaller
  element or $0$ if $\tau$ is smaller than the first element of $S$. %
  We observe that having \da{} for a set of tuples \(S\) in time \(t\)
  implies having \emph{ranking} on \(S\) in time \(\bigo(\log\size{S} \cdot
  t)\). %
  This holds since we can simply do a binary search on \(S\) to find the
  correct rank for a given tuple. %

  Next, we claim that, given an element \(\tau\) of rank \(r_2\) in \(S_2\), we
  can find its rank in the subtraction of the sets \(S_2 \setminus S_1\) in
  time $\bigo(\log\size{S_1} \cdot t_1)$. %
  We use ranking for $\tau$ to find its rank $r_1$ in $S_1$. Then, we deduce
  that its rank in \(S_2 \setminus S_1\) is $r_2-r_1$. %
  Now, we can simply do a binary search over the ranks of $S_2$. %
  For each rank $r_2$, we access its element $\tau$ in time $t_2$, and check its
  ranking in $S_2\setminus S_1$ as described above. %
  This leads to \da{} for \(S_2 \setminus S_1\) in time \(\bigo((t_2 + t_1
  \cdot \log\size{S_1}) \cdot \log\size{S_2})\). %
  % The idea is to, given an element \(\tau \in S_2\), find out how many elements of \(S_1\) are smaller by the use of ranking. %
  % From there, we can deduce the index of \(\tau\) in the subtraction of the sets \(S_2 \setminus S_1\). %
  % The idea is to show that one can have direct access by looking for
  % the \(\frac{\size{S_2}}{2}\)-th element of \(S_2\) and finding out how
  % many answers \(n_{S_1}\) of \(S_1\) are smaller by the use of ranking. %
  % The index we currently have in \(Q_2 \setminus Q_1\) is thus \(\frac{\size{S_2}}{2}
  % - n_{S_1}\). %
  % Then it's a binary search until the correct index is found, leading to
  % \da{} for \(S_2 \setminus S_1\) in time \(\bigo((t_2 + t_1 \cdot \log\size{S_1}) \cdot
  % \log\size{S_2})\). %
\end{proof}

\begin{lem}
  \label{thm:da-from-subqueries}
  Let \(Q\) be a signed join query with \(n\) variables. %
  Assume that for all $N \subseteq \atomn[Q]$ we have \da{} for $Q_{N,\emptyset}$ with
  preprocessing time \(t_p\) and access time \(t_a\). %
  Then, we have \da{} to \qnn{} for any disjoint
  pair \(N_1, N_2\) of subsets of \(\atomn[Q]\) with preprocessing time \(2^{|\atomn[Q]|} \cdot t_p\) and
  access time \( (A n \cdot \log |D|)^{2|\atomn[Q]|} \cdot t_a\) for some constant $A$. %
\end{lem}
\begin{proof}
  % We prove this claim by showing that we have efficient \da{} to \qnn{} for
  % any disjoint pair \(N_1, N_2\) by induction over the size of \(N_2\). %
  The lemma follows by proving the following statement by induction over the size of \(N_2\): 
  for any subset \(N_1 \subseteq \atomn[Q]\) such that \(N_1\) and \(N_2\)
  are disjoint, we have \da{} for \qnn{} with preprocessing time
  \(2^{|N_2|} \cdot t_p\) and access time \((2Cn\cdot \log |D|)^{2|N_2|} \cdot t_a\) where $C$ is the constant from \cref{lem:substraction}. %
  The base case is when \(N_2 = \emptyset\), and it is given by the hypothesis of
  the lemma statement. %

  Now let us consider \(N_2\) to be a non-empty set. Since it is not empty,
  we can write it as \(N_2 = N_2' \wedge \lnot R(\tup)\), where $\lnot R(\tup)$ is one arbitrarily chosen atom from $N_2$. 
  From the induction hypothesis, since \(|N_2'| < |N_2|\), 
  for any subset \(N_1' \subseteq \atomn[Q]\) such that \(N_1' \cap N_2' =
  \emptyset\), we have \da{} for the query \(\qnn[N_1'][N_2']\) with
  preprocessing time \(2^{|N_2| - 1} \cdot t_p\) and access time
  \((2Cn \cdot \log |D|)^{2|N_2| - 1} \cdot t_a)\). %

  We can rewrite \qnn{} as:
  \begin{align*}
    \qnn &= P \wedge \overline{N_1} \wedge N_2' \wedge \lnot R(\tup) \\
         &= (P \wedge \overline{N_1} \wedge N_2') \setminus (P \wedge \overline{N_1} \wedge N_2' \wedge R(\tup)) \\
         &= \qnn[N_1][N_2'] \setminus \qnn[N_1 \cup \{R(\tup)\}][N_2']
  \end{align*}

  % Since \(N_2' \subset N_2\), we know that we have \da{} to the answers
  % of \qnn[N_1][N_2'] by the induction
  % hypothesis. % For all \(N_1' \subseteq N\), we have assumed to have \da{} to the answers of \qnn[N_1'][N_2']. %
  % Let \(N_1'' = N_1 \cup \{R\}\). %
  \noindent % FIXED Indent
  Notice that \((N_1 \cup \{R(\tup)\}, N_2')\) is a disjoint pair since \((N_1, N_2' \cup
  \{R(\tup)\})\) is a disjoint pair, and $R(\tup)$ does not appear in $N_2$ more
  than once since we consider self-join free queries. %
  We know from the induction hypothesis that we have direct access to the
  answers of both \qnn[N_1][N_2'] and \qnn[N_1 \cup \{R(\tup)\}][N_2']. %
  Moreover, since \(N_1 \subset N_1 \cup \{R(\tup)\}\) and adding positive atoms can only restrict
  the answer set of a query, we have that \(\ans[\qnn[N_1 \cup \{R(\tup)\}][N_2']]
  \subseteq \ans[\qnn[N_1][N_2']]\). %
  We apply \cref{lem:substraction} to get \da{} to the answers of
  \qnn[N_1][N_2]. Preprocessing consists in doing the preprocessing for $\qnn[N_1 \cup \{R(\tup)\}][N_2']$ and $\qnn[N_1][N_2']$. By induction, both of them takes \(2^{|N_2|-1} \cdot t_p\) hence a preprocessing time of \(2^{|N_2|} \cdot t_p\) time. For the access time, we apply \cref{lem:substraction} with $t_1=t_2=(2Cn \cdot \log |D|)^{2(|N_2|-1)} \cdot t_a$ and bound $S_1,S_2$ with $|D|^n$ and get:  $$2 \cdot (2Cn \cdot \log |D|)^{2(|N_2|-1)} \cdot t_a \cdot (Cn \cdot \log |D|)^2 \leq (2Cn \cdot \log |D|)^{2|N_2|} \cdot t_a$$
   This concludes our   induction step. %
\end{proof}

We show the lower bound with the following statement:

\begin{lem}
  \label{thm:sjq-lowerbound}
  Let \(Q\) be a signed join query with \(n\) variables and without
  self-joins. %
  Assume we have \da{} for \(Q\) with preprocessing time \(t_p\) and access
  time \(t_a\). %
  Then, for any disjoint pair \((N_1, N_2)\) of $\atomn$, we have \da{} to
  \(\qnn\) with preprocessing time \(2^{|\atomn[Q]|} \cdot t_p\) and
  access time \((An \cdot \log |D|)^{2|\atomn[Q]|} \cdot t_a\) for some constant $A$. %
\end{lem}
\begin{proof}
  We prove by induction over the size of \(N_1\) that, for any disjoint pair
  \((N_1, N_2)\), we have \da{} for \qnn{} with preprocessing time
  \(2^{|N_1|} \cdot t_p\) and access time \((2Cn \cdot \log |D|)^{2|N_1|} \cdot t_a\) where $C$ is the constant from \cref{lem:substraction}. %
  The base case is when \(N_1 = \emptyset\). %
  If \(N_2 = \atomn\), then since \(\qnn[\emptyset][\atomn] = Q\), it is given by
  the hypothesis of the theorem statement. %
  If \(N_2 \subset \atomn\), then we can use the algorithm of \(\qnn[\emptyset][\atomn]\) for
  answering \(Q\) by setting the negated relations that are not in $N_2$ to
  be empty. %
  Notice that we can do so freely because the query has no
  self-joins. %

  Now, let us consider \(N_1\) to be a non-empty set. %
  We can then denote it as \(N_1=N_1' \wedge \lnot R\). %
  By induction, we have \da{} for the answers of \qnn[N_1'][N_2] for every
  \(N_2\) disjoint from \(N_1\) with preprocessing time \(2^{|N_1|-1}
  \cdot t_p\) and access time \((2Cn \cdot \log |D|)^{2(|N_1|-1)} \cdot
  t_a\). %

  We can rewrite \qnn{} as :
  \begin{align*}
    \qnn &= P \wedge \overline{N_1} \wedge N_2 \\
         &= P \wedge \overline{N_1'} \wedge N_2 \wedge R \\
         &= (P \wedge \overline{N_1'} \wedge N_2) \setminus (P \wedge \overline{N_1'} \wedge N_2 \wedge \lnot R)\\
         &= \qnn[N_1'][N_2] \setminus \qnn[N_1'][N_2 \cup \{\lnot R\}]
  \end{align*}
  \noindent %FIXED Indent
  We can use the induction hypothesis on both \qnn[N_1'][N_2] and
  \qnn[N_1'][N_2 \cup \{\lnot R\}]. %
  Since having more negated atoms implies that the answer set is more
  restricted, we have that \(\ans[\qnn[N_1'][N_2 \cup \{\lnot R\}]]
  \subseteq \ans[\qnn[N_1'][N_2]]\). %
  We apply \cref{lem:substraction} to get direct access to the answers of
  \qnn[N_1][N_2]. %
  Preprocessing takes \(2\cdot 2^{|N_1|-1} \cdot t_p= 2^{|N_1|} \cdot t_p\)
  time. For the access time, we apply \cref{lem:substraction} by bounded $S_1$ and $S_2$ by $|D|^n$ and by using $t_1=t_2= (2Cn \cdot \log |D|)^{2(|N_1|-1)}$ that we have by induction. It gives a total  access time bounded by   \((2Cn \cdot \log |D|)^{2(|N_1|-1)} \cdot (Cn \log |D|)^2\) which  is bounded by \((2Cn \cdot \log |D|)^{2|N_1|}\) . This  concludes our induction step. %
\end{proof}

In the following, we want to transfer the knowledge we have about positive queries to
conclude the complexity of signed queries.

\begin{thm}
  \label{thm:qN}
  Let \(Q\) be a self-join free signed join query with \(n\) variables. %
  \begin{itemize}
      \item If \(Q\) has direct access with
  preprocessing time \(t_p\) and access time \(t_a\), then for all $N \subseteq \atomn[Q]$ we have \da{} for $Q_{N,\emptyset}$ with preprocessing time
\(2^{|\atomn[Q]|} \cdot t_p\) and
  access time \((An \cdot \log |D|)^{2|\atomn[Q]|} \cdot t_a\).
    \item If for all $N \subseteq \atomn[Q]$ we have \da{} for $Q_{N,\emptyset}$ with
  preprocessing time \(t_p\) and access time \(t_a\), then \(Q\) has direct access with preprocessing time \(2^{|\atomn[Q]|} \cdot t_p\) and
  access time \((An \cdot \log |D|)^{2|\atomn[Q]|} \cdot t_a\),
\end{itemize}
where $A$ is a constant.
\end{thm}
\begin{proof}
The first part of the statement follows from applying \cref{thm:sjq-lowerbound} with $N_2 = \emptyset$. The second part follows from \cref{thm:da-from-subqueries} applied with $N_1 = \emptyset$ and $N_2 = \atomn$ since $Q_{\emptyset, \atomn} = Q$. 

\end{proof}

%%% Local Variables:
%%% mode: LaTeX
%%% TeX-master: "main"
%%% End:

\subsection{Optimally solving signed join queries}
\label{sec:optimally-solving-sjq}

\cref{thm:qN} suggests that the (data) complexity of answering direct access tasks on a signed join query $Q$ is as hard as the hardest positive query that can be obtained by keeping only a subset of negative atoms and turning them positive. The complexity of answering direct access tasks on positive join queries is well understood from~\cite{bringmann2022tight}. In that paper, the following is shown:

\begin{thmC}[\cite{bringmann2022tight}, Theorem 44]
  \label{thm:complexity-posqueries} There exists a polynomial $p$ such that for every self-join free positive join query $Q$ (considered constant) on variables set $X$ and an order $\prec$ on $X$, there exists $\iota(Q,\prec) \in \mathbb{Q}_+$ such that:
  \begin{itemize}
  \item  for every database $\db$, direct access tasks for $\ans$ with order $\lexprec$ can be answered with preprocessing time $\bigo((p(|Q|) \cdot |\db|)^{\iota(Q,\prec)})$ and access time $\bigo(p(|Q|) \cdot \log(|\db|))$. 
  \item if there exists $f \from \N \to \N$ and $\varepsilon > 0$ such that for every database $\db$ and constant $\delta > 0$, direct access tasks for $\ans$ with order $\lexprec$ can be solved with preprocessing time $\bigo(f(|Q|) \cdot |\db|^{\iota(Q,\prec)-\varepsilon})$  and access time $\bigo(f(|Q|) \cdot |\db|^\delta)$, then the Zero-Clique Conjecture is false. 
  \end{itemize}
\end{thmC}

\noindent %FIXED Indent
The value $\iota(Q,\prec)$ is a function that only depends on $Q$ and the chosen order and characterizes the optimal preprocessing time needed. We do not need the precise definition of this function yet as only its existence is necessary for the lower bound. More details on this parameter will be presented in \cref{sec:hyperorder-width}. The Zero-Clique Conjecture is a widely believed conjecture in the domain of fine-grained complexity saying that for every $k$ and $\varepsilon>0$, there is no randomized algorithm with complexity $\bigo(n^{k-\varepsilon})$ able to decide whether there exists a $k$-clique whose edge weights sum to $0$ in a edge-weighted graph. In other words, unless such algorithm exists, the best preprocessing one can hope for to solve direct access tasks of positive join queries is $\bigo(|\db|^{\iota(Q,\prec)})$. In \cite{bringmann2022tight}, complexities in \cref{thm:complexity-posqueries} are given in terms of data complexity but it is not hard to see that the dependency on $|Q|$ is polynomial for constant $\iota(Q,\prec)$. 

We now extend the definition of $\iota$ to signed join queries as follows, inspired by \cref{thm:qN}: let $Q = P \wedge N$ be a signed join query, where $P$ are the positive atoms of $Q$ and $N$ the negative atoms of $Q$. Let $X$ be the variables set of $Q$  and $\prec$ an order on $X$. We define $\iota^s(Q,\prec) = \max_{N \subseteq \atomn} \iota(Q_{N,\emptyset},\prec)$, where $\iota(\cdot,\cdot)$ is the incompatibility number defined in~\cite{bringmann2022tight}. Observe that by definition, if $Q$ is a positive join query, then $\iota(Q,\prec) = \iota^s(Q,\prec)$. Moreover, \cref{thm:qN,thm:complexity-posqueries} directly imply this generalisation of \cref{thm:complexity-posqueries} to signed join queries:

\begin{thm}
  \label{thm:scq-optimal}  There exists a function $g \from \N \to \N$ such that, given a self-join free signed join query $Q$ on variables set $X$ and an order $\prec$ on $X$, we have:
  \begin{itemize}
  \item  for every database $\db$, direct access tasks for $\ans$ and order $\lexprec$ can be answered with preprocessing time $g(|Q|) \cdot |\db|^{\iota^s(Q,\prec)}$ and access time $g(|Q|) \cdot \log^{2|\atomn[Q]|+1}(|\db|)$.
  \item  if there exists $f \from \N \to \N$ and $\varepsilon > 0$ such that for every database $\db$ and constant $\delta > 0$, direct access tasks for $\ans$ with order $\lexprec$ can be solved with preprocessing time $\bigo(f(|Q|) \cdot |\db|^{\iota^s(Q,\prec)-\varepsilon})$  and access time $\bigo(f(|Q|) \cdot |\db|^\delta)$, then the Zero-Clique Conjecture is false. 
  \end{itemize}
\end{thm}
\begin{proof}
  Let $Q$ be a signed join query, $m$ be the number of negative atoms of $Q$ and let $k=\iota^s(Q,\prec)$. Moreover, let $p$ be the polynomial from \cref{thm:complexity-posqueries}. By definition, we have that for every $N \subseteq \atomn$, $\iota(Q_{N,\emptyset}, \prec) \leq k$. Hence by \cref{thm:complexity-posqueries}, we can solve the direct access task for  $\ans[Q_{N,\emptyset}]$ with order $\lexprec$ with preprocessing time $\bigo((p(|Q|) \cdot |\db|)^k)$ and access time $\bigo(p(|Q|) \cdot \log(|\db|))$. By \cref{thm:da-from-subqueries}, we then have direct access for $Q$ on $\db$ with preprocessing time $\bigo(2^{|\atomn[Q]|} (p(|Q|)|\db|)^k)$ and access time $\bigo(p(|Q|) \cdot |X| \cdot (A|X|)^{2|\atomn[Q]|} \cdot \log^{2|\atomn[Q]|+1}(|\db|))$. Observe that $A$ is a constant and $|X| \leq |Q|$, hence we can define $g(m) = K (p(m) \cdot m \cdot A|X|^{2m})$ for some large enough constant $K$ and we have the desired bound: the preprocessing time can be done in $g(|Q|) |\db|^k$ and access time $g(|Q|) \cdot \log^{2|\atomn|+1}(|\db|)$

  Now assume that there exists $f \from \N \to \N$ and $\varepsilon > 0$ such that for every $\delta' > 0$, we have a direct access scheme for $Q$ with preprocessing time $\bigo(f(|Q|) \cdot |\db|^{k-\varepsilon})$ and access time $\bigo(f(|Q|) \cdot |\db|^{\delta'})$. Let  $N \subseteq \atomn$ be such that $\iota(Q_{N,\emptyset},\prec)=k$.
%  Observe that $Q_{N,\emptyset}$ is a positive query maximizing $N' \mapsto \iota(Q_{N',\emptyset},\prec)$.
Let $\delta > 0$ and set $\delta' = \delta/2$. By \cref{thm:sjq-lowerbound}, we have direct access to $\ans[Q_{N,\emptyset}]$ with preprocessing time $\bigo(2^{|\atomn[Q]|} f(|Q|) \cdot |\db|^{k-\varepsilon}) = \bigo(h(|Q_{N,\emptyset}|) \cdot |\db|^{k-\varepsilon})$  and access time $\bigo(f(|Q|) \cdot \log^{2|\atomn[Q]|+1}(|\db|) |\db|^{\delta \over 2}) = \bigo(h(|Q_{N,\emptyset}|) \cdot |\db|^\delta)$ for some $h \colon \N \to \N$. But then by \cref{thm:complexity-posqueries}, we have an algorithm for $Q_{N,\emptyset}$ whose complexity implies that the Zero-Clique Conjecture is false.  
\end{proof}

Observe that the complexity obtained to preprocess a signed query has an exponential dependency on the size of the query since it is obtained using \cref{thm:da-from-subqueries} which intuitively does a direct access preprocessing for every subquery of $Q$. This exponential dependency in $|Q|$ is not present for positive queries in~\cite{bringmann2022tight}. \cref{sec:ocircuits,sec:cqtocircuits} are dedicated to design a better algorithm, achieving the same data complexity as in~\cref{thm:scq-optimal} but with a better dependency on $|Q|$ and a more direct algorithm.

Let us conclude this section with a final observation regarding \cref{thm:scq-optimal}. It is proven in \cite{bringmann2022tight} that \cref{thm:complexity-posqueries} also holds when $Q$ contains self-joins. This is not true for signed join queries which is why we explicitly assumed $Q$ to be self-join free. Indeed, for any join query $Q$ containing an atom $R(\vec{x})$ and order $\prec$, the signed query $Q' = Q \wedge \neg R(\vec{x})$ is such that $\iota^s(Q,\prec) = \iota^s(Q',\prec)$ but is not self-join free. That said, we clearly have $\ans[Q'] = \emptyset$ for every database $\db$, hence direct access for $Q'$ can easily be answered with preprocessing and access time independent on $|\db|$, which shows that the lower bound from \cref{thm:scq-optimal} does not hold for signed join queries with self-joins (the upper bound holds since we can always consider each occurrence of a relation as a unique relation). It is an interesting research direction to understand the complexity of signed conjunctive queries with self-join that we leave open for future work. 

\subsection{Hypergraph decompositions and incompatibility numbers}
\label{sec:hyperorder-width}

In this section, we give the definition of the incompatibility number $\iota(\cdot,\cdot)$ from \cite{bringmann2022tight} and connect $\iota$ and $\iota^s$ with existing and new hypergraph decomposition techniques. 

%The link between the incompatibility number of $Q$ and fractional hypertree width of $Q$ has already been observed in \cite{bringmann2022tight}. In this section, we revisit this connection and extend this analysis to $\iota^s$ and signed hypergraphs. We show that $\iota^s$ can be framed as a hypergraph measure that we call fractional signed-hyperorder width and that the case where $\iota^s$ is $1$ corresponds to the notion of signed acyclicity. We also compare this new hypergraph measure to existing hypergraph measures such as nest set width.

\paragraph{Elimination orders.} Given a hypergraph $H=(V,E)$ and an order $\prec$ such that $V = \{v_1, \dots, v_n\}$ with $v_1 \prec \dots \prec v_n$, we define a series of hypergraphs as $H^\prec_1, \dots, H^\prec_{n+1}$ defined as follows: $H^\prec_1 = H$ and $H^\prec_{i+1} = H^\prec_i/v_i$. The \emph{hyperorder width of the elimination order $\prec$ for $H$, denoted by  $\how{H}[\prec]$ } is defined as $\max_{i\leqslant n} \rho(\neigh[v_i][H^\prec_i], E)$. The \emph{hyperorder width $\how{H}$ of $H$} is defined as the best possible width using any elimination order, that is, $\how{H}=\min_\prec \how{H}[\prec]$. We similarly define the \emph{fractional hyperorder width of $\prec$ for $H$, denoted by $\fhow{H}[\prec]$} as $\max_{i\leqslant n} \rho^*(\neigh[v_i][H^\prec_i], E)$ and the \emph{fractional hyperorder width $\fhow{H}$ of $H$} as $\fhow{H}=\min_\prec \fhow{H}[\prec]$.

The definition of fractional hyperorder width of an elimination order $\prec$ matches the definition of the $\iota(Q,\succ)$ from \cite{bringmann2022tight} and used in \cref{thm:complexity-posqueries,thm:scq-optimal}. However, the incompatibility number is stated for the reversed of the elimination order. In this paper, we choose to make the connection with the existing literature on elimination order for hypergraphs more explicit and hence use the reversed order. We hence have:

\begin{thm}
  \label{thm:iota-is-fhow}
  For every positive join query $Q$ on variables set $X$ and order $\prec$ on $X$, it holds that $\iota(Q,\succ) = \fhow{H(Q)}[\prec]$. 
\end{thm}

\noindent
It has already been observed many times (\cite[Appendix C]{abo2015faq_arxiv} or \cite{fichte2018smt, ganian2022threshold, abo2016faq} but also in Proposition 14 of \cite{bringmann2022tight}, or via the notion of static width in~\cite{kara2025conjunctive}) that $\how{H}$ and $\fhow{H}$ are respectively equal to the generalised hypertree width and the fractional hypertree width of $H$ and that there is a natural correspondence between a tree decomposition and an elimination order having the same width. However, to be able to express our tractability results as a function of the order, it is more practical to define the width of orders instead of hypertree decompositions\footnote{Strictly speaking, the definition of $\how{\cdot}$ and $\fhow{\cdot}$ in \cite{abo2016faq} differ slightly in that the elimination step of $v$ removes every edge containing $v$ and replace it by the neighborhood of $v$, where in our definition, we keep them. This does not change the notion of neighborhood at each step so it results in the same widths but with a slightly different definition.}.

The case $\iota(Q,\succ)=1$ hence corresponds to $\fhow{H(Q)}[\prec] = 1$, that is, $H(Q)$ is $\alpha$-acyclic and $\prec$ is an $\alpha$-elimination order witnessing this acyclicity, as defined in~\cite{brault2016hypergraph}. (Reversed) orders witnessing this has been also previously called orders without disruptive trio~\cite{Carmeli2023} which exactly corresponds to the classical notion of elimination ordering witnessing $\alpha$-acyclicity (see~\cite{brault2016hypergraph} for a survey, where they are called $\alpha$-elimination order). In this paper, we will now mostly use a terminology based on hypergraph decomposition of the query rather than incompatibility number since we also aim at comparing our results with other results on negative join queries stated in terms of hypergraphs.

\paragraph*{Signed hyperorder width.} In the case of signed join queries, we can naturally generalise the notion of hyperorder width so that it matches our extended definition of incompatibility number of signed join query $\iota^s(Q,\succ)$.

% one can deal with positive and negative atoms differently, which is not reflected by the definition of $\bfhow{\cdot}$. We generalise these widths to signed hypergraphs by maximising over every subhypergraphs obtained by removing negative atoms, generalising a notion of acyclicity introduced by Brault-Baron in~\cite{brault2013pertinence} that mixes $\beta$- and $\alpha$-acyclicities for signed hypergraphs.

Let $H=(V,E_+,E_-)$ be a signed hypergraph. We say that $H'$ is \emph{a negative subhypergraph of $H$} and write $H' \subseteq^- H$ if it is a hypergraph of the form $H' = (V,E_+\cup E')$ for some $E' \subseteq E_-$. Given an order $\prec$ on $V$, we define:

\begin{itemize}
\item the \emph{signed hyperorder width $\show{H}[\prec]$ of $\prec$ for $H$} as $$\show{H}[\prec] = \max_{H' \subseteq^- H} \how{H}[\prec]$$ and the corresponding \emph{signed hyperorder width $\show{H}$ of $H$}, defined as $\show{H} = \min_{\prec} \show{H}[\prec]$,
  \item the \emph{fractional signed hyperorder width $\sfhow{H}[\prec]$ of $\prec$ for $H$} as $$\sfhow{H}[\prec] = \max_{H' \subseteq^- H} \fhow{H}[\prec]$$ and the corresponding \emph{fractional signed hyperorder width $\sfhow{H}$ of $H$} defined as $\sfhow{H} = \min_{\prec} \sfhow{H}[\prec]$.
\end{itemize}
\noindent %FIXED Indent
By definition and by \cref{thm:iota-is-fhow}, we immediately have:

\begin{thm}
  For every signed join query $Q$ on variables set $X$ and $\prec$ an order on $X$, we have $\iota^s(Q, \succ) = \sfhow{H(Q)}[\prec]$. 
\end{thm}

\noindent
From now on, we will be working with hypergraph measures instead of incompatibility number when dealing with join queries. For future comparison, let us restate \cref{thm:scq-optimal} in terms of hypergraph measure:

\begin{thm}
  \label{thm:re-scq-optimal}  There exists a polynomial $p$ such that given a self-join free signed join query $Q$ on variables set $X$ and an order $\prec$ on $X$, we have:
  \begin{itemize}
  \item  for every database $\db$, direct access tasks for $\ans$ and order $\lexprec$ can be answered with preprocessing time $\bigo(2^{|\atomn[Q]|}(p(|Q|) \cdot |\db|)^{\sfhow{H(Q)}[\succ]})$ and access time $\bigo(p(|Q|) \cdot |X| \cdot \log^{2|\atomn[Q]|+1}(|\db|))$.
  \item  if there exists $f \from \N \to \N$ and $\varepsilon >  0$ such that for every database $\db$ and constant $\delta > 0$, direct access tasks for $\ans$ with order $\lexprec$ can be solved with preprocessing time $\bigo(f(|Q|) \cdot |\db|^{\sfhow{H(Q)}[\succ]-\varepsilon})$  and access time $\bigo(f(|Q|) \cdot |\db|^\delta)$, then the Zero-Clique Conjecture is false. % removed space behind comma
  \end{itemize}
\end{thm}

\noindent %FIXED Indent
Observe again that the dependency on $|Q|$ in the preprocessing phase given by \cref{thm:re-scq-optimal} is exponential. In the next two sections, we propose another, more direct, approach for solving direct access on signed conjunctive queries which also reduces the combined complexity to a polynomial in both $|Q|$ and $|\db|$ when parametrizing by the non-fractional version of hyperorder width of $Q$ (that is $\show{H(Q)}[\prec]$). Moreover, the algorithm presented in the next two sections has a better access time than the one we obtained in \cref{thm:re-scq-optimal}.

%%% Local Variables:
%%% mode: latex
%%% TeX-master: "main"
%%% End:

\section{Direct Access and Ordered relational circuits}
\label{sec:ocircuits}

In this section, we introduce a data structure that can be used to
succinctly represent relations. This data structure is an example of
factorised representation, such as d-representations~\cite{olteanu2015},
but does not need to be structured along a tree, which will allow us to
handle more queries, and especially queries with negative atoms -- for
example $\beta$-acyclic signed conjunctive queries, a class of queries that
cannot be represented by polynomial size d-representations~\cite[Theorem
9]{Capelli17}.

%\subsection{Definitions}

\subsection{Relational circuits.}

A \jcircuit{} $C$ on variables set $X = \{x_1, \dots, x_n\}$ and domain $D$ is 
a multi-directed\footnote{That is, there may be more than one
  edge between two nodes $u$ and $v$.} acyclic graph (DAG), where we refer to the vertices as gates. For a given gate $g$, every $g'$ with a directed edge $g' \rightarrow g$ will be called \emph{an input of $g$}. We denote by $\mathsf{input}(g)$ the set of inputs of $g$. The circuit has one distinguished gate $\out(C)$ called the \emph{output} of $C$.  Moreover, the circuit is labelled as follows:

\begin{itemize}
\item every gate of $C$ with no ingoing edge, called an \emph{input of
    $C$}, is labelled by either $\bot$ or
  $\top$; % decision nodes oriented toward the variables; not classical but %
       % preserves the "circuit" way of doing it.
\item a gate $v$ labelled by a variable $x \in X$ is called a
  \emph{decision gate}. Each ingoing edge $e$ of $v$ is labelled by a value
  $d \in D$ and for each $d \in D$, there is at most one ingoing edge of
  $v$ labelled by $d$. This implies that a decision gate has at most $|D|$
  ingoing edges; and
\item every other gate is labelled by $\bowtie$.
\end{itemize}
\noindent %FIXED Indent
The set of all the decision gates in a circuit $C$ is denoted by
$\mathsf{decision}(C)$. Given a gate $v$ of $C$, we denote by $C_v$ the
\emph{subcircuit of $C$ rooted in $v$} to be the circuit whose output is $v$ and which contains every gate $g$ such that there is a directed path from $g$ to $v$. We define
the \emph{variable set of $v$}, denoted by $\var(v) \subseteq X$, to be the
set of variables $x$ labelling a decision gate in $C_v$. The variable
labeling a decision gate $v$ is denoted by $\decvar(v)$. The
\emph{size} $|C|$ of a \jcircuit{} is defined to be the number of edges of
its underlying DAG.

The intended semantics of a \jcircuit{} is that each gate of the circuit computes a relation $\rel(v)$ on variables set $\var(v)$. Intuitively, join nodes computes the joins of the relations computed by its input and decision nodes computes the union of the relation computed by its input with an additional variable $x$ whose value depends on the label of the edge between the decision node and its input.

More formally, we define the \emph{relation $\rel(v) \subseteq D^{\var(v)}$ computed at
  gate $v$} inductively as follows: if $v$ is an input labelled by $\bot$,
then $\rel(v) = \emptyset$. If $v$ is an input labelled by $\top$, then
$\rel(v) = D^\emptyset$, that is, $\rel(v)$ is the relation containing only
the empty tuple. Otherwise, let $v_1, \dots, v_k$ be the inputs of $v$. If
$v$ is a $\bowtie$-gate, then $\rel(v)$ is defined to be $\rel(v_1) \bowtie
\dots \bowtie \rel(v_k)$. If $v$ is a decision gate labelled by a variable
$x$, $\rel(v) = R_1 \cup \dots \cup R_k$ where $R_i = [x \gets d_i] \bowtie \rel(v_i) \bowtie D^{\var(v) \setminus (\var(v_i) \cup \{x\})}$ and $d_i$ is the label of the incoming edge $(v_i,v)$. It is readily verified that $\rel(v)$ is a
relation on domain $D$ and variables set $\var(v)$. The \emph{relation computed
  by $C$ over a set of variables $X$} (assuming $\var(C) \subseteq X$),
denoted by $\rel_X(C)$, is defined to be $\rel(\out(C)) \times D^{X
  \setminus \var(\out(C))}$.

To ease notation, we use the following convention: if $v$ is a decision-gate and $d \in D$, we denote by $v_d$ the gate of $C$ that is connected to $v$ by an edge $(v_d,v)$ labelled by $d$. 

Deciding whether the relation computed by a \jcircuit{} is non-empty
is $\NP$-hard by a straightforward reduction from model checking of
conjunctive queries~\cite{chandra77}. Such circuits are hence of
little use to get tractability results. We are therefore more
interested in the following restriction of \jcircuit{s}: a \dcircuit{}
$C$ is a \jcircuit{} such that: (i) for every $\bowtie$-gate $v$ of
$C$ with inputs $v_1,\dots,v_k$ and $i < j \leqslant k$, it holds that
$\var(v_i) \cap \var(v_j) = \emptyset$, (ii) for every decision gate
$v$ of $C$ labelled by $x$ with inputs $v_1,\dots,v_k$ and
$i \leqslant k$, it holds that $x \notin \var(v_i)$. This restriction
is akin to the one found in the literature of knowledge compilation
and \dcircuit{s} can actually be seen as a generalization of
decision-DNNF to non-Boolean domains~\cite{kcmap}. Cartesian product corresponds to
decomposable $\wedge$-gates. Checking whether the relation computed by
a \dcircuit{} $C$ is non-empty can be done in time $\bigo(|C|)$ by a
dynamic programming algorithm propagating in a bottom-up fashion
whether $\rel(v)$ is empty.  Similarly, given a \dcircuit{} $C$, one
can compute the size of $\rel(C)$ in polynomial time in $|C|$ by a
dynamic programming algorithm propagating in a bottom-up fashion
$|\rel(v)|$. Proof of this fact can straightforwardly be adapted from existing literature on restricted Boolean circuits~\cite{kcmap}.  

\begin{figure}[ht]
  \centering
  \includegraphics[scale=0.75]{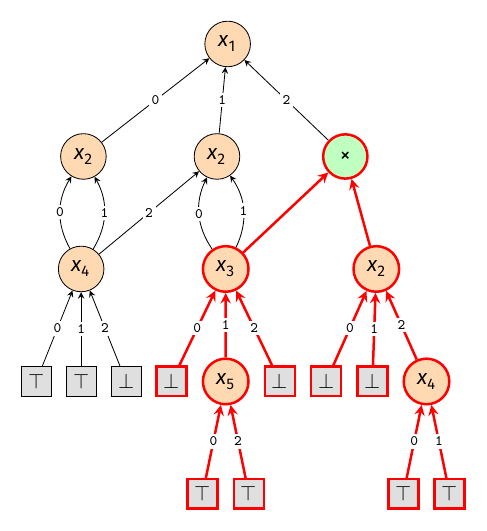}
  \caption[Example of a simple \ocircuit{}]{Example of a
    \dcircuit{}. The domain used is $\{0, 1, 2\}$ and the variable set is
    $\{x_1,x_2,x_3,x_4\}$.}
  \label{fig:circuit1}
\end{figure}

\begin{exa}
  \cref{fig:circuit1} shows an example of a \dcircuit{} on variables $\{x_1,\dots,x_5\}$ and domain $\{0,1,2\}$. Observe that the Cartesian product has variables $\{x_3,x_5\}$ on its left side and variables $\{x_2,x_4\}$ on its right side, which are indeed disjoint variable sets. The relation computed by the circuits includes for example the tuples $\langle x_1\gets 2, x_2 \gets 2, x_3 \gets 1, x_4 \gets 0, x_5 \gets 0 \rangle$, $\langle x_1\gets 0, x_2 \gets 0, x_3 \gets 0, x_4 \gets 0, x_5 \gets 0 \rangle$, but no tuple setting both $x_1$ to $0$ and $x_2$ to $2$. Indeed, every tuple with $x_1 \gets 0$ in the relation computed by $C$ must be extended by a tuple computed by the leftmost decision gate on $x_2$ but this decision gate has no edge labeled with $2$.

  If $v$ is the only Cartesian product in \cref{fig:circuit1}, one can observe that $C_v$ is the circuit highlighted with thick red border and $\rel(v)$ is the relation over variables $\{x_2,x_3,x_4,x_5\}$ and  domain $\{0,1,2\}$ containing tuples $\langle x_2 \gets 2, x_3 \gets 1, x_4 \gets 0, x_5 \gets 0 \rangle, \langle x_2 \gets 2, x_3 \gets 1, x_4 \gets 0, x_5 \gets 2 \rangle, \langle x_2 \gets 2, x_3 \gets 1, x_4 \gets 1, x_5 \gets 0 \rangle, \langle x_2 \gets 2, x_3 \gets 1, x_4 \gets 1, x_5 \gets 2 \rangle$.
\end{exa}

% A \dcircuit{} $C$ is said to be \emph{complete} if for every decision gate $v$
% on variable $x$ with inputs $v_1, \dots, v_k$, we have that $\var(v_i) =
% \var(v) \setminus \{x\}$ for every $i \leqslant k$. Observe that it implies
% that $\rel(v)$ is defined from $\rel(v_i)$ as $([x \gets \ell(e_1)] \times
% \rel(v_1)\big) \cup \dots \cup \big([x \gets \ell(e_k)] \times
% \rel(v_k)\big)$, that is, $\ecup$ has been replaced by $\cup$ since we are now
% guaranteed that $\rel(v_i)$ and $\rel(v_j)$ are defined on the same variable
% set. Moreover, $C$ is said to be \emph{complete with respect to $X$} if it is
% complete and if $\var(C) = X$.

\paragraph*{Ordered Relational Circuits.}

Let $X$ be a set of variables and $\prec$ an order on $X$. We say that a
\dcircuit{} $C$ on domain $D$ and variables set $X$ is a
\emph{$\prec$-\ocircuit{}} if for every decision gate $v$ of $C$ labelled
with $x \in X$, it holds that for every $y \in \var(v) \setminus \{x\}$, $x
\prec y$. We simply say that a circuit $C$ is an \ocircuit{} if there exists
some order $\prec$ on $X$ such that $C$ is a $\prec$-\ocircuit{}.
Observe that the example from \cref{fig:circuit1} is ordered for the order $(x_1,x_2,x_3,x_4,x_5)$. Observe that, according to the definition, nothing prevents variables under a Cartesian product to be interleaved as depicted on the example: $x_2$ and $x_4$ are on one side while $x_3$ and $x_5$ are on the other. 

% Todo: example and figures

\subsection{Direct access for ordered relational circuits}
\label{sec:ra-ocircuits}

Given a relational circuit $C$, we can define a notion of direct access
tasks for circuits as follows: given a relational circuit $C$ on variables
$X$ and an order $\prec$ on $X$, an algorithm solves direct access for $C$ for
order $\lexprec$ with precomputation time $t_p$ and access time $t_a$ if
there exists an algorithm that runs in time $t_p$ which constructs a
data structure $C'$ and then, using $C'$, there is an algorithm that runs in
time $t_a$ such that, on input $k$, returns the $k^{th}$ tuple of $\rel(C)$
for the order $\lexprec$. %

The main result of this section is an algorithm that allows for direct
access for an \ocircuit{} on domain $D$ and variables set $X$. More precisely,
we prove the following: %

\begin{thm}
  \label{thm:rankedcircuit}
  Let $\prec$ be an order on $X$ and $C$ be a $\prec$-\ocircuit{} on domain
  $D$ and variables set $X$, then we can solve direct access tasks on $\rel(C)$ for
  order $\lexprec$ with access time 
  $\bigo(\size{X}^3\log(\size{X}) + |X|^2 \log\size{D})$ and precomputation time
  $\bigo(\size{X} \log(\size{X}) \cdot \size{C})$.
\end{thm}

\subsubsection{Precomputation}

In this section, we assume that $C$ is a $\prec$-\ocircuit{} with respect
to $X$. %
Moreover, for every decision gate $v$, we assume that its ingoing edges are
stored in a list sorted by increasing value of their label. %
More precisely, we assume that the ingoing edges of $v$ are a list
$[e_1,\dots, e_k]$ such that $d_1 < \dots < d_k$ where $d_i$ is the label of $e_i$. %
Since we are working in the unit-cost model, we can always preprocess $C$
in linear time to sort the ingoing edges of $v$. %

\paragraph{The $\nrel(\cdot,\cdot)$ values.} We are interested in the following values: for every decision gate $v$ of
$C$ labelled with variable $x$ and ingoing edges $[e_1,\dots, e_k]$ with respective labels $[d_1,\dots,d_k]$, we define for every $i \leq k$, 
$\nrel(v,d_i)$ as $\#\sigma_{x \leqslant d_i}(\rel(v))$. %
That is, $\nrel(v,d_i)$ is the number of tuples from $\rel(v)$ that assign a
value on $x$ smaller or equal than $d_i$. %
The precomputation step aims to compute $\nrel$ so that we can access
$\nrel(v,d_i)$ quickly for every $v$ and $d_i$. %

\begin{exa}
%Added Centering to maintain total centering
\begin{figure}[htp]
  \centering
  \begin{subfigure}[c]{0.45\linewidth}
  \centering
    \includegraphics[scale=1]{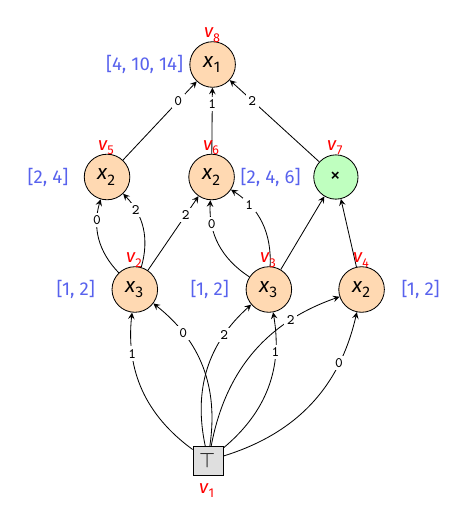}
  \end{subfigure}
  \hspace{1mm}
  \begin{subfigure}[c]{0.45\linewidth}
  \centering
    \begin{tabular}{ccc}
      $x_1$ & $x_2$ & $x_3$ \\ \toprule %Removed Horizontal lines for cleanliness
      0 & 0 & 0 \\
      0 & 0 & 1 \\
      0 & 2 & 0 \\
      0 & 2 & 1 \\
      1 & 0 & 1 \\
      1 & 0 & 2 \\
      1 & 1 & 1 \\
      1 & 1 & 2 \\
      1 & 2 & 0 \\
      1 & 2 & 1 \\
      2 & 0 & 1 \\
      2 & 2 & 2 \\
      2 & 0 & 1 \\
      2 & 2 & 2 \\ \bottomrule
    \end{tabular}
      
  \end{subfigure}
  \caption[Example of an annotated circuit]{Example of a \dcircuit{}
    annotated with $\nrel{}$ values. The domain used is $\{0, 1, 2\}$
    for variables $x_1, x_2$ and $x_3$. The lists shown to the left of
    the decision gates represent the values of \nrel{} for those
    gates. The relation computed by the circuit is given on the right.
    The full annotation of the circuit needed for direct access is a
    bit more complicated and is described later in this section.}
  \label{fig:annotated_circuit}
\end{figure}

We give an example of an annotated circuit on~\cref{fig:annotated_circuit} where each decision-gate is labeled with a list. For a decision-gate $v$ in the circuit with incoming edge $e_1,\dots,e_k$ labeled with $d_1,\dots,d_k$ with $d_1 < \dots < d_k$, the $i^{\mathsf{th}}$ entry of the list is $\nrel(v,d_i)$. %
In this example, gate $v_2$ is labelled by \([1, 2]\). %
This is because the relation computed by $v_2$ is on variable $\{x_3\}$ and contains two tuples: $\langle x_3 \gets 0\rangle$ and $\langle x_3 \gets 1\rangle$. Hence, there is one tuple in this relation which sets $x_3$ to a value smaller than $0$, two tuples which set $x_3$ to a value smaller than $1$. Since there is no incoming edge labeled by $2$, we do not need to compute $\nrel(v_2,2)$. That is, $\nrel(v_2,0)=1$ and $\nrel(v_2,1)=2$.  Observe that gate $v_3$ is also labeled by $[1,2]$ but this corresponds to $\nrel(v_3,1)=1$ and $\nrel(v_3,2)=2$ since there is no incoming edge labeled by $0$.

The gate $v_6$ is labeled by $[2,4,6]$. Indeed, it can be seen that there are two tuples with $x_2=0$ in $\rel(v_6)$. Both tuples come from following the $0$-labeled incoming edge of $v_6$ to $v_3$. The size of the relation computed by $v_3$ is $2$, which can be read as value $\nrel(v_3,2)$, that is, $\nrel(v_6,0)=2$. Similarly, by following the $1$-labeled incoming edge to $v_3$, we know that there are $2$ tuples with $x_2=1$ in $\rel(v_6)$, hence $4$ tuples with $x_2 \leq 1$, that is, $\nrel(v_6,1) = 2+\nrel(v_6,0) = 4$. We finally get $2$ tuples with $x_2=2$ from $v_2$, giving $\nrel(v_6,2) = 2+\nrel(v_6,1)=6$. 
\end{exa}

%\newpage

\paragraph*{Dynamic programming for computing $\nrel(\cdot)$.} As hinted in the previous example, our algorithm performs a bottom-up computation to compute $\nrel(\cdot,\cdot)$ values. However, some other values need to be computed along them to ease later computation. The first one we need is the number of tuples in $\rel(v)$ for every gate $v$ of $C$. We get this value inductively as follows.
If $v$ is an input gate labelled by $\bot$ then we obviously have
$|\rel(v)|=0$ and if it is an input gate labelled by $\top$, we have
$|\rel(v)|=1$. %
Now, if $v$ is a decision-gate on variable $x$ with (sorted) ingoing edges
$e_1=(v_1,v),\dots,e_k=(v_k,v)$ labelled respectively by
$d_1=\ell(e_1) < \dots < d_k = \ell(e_k)$, then $|\rel(v)|$ can inductively
be computed as follows: %

\begin{equation}
  \label{eq:rel}
  |\rel(v)| = \sum_{i=1}^k |\rel(v_i)| \times |D|^{|\Delta(v,v_i)|}
  \text{ where } \Delta(v,w)=\var(v) \setminus (\{x\} \cup \var(w)).
\end{equation}

\noindent
Similarly, $\nrel(v, d_i)$ can be computed by restricting the previous
relation on the inputs of $v$:

\begin{equation}
  \label{eq:nrel}
  \nrel(v,d_{i})  = \sum_{j=1}^i |\rel(v_j)| \times |D|^{|\Delta(v,v_j)|}
\end{equation}

In particular, observe that $\nrel(v,d_1) = |\rel(v_1)| \times |D|^{|\Delta(v,v_1)|}$ and
$\nrel(v, d_{i+1}) = \nrel(v,d_i) + |\rel(v_{i+1})| \times |D|^{|\Delta(v,v_{i+1})|}$. %

Finally, if $v$ is a Cartesian product, we clearly have
$|\rel(v)| = \prod_{w \in \mathsf{input}(v)} |\rel(w)|$. %
Hence, one can compute $\nrel$ using a dynamic programming algorithm that inductively
computes $\nrel(v,d)$ for each decision-gate $v$ of the circuit with an
ingoing edge labelled by $d$. Observe however that in order to efficiently compute these values, we also need to have easy access to $|\rel(v)|$ and $\var(v)$ for every gate $v$. Fortunately, as hinted above, these values are also easy to compute inductively.

More precisely, the dynamic programing algorithm works as follows: we start
by performing a topological ordering $(v_1, \dots, v_N)$ of the gates of $C$ that is compatible
with the underlying DAG of the circuit. In particular, it means that for a
gate $v$ and an input $w$ of $v$, the topological ordering has to place $w$
before $v$. This can be computed in time $\bigo(|C|)$. % Moreover, we also add the following constraint: if $w$ and $w'$
% are both inputs of a decision gate $v$ and if the edge $(w,v)$ is labelled
% by $d \in D$ and the edge $(w',v)$ is labelled by $d' \in D$ such that $d <
% d'$, then we ask for the topological order to place $w$ before $w'$. It
% is easy to construct such an ordering by simply doing a topological order
% of the DAG of $C$ augmented by the following edges: for every decision gate $v$ with sorted ingoing edges $e_1=(v_1,v),\dots,e_k=(v_k,v)$ (recall we assume  $d_1=\ell(e_1) < \dots < d_k=\ell(e_k)$), we add edges $(w_i, w_{i+1})$. This modified DAG has size at most $2|C|$ and since computing a topological ordering of a DAG can be done in linear time, we can construct
% it in time $\bigo(|C|)$. 
 
We dynamically compute, for every gate $v$, the values $|\rel(v)|$, the sets $\var(v)$, and if $v$ is a decision gate with an ingoing edge labelled by $d \in D$, we also compute $\nrel(v,d)$.

We proceed as follows: we allocate a table $T_{\rel}$ of size $N$ such that for $i \leq N$, $T_{\rel}[i]$ is initialized to $0$. At the end of the algorithm, $T_{\rel}[i]$ will contain $|\rel{v_i}|$. We also allocate a table $T_{\var}$ such that for $i \leq N$, $T_{\var}[i]$ is initialized with a $|X|$-bitvector containing on $0$. At the end of the algorithm, $T_{\var}[v_i][k]$ is $1$ if and only if $x_k \in \var{v_i}$.   This initialization step takes $\bigo(N \cdot |X|) = \bigo(|C| \cdot |X|)$.

We now initialize $T_{\nrel}$ of size $N$ where for $i \leq N$, the entry $T_{\nrel}[i]$ is intialized
with an array of size $|\mathsf{input}(v_i)|$ containing only $-1$ values. Clearly, $T_{\nrel}$ has size $\bigo(|C|)$ since it contains one entry per edge of $C$. At the end of the algorithm, for $i \leq N$, if $v_i$ is a decision-gate, then $T_{\nrel}[i][p]$ contains $\nrel(v_i,d_p)$ where $d_p$ is the label of the $p^{\mathsf{th}}$ input of $v_i$. If $v_i$ is a Cartesian product or an input, $T_{\nrel}[i]$ is not relevant and is not modified after the initialization. 

We then populate each entry of these tables following the previously
constructed topological ordering and using the relations written above
(see (\ref{eq:rel}) and (\ref{eq:nrel})) and the fact that
$\var(v) = \bigcup_{w \in \mathsf{input}(v)} \var(w)$. To compute
$\nrel$ for a decision-gate $v$, we let $(w_1, v), \dots, (w_k,v)$ be
the incoming edges of $v$ ordered by increasing labels
$d_1 < \dots < d_k$. We set $T_{\nrel}[v,d_1]$ to
$\nrel(v,d_1)$ which is equal to $|\rel(w_1)|\cdot|D|^{|\Delta(v,w_1)|}$. We then compute
$T_{\nrel}[v,d_{i+1}]$ as
$T_{\nrel}[v,d_i]+|\rel(w_{i+1})|\cdot|D|^{|\Delta(v,w_{i+1})|}$ using
relation~(\ref{eq:nrel}).

% An example of an \ocircuit{} that has been annotated by our algorithm is
% presented in \cref{fig:annotated_circuit}.

It is clear from the relations (\ref{eq:nrel}) that at the end of this
precomputation, we have $T_{\nrel}[v,d]$ contains $\nrel(v,d)$ if $d$
labels an incoming edge of $v$. Pseudocode for this algorithm can be found in \cref{alg:annotate}.

\begin{algorithm}[htp]
  \caption[Annotation algorithm]{An algorithm to annotate a circuit $C$ on variables $X$ and domain $D$}
  \label{alg:annotate}
  \begin{algorithmic}[1]
    \Procedure{Annotate}{$C$, $X$, $D$}
    \State $v_1, \dots, v_N \gets $ a topological ordering of the gates of $C$\;
    \State $T_\rel \gets $ table of size $N$, initialized to $0$\;
    \State $T_\var \gets $ table of size $N$, initialized with $0$-bitvectors of size $|X|$\;
    \State $T_{\nrel} \gets $ table of size $N$; 

    \For{$i = 1$ to $N$}
    \eIf{$v_i$ is $\top$-input}{$T_\rel[i] \gets 1$; \textbf{continue}}
    \eIf{$v_i$ is $\bot$-input}{$T_\rel[i] \gets 0$; \textbf{continue}}
    \If{$v_i$ is a $\times$-gate}
    \State $T_\rel[i] \gets \prod_{v_j \in \mathsf{input}(v_i)}T_\rel[j]$ \;
    \State $T_\var[i] \gets \bigcup_{v_j \in \mathsf{input}(v_i)}T_\var[j]$ \;
    \ElsIf{$v_i$ is a decision gate on $x_k$}
    \State $T_\var[i] \gets \{x_k\} \cup \bigcup_{v_j \in \mathsf{input}(v_i)}T_\var[j]$ \;
    \State $T_\rel[i] \gets \sum_{v_j \in \mathsf{input}(v_i)}T_\rel[j] \times |D|^{|T_\var[j] \setminus T_\var[i]|-1}$ \;
    \State Order incoming edges $(v_{a_1},v_i), \dots, (v_{a_k},v_i)$ of $v_i$ by increasing label\;
    \State $T_{\nrel}[i] \gets $ new array of size $k$ initialized to $0$\;
    \State $T_{\nrel}[i][1] \gets T_{\rel}[a_1]$\;
    \For{$j=2$ to $k$}
    \State $T_{\nrel}[i][k] \gets T_{\nrel}[i][j-1] + T_{\rel}[a_j]$\;
    \EndFor
    \EndIf
    \EndFor
    
    \State \Return $T_\rel, T_\var, T_{\nrel}$\;
    \EndProcedure
  \end{algorithmic}
\end{algorithm}

\begin{exa}
  From the circuit in \cref{fig:annotated_circuit}, assume we have populated $T_\rel$, $T_\var$ and $T_{\nrel}$ up to $i=6$. We now need to compute the seventh entry of each table. Since $v_7$ is a $\times$-gate, we do not need to do anything for $T_{\nrel}[7]$. Now $T_\rel[7]$ must contain the size of the relation computed by $v_7$. Since $v_7$ is a $\times$-gate, this is $|\rel(v_3)| \times |\rel(v_4)|$ which is equal to $T_\rel[3] \times T_\rel[4]$ by induction. So we set $T_\rel[7] = T_\rel[3] \times T_\rel[4]$. Similarly, $T_\var[7] = T_\var[3] \cup T_\var[4]$. Since $T_\var[7]$ is encoded as a bitvector, the union boils down to making a bitwise OR operation.

  Finally, we need to compute $T_\rel[8], T_\var[8], T_{\nrel}[8]$. Since $v_8$ is a decision gate, $T_\var[8] = T_\var[5] \cup T_\var[6] \cup T_\var[7] \cup \{x_1\}$. Now $T_\rel[8] = T_\rel[5]+T_\rel[6]+T_\rel[7]$. Observe that in this case, we do not need to add corrective factor since $v_5,v_6$ and $v_7$ define relations on the same variables set. If the variables sets were different, we would have to adjust the value following \cref{eq:rel}. Now, it remains to compute $T_{\nrel}[8]$. Since $v_8$ has three incoming edges, we will have to compute $T_{\nrel}[8][1],T_{\nrel}[8][2], T_{\nrel}[8][3]$. We order the edges by increasing label and find that: $T_{\nrel}[8][1] = T_{\rel}[5]=4$, $T_{\nrel}[8][2]=T_{\nrel}[8][1]+T_{\rel}[6] = 4+6=10$ and $T_{\nrel}[8][3]=T_{\nrel}[8][2]+T_{\rel}[7] = 10+4=14$. 
\end{exa}

The following lemma follows directly from the previously described algorithm:
\begin{lem}[Precomputation complexity]
  \label{lem:precomputation-complexity}
  Given a $\prec$-\ocircuit{} $C$, we can compute a data structure in time
  $\bigo(\size{X} \log(\size{X}) \cdot \size{C})$ that allows us to
  access in constant time to the following:
  \begin{itemize}
  \item  $\var(v)$, $|\rel(v)|$ for every gate $v$ of $C$,
  \item  $\nrel(v,d)$ for every decision gate $v$ having an ingoing edge labelled by $d \in D$.
  \end{itemize}
\end{lem}

\begin{proof}
  The data structure simply consists of the three tables $T_{\nrel},
  T_{\rel}$ and $T_{\var}$. It is easy to see that each entry of $T_{\var}$
  can be computed in time $\bigo(|X|)$ since we only have to compute the
  union of sets of elements in $X$ and we can use a bitmask of size $|X|$ to do it efficiently. Indeed, we can represent $\rel(v)$ as a bitvector $T_{\var}[v]$ with $|X|$ entries for each $v$ such that $T_{\var}[v][i]=1$ if and only if $x_i \in \rel(v)$. Now computing $T_{\var}[v][i]$ can be done by looking at every input $w$ of $v$ and check whether $T_{\var}[w][i]=1$. For each gate $w$, we will have one such look up for each incoming edge and each variable, hence one can compute $T_{\var}$ in time  $\bigo(|C| \cdot |X|)$.

  Now observe that to compute $T_{\nrel}$, one has
  to perform at most two arithmetic operations for each edge of $C$.
  Indeed, to compute $T_{\rel}[v]$, one has to perform at most one addition
  and one multiplication for each $w \in \mathsf{input}(v)$, whose cost can
  be associated to the edge $(w,v)$. Similarly, to compute $T_{\nrel}[v,d]$
  as described in the previous paragraph, we do one addition and one
  multiplication for each edge $(w,v)$ in the circuit. Hence, we perform at
  most $\bigo(|C|)$ arithmetic operations. Now, the cost of these
  arithmetic operations is at most $\bigo(|X|\log(|X|))$ in the unit-cost RAM model (see \cref{sec:preliminaries} for a discussion). Indeed, all operations are performed on integers representing sizes of relations on domain $D$
  and variables set $Y \subseteq X$. Hence, their value cannot exceed $|D|^{|X|}$.

  Hence, we have a total complexity of $\bigo(|X|\log(|X|) \cdot |C|)$.
\end{proof}

\subsubsection{Direct access}

We now show how the precomputation from
\cref{lem:precomputation-complexity} allows us to get direct access for
\ocircuit{s}. We first show how one can solve a direct access task for any
relation as long as we have access to very simple counting oracles. We then
show that one can quickly simulate these oracle calls in \ocircuit{s} using
precomputed values.

\begin{exa}
An example of how the algorithm works is shown in \cref{fig:da}. %
\cref{sfig:da7} shows the paths taken in the circuit to evaluate the
\(7^{\mathsf{th}}\) tuple in the circuit. %
The main idea resides in the fact that we can connect \da{} with counting
tasks. %
If we want the \(7^{\mathsf{th}}\) tuple in the circuit, we cannot assign \(x_1
\gets 0\), since we know that there are only \(4\) possible extensions in
that case. %
We should however assign \(x_1 \gets 1\), since there are \(10\) possible
assignments with \(x_1 \leq 1\). %
We then move down the circuit and search in the new gate, but we have to
adapt the index since \(4\) of the possible extensions have been discarded
(because they involve assigning \(x_1 \gets 0\)). %
We are now therefore looking for the \(3^{\mathsf{rd}}\) tuple in this
subcircuit. %
Again, assign $x_2 \gets 0$ will only give $2$ possible extensions, which is not enough to find the third solution. Assign $x_2 \gets 1$ however is enough since there are $4$ assignments with $x_2 \leq 1$. We hence set $x_2 \gets 1$ and proceed with the last decision-gate. By assigning $x_2 \gets 1$, we discarded $2$ extensions, so we are now looking for the first answer in this last decision-gate, which is obtained by setting $x_3 \gets 1$.
We continue the search in this manner until reaching a terminal gate. %

A more involved example can be found in \cref{sfig:da13}. %
In this case, we are looking for the \(13^{\mathsf{th}}\) tuple of the
circuit. %
However, after assigning (with the same method as before) \(x_1 \gets 2\), we
reach a \(\times\)-gate, and therefore do not assign a value to a given
variable. %
What happens is we are now looking for the \(13 - 10 = 3^{\mathsf{rd}}\) solution of a
set of circuits comprised of the bottom-centre gate and the bottom-right gate. %
We choose which of the variables from the roots of the set of circuits to
assign first based on the order, so in this case, we start with \(x_2\). %
We have to remember that for each possible solution of the circuit rooted in
\(x_2\), there are two possible extensions from the circuit rooted in \(x_3\)
(see \cref{lem:rel-sink-times}). %
This is the same thing as imagining the gate being virtually relabelled by \([2,
2, 4]\). %
In other words, $x_2 \gets 0$ provides two tuples for the Cartesian product, which is not enough to get the third answer. The same happens for $x_2 \gets 1$ but  $x_2 \gets 2$ has enough extension since there are $4$ assignments for the Cartesian product that have $x_2 \leq 2$. We hence assign  $x_2 \gets 2$. We discarded $2$ assignments with this choice, we are hence looking for the first assigmnent for $x_3$ which is found by setting $x_3$ to \(1\). %

\begin{figure}[htp]
  \hspace*{-1cm}
  \centering
  \begin{subfigure}[c]{0.48\linewidth}
    \includegraphics[scale=1]{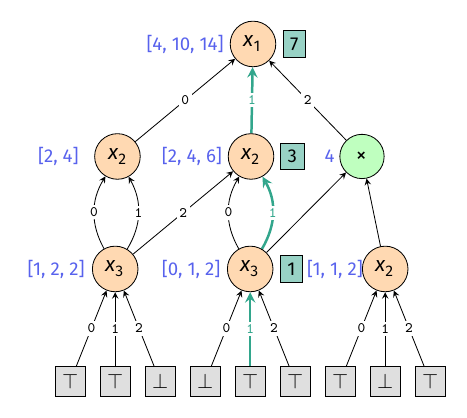}
    \caption{$k = 7$}
    \label{sfig:da7}
  \end{subfigure}
  \hspace{1mm} % Changed to static miniwidth to not exeed margins
  \begin{subfigure}[c]{0.48\linewidth}
    \includegraphics[scale=1]{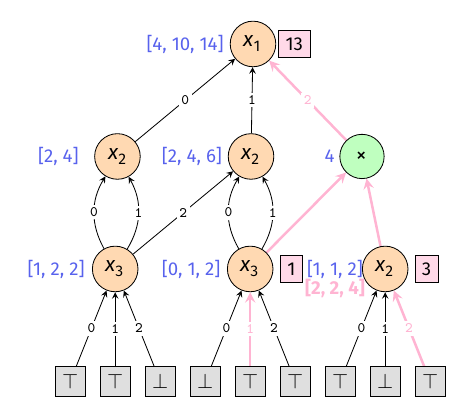}
    \caption{$k = 13$}
    \label{sfig:da13}
  \end{subfigure}
  \caption[Example of DA tasks]{Examples of the paths followed during
    different direct access tasks on the same annotated \ocircuit{} for order $(x_1,x_2,x_3)$.}
  \label{fig:da}
\end{figure}
\end{exa}
\noindent
The following lemma connects direct access tasks with counting tasks. A similar claim can be found in~\cite[Proposition 3.5]{bringmann2022tight}:

\begin{lem}
  \label{lem:compl-rk}
  Assume that we are given a relation $R \subseteq D^X$ with $X = \{x_1,
  \dots, x_n\}$ and an oracle $\mathcal{A}$ such that for every prefix assignment $\tau \in D^{\{x_1,\dots,x_p\}}$ and $N \in \N$, it returns the smallest value $d \in D$ such that $\#\sigma_{x_{p + 1} \leqslant d}(\sigma_{\tau}(R)) \geq N$ and the value $\#\sigma_{x_{p + 1} < d}(\sigma_{\tau}(R))$.  Then, for any $k \leq |R|$, we can compute $R[k]$ using $\bigo(|X|)$  calls to oracle $\mathcal{A}$.
\end{lem}

\begin{proof}
Let $\alpha = R[k]$ be the answer we are looking for. We iteratively find the value of $\alpha$ on $X$ by applying \cref{lem:assignment-of-tau-x}. More precisely, assume that we already know $\alpha(x_1), \dots, \alpha(x_p)$ for some $p < n$. Moreover, we know $k_p$ such that $\alpha = \sigma_\beta(R)[k_p]$ where $\beta$ is the prefix assignment of $\{x_1,\dots,x_p\}$ such that $\beta(x_i) = \alpha(x_i)$. We know by \cref{lem:assignment-of-tau-x} that $\alpha(x_{p+1}) =  \min \{d \mid \#\sigma_{x_{p + 1} \leqslant d}(\sigma_{\beta}(R)) \geq k_p\}$ and $k_{p+1} = k_p - \#\sigma_{x_{p + 1} < d}(\sigma_{\beta}(R))$, which we can compute with one call to oracle $\mathcal{A}$. Hence, by induction, we can fully reconstruct $\alpha$ with $\bigo(|X|)$ calls to oracle $\mathcal{A}$. 
\end{proof}

In order to evaluate the true complexity of answering direct access tasks, we
now also have to evaluate the complexity of a single oracle call. To do that, we need to understand how an assignment of the form $D^{\{x_1,\dots,x_p\}}$ for some $p$ behaves with respect to the circuit.  We hence define a \emph{prefix assignment} of size $p$ to be an assignment $\tau
\in D^{\{x_1,\dots,x_p\}}$ with $p \leqslant n$. Moreover, for a gate $v$ in $C$, we define $\sink(v)$ as:

\[
  \sink(v) =
  \begin{cases*}
    \bigcup_{w \in \mathsf{inputs}(v)} \sink(w) & if $v$ is a $\times$-gate \\
    \{v\} & otherwise (that is, $v$ is an input or a decision gate)
  \end{cases*}
\]

We have the following property:
\begin{lem}
  \label{lem:rel-sink-times}
  For any gate $v$, we have  $\rel(v) = \bigtimes_{w
    \in \sink(v)}\rel(w)$.
\end{lem}
\begin{proof}
  We prove this by induction on the circuit. If $v$ is a decision gate or an
  input, then, as $\sink(v) = \{v\}$, the property is trivial. If $v$ is a
  $\times$-gate, then by definition, $\rel(v) =
  \bigtimes_{w\in\mathsf{inputs}(v)}\rel(w)$. By our induction hypothesis,
  $\rel(w) = \bigtimes_{g \in \sink(w)}\rel(g)$ for all inputs $w$ of $v$. Therefore, by associativity
  and commutativity of the Cartesian product, $\rel(v) = \bigtimes_{w \in \mathsf{inputs}(v)}\bigtimes_{g \in
    \sink(w)} \rel(g) = \bigtimes_{g \in \bigcup_{w \in \mathsf{inputs}(v)} \sink(w)}\rel(g) = \bigtimes_{g \in \sink(v)} \rel(v)$.
\end{proof}

Given a prefix assignment $\tau$, we want to characterize the set of tuples in $C$ that are compatible with $\tau$. To do so, we can follow every decision-gate whose variable are set by $\tau$ and follow every input of $\times$-gates. This gives a set of gates in the circuit that we call the frontier of $\tau$ such that any assignment of $\rel(C)$ extending $\tau$ must be made from assignments of each gate in the frontier.  More formally, we introduce the \emph{frontier $f_\tau$ of $\tau$} in $C$ as follows:

\begin{enumerate}
\item instantiate a set $F = \{\out(C)\}$, containing only the root of the circuit
\item as long as it is possible, transform $F$ as follows:
  \begin{itemize}
  \item if $v \in F$ is a $\times$-gate, remove $v$ from $F$ and add $\sink(v)$ to $F$,
  \item if $v \in F$ is a decision gate and the variable $x$ labelling $v$ is
    assigned in the prefix, \emph{i.e.} $x \in \{x_1,\dots,x_p\}$, remove $v$ from $F$ and add $v_d$ to $F$, where $v_d$ is the input of $v$ such that edge $(v_d,v)$ is labelled by $\tau(x)$.
  \end{itemize}
\item if $F$ contains a $\bot$-gate, then $f_\tau = \emptyset$, otherwise
  $f_\tau = F$.
\end{enumerate}
\noindent
If, for a given gate $v$, the set $\sink(v)$ contains a $\bot$-gate, then the
circuit is no longer satisfiable, which is why we return $\emptyset$ in this
case. Note that this should not happen while building the \kso{} for $C$.

Frontiers are particularly useful because they can represent the following relation: the \emph{frontier relation} denoted by $\rel(f_\tau)$ is defined as the relation  $\bigtimes_{v \in f_\tau} \rel(v)$. It is a relation defined on variables set $\var(f_\tau) :=
\bigcup_{v \in f_\tau}\var(v)$.

For a prefix $\tau$ on variables set $\{x_1,\dots,x_p\}$, we denote by $\sigma_\tau(R)$
the relation $\sigma_{x_1 = \tau(x_1),\dots,x_p}$ ${= \tau(x_p)}(R)$. We have the following connection between the relation of a frontier and prefix assignments: %Split Equation into two to maintain margins

\begin{lem}
  \label{lem:sigma-for-prefix}
  Let $C$ be an \ocircuit{} on variables set $X = \{x_1,\dots,x_n\}$ and $\tau$ be a prefix assignment on variables set $\{x_1,\dots,x_p\}$. Then $\sigma_\tau(\rel_X(C)) = \{\tau\} \times \rel(f_\tau) \times D^{\{x_{p+1},\dots,x_n\} \setminus
    \var(f_\tau)}$.
\end{lem}
\begin{proof}
  We prove the lemma by induction on the size of the prefix. For an empty
  prefix $\tau = \langle\rangle$, we have $f_\tau = \sink(\out(C))$.
  Indeed, if $\out(C)$ is a decision gate or input, then it is trivial since $\sink(\out(C)) = \{\out(C)\}$.
  Otherwise we simply sink through the $\times$-gate since no variable is
  assigned by $\tau$. We have that $\sigma_{\langle\rangle}(\rel_X(C)) = \rel_X(C)$,
  which is itself by definition equal to $\rel(\out(C)) \times
  D^{X\setminus \var(\out(C))}$. From \cref{lem:rel-sink-times}, we know
  that $\rel(\out(C)) = \bigtimes_{w\in\sink(\out(C))}\rel(w)$. We know
  that $\var(\out(C)) = \var(f_{\langle\rangle})$. Thus, we have that
  $\sigma_{\langle\rangle}(\rel_X(C)) =
  \bigtimes_{w\in\sink(\out(C))}\rel(w) \times D^{\{x_1,\dots,x_n\}
    \setminus \var(f_{\langle\rangle})}$.

  Now suppose the property holds for any prefix $\tau$ of size $p$. We now
  show that it also holds for a prefix $\tau' = \tau \times [x_{p+1} \gets
  d]$.

  We can rewrite $\sigma_{\tau'}(\rel_X(C))$ as $\sigma_{x_{p+1} =
    d}(\sigma_\tau(\rel_X(C)))$. Applying the induction hypothesis to this equality gives:
  \[
    \sigma_{\tau'}(\rel_X(C)) = \sigma_{x_{p+1} = d}\left(\{\tau\} \times
      \rel(f_\tau) \times
      D^{\{x_{p+1},\dots,x_n\} \setminus \var(f_\tau)}\right)
  \]

  From here, we have two possibilities: either there exists a decision gate
  $v \in f_\tau$ such that $\decvar(v) = x_{p+1}$ or not.\\

  \noindent\textbf{First case: there exists $v \in f_{\tau}$ such that $\decvar(v) = x_{p+1}$.} In this
  case, we have by definition: $f_{\tau'} = f_\tau \setminus \{v\} \cup
  \sink(v_d)$. 

We start by pointing out that for a decision gate $v$ with $x_{p+1} = \decvar(v)$
and $d \in D$, we have $\sigma_{x_{p+1} = d}(\rel(v)) = \{[x_{p+1} \gets d]\} \times
\rel(v_d) \times D^{\var(v) \setminus (\{x_{p+1}\} \cup \var(v_d))}$. In other words, every tuple in the relation computed by $v$ where $x_{p+1} = d$ is of the form $[x_{p+1}\gets d] \times \tau' \times \sigma'$ where $\tau' \in \rel(v_d)$ and $\sigma'$ is any tuple on domain $D$ and variables set $\var(v) \setminus (\{x_{p+1}\} \cup \var(v_d))$.
We can therefore write:

\begin{align*}
  \allowdisplaybreaks
  \sigma_{\tau'}(\rel_X(C)) &= \sigma_{x_{p+1} = d}\left(\{\tau\} \times
                              \rel(f_\tau) \times D^{\{x_{p+1},\dots,x_n\}
                              \setminus \var(f_\tau)}\right) \\
                            &\text{since $x_{p+1}$ only appears in the frontier}:\\
                            &= \{\tau\} \times D^{\{x_{p+1},\dots,x_n\}
                              \setminus \var(f_\tau)} \times \sigma_{x_{p+1} = d}(\rel(v)) \times \bigtimes_{w \in f_\tau \setminus \{v\}} \rel(w) \\
                            &\text{from the previous relation:} \\
                            &= \{\tau\} \times D^{\{x_{p+1},\dots,x_n\}
                              \setminus \var(f_\tau)} \times  \{[x_{p+1} \gets d]\} \times
                              \rel(v_d) \times D^{\var(v) \setminus (\{x_{p+1}\} \cup \var(v_d))}\\
                            &\qquad\times \bigtimes_{w \in f_{\tau}\setminus\{v\}} \rel(w) \\
                            &\text{from \cref{lem:rel-sink-times}:}\\
                            &= \{\tau\} \times D^{\{x_{p+1},\dots,x_n\}
                              \setminus \var(f_\tau)} \times  \{[x_{p+1} \gets d]\} \times
                              \bigtimes_{w \in \sink(v_d)}\rel(w) \\
                            &\qquad\times D^{\var(v) \setminus (\{x_{p+1}\} \cup \var(v_d))}\times \bigtimes_{w \in f_{\tau}\setminus\{v\}} \rel(w) \\
                            &= \{\tau\} \times  \{[x_{p+1} \gets d]\} \times D^{\{x_{p+1},\dots,x_n\}\setminus \var(f_\tau)}\times D^{\var(v) \setminus (\{x_{p+1}\} \cup \var(v_d))} \\
                            &\qquad\times \bigtimes_{w \in \sink(v_d)}\rel(w) \times \bigtimes_{w \in f_{\tau}\setminus\{v\}} \rel(w) \\
                            &= \{\tau'\} \times \bigtimes_{w\in f_{\tau'}}\rel(w) \times D^{\{x_{p+2},\dots,x_n\}\setminus \var(f_\tau')}
\end{align*}

\noindent\textbf{Second case: there are no $v \in f_{\tau}$ such that $\decvar(v) = x_{p+1}$.} In this case, since the circuit is ordered, it means that $x_{p+1} \notin \var(f_\tau)$. Moreover $f_\tau = f_{\tau'}$.  We can therefore write:

\begin{align*}
  \sigma_{\tau'}(\rel_X(C)) &= \{\tau\} \times \bigtimes_{w \in f_\tau}\rel(w) \times  \sigma_{x_{p+1} = d}(D^{\{x_{p+1},\dots,x_n\}\setminus \var(f_\tau)}) \\
                            &\text{since $x_{p+1}$ does not appear in the frontier or  $\tau$:} \\
                            &= \{\tau\} \times \bigtimes_{w \in f_\tau}\rel(w) \times \{[x_{p+1} \gets d]\} \times D^{\{x_{p+2},\dots,x_n\}\setminus \var(f_\tau)} \\
                            &= \{\tau'\} \times \bigtimes_{w \in f_{\tau'}}\rel(w) \times D^{\{x_{p+2},\dots,x_n\}\setminus \var(f_{\tau'})} \\
                            &\text{since $f_\tau = f_{\tau'}$.}
\end{align*}
\noindent
Since the property is true for the empty prefix and inductively true, we
conclude that it is true for any prefix $\tau$.
\end{proof}

In order to be useful in practice, building and using the frontier of a
prefix assignment $\tau$ cannot be too expensive. We formulate the
following complexity statement:

\begin{lem}
  \label{lem:building-ftau}
  Let $\tau$ be a prefix assignment over the set of variables set $X =
  \{x_1,\dots,x_p\}$. We can compute $f_\tau$ in time $\bigo(\size{X})$.
\end{lem}
\begin{proof}
  Let $\tau$ be a prefix assignment of size $p$. The frontier $f_\tau$ is
  built in a top-down fashion, by following the edges corresponding to the
  variable assignments in $\tau$. For each variable $x$ assigned by $\tau$,
  we follow at most one edge from a decision gate $v$ such that $\decvar(v)
  = x$ and the edge is labelled $\tau(x)$. This means $p$ edges are
  followed for the assignments. Moreover, to get to $v$, we might have to
  follow edges from $\times$-gates. Since the variable sets underneath
  $\times$-gates are disjoint from one another, we have that the number of
  such edges is bounded by $\size{X}$. This implies the total cost of
  building the frontier $f_\tau$ is $\bigo(\size{X} + p)=\bigo(\size{X})$.
\end{proof}

We are now ready to prove the main connection between frontiers and \cref{lem:compl-rk}:

\begin{lem}
  \label{lem:oracle-complexity}
  Let $C$ be an \ocircuit{} on variables set $X=\{x_1,\dots,x_n\}$ such that $\nrel(v,d)$ and $\var(v)$ have been precomputed as in \cref{lem:precomputation-complexity}. Let $\tau$ be a prefix assignment of $D^{\{x_1,\dots,x_p\}}$. Then, for every $N$, one can compute the smallest value $d \in D$ such that $\#\sigma_{x_{p+1} \leqslant d}(\sigma_\tau(\rel(C))) \geq N$ and the value $\#\sigma_{x_{p+1} < d}(\sigma_\tau(\rel(C)))$ in time $\bigo(|X|^2\log |X| + |X|\log(|D|))$.
\end{lem}

\begin{proof}
  We start by building the frontier $f_\tau$ associated with the prefix
  assignment $\tau$. From \cref{lem:building-ftau}, we know this can be
  done in time $\bigo(|X|)$. By \cref{lem:sigma-for-prefix}:

  \noindent
  We can rewrite:
  \begin{align*}
    \sigma_{x_{p+1} \leq d}(\sigma_\tau(\rel(C))) &= \sigma_{x_{p+1} \leq d}(\{\tau\} \times \rel(f_\tau) \times D^{\{x_{p+1},\dots,x_n\}\setminus \var(f_\tau)}) \\
    &= \{\tau\} \times \sigma_{x_{p+1} \leq d}(\rel(f_\tau) \times D^{\{x_{p+1},\dots,x_n\}\setminus \var(f_\tau)})
     % \intertext{There are now two possible outcomes: either $x_{p+1}$ is tested by $f_\tau$ or not. In the first case, since $x_{p+1}$ only appears in the frontier, tested by a gate $v$, we have:} 
     %                                                     &= \#\{\tau\} \cdot \#(D^{\{x_{p+2},\dots,x_n\}\setminus\var(f_\tau)}) \cdot \#\sigma_{x_{p+1}\leq d}\left(\bigtimes_{w\in f_\tau}(\rel(w))\right)\\
     %                                                     &= \size{D}^{\size{\{x_{p+2},\dots,x_n\}\setminus\var(f_\tau)}} \cdot \#(\bigtimes_{w\in f_\tau \setminus \{v\}}(\rel(w))) \cdot\#\sigma_{x_{p+1}\leq d}(\rel(v)) \\
     %                                                     &= \size{D}^{\size{\{x_{p+2},\dots,x_n\}\setminus\var(f_\tau)}} \cdot (\prod_{w\in f_\tau \setminus \{v\}}\#\rel(w)) \cdot \nrel(v, d)
  \end{align*}
  \noindent
  There are now two possible outcomes: either there is a decision gate $v$ labelled by variable $x_{p+1}$ in $f_\tau$ or not. In the first case, since $x_{p+1}$ only appears in the frontier, we have for every $d \in D$:
  \begin{align*}
    \sigma_{x_{p+1} \leq d}(\sigma_\tau(\rel(C))) &= \{\tau\} \times \sigma_{x_{p+1} \leq d}(\rel(v)) \times \bigtimes_{w \in f_\tau \setminus \{v\}} \rel(w) \times D^{\{x_{p+2},\dots,x_n\}\setminus \var(f_\tau)}
  \end{align*}
Hence we have:
  \begin{align*}
    \#\sigma_{x_{p+1} \leq d}(\sigma_\tau(\rel(C))) &= \#\sigma_{x_{p+1} \leq d}(\rel(v)) \times \prod_{w \in f_\tau \setminus \{v\}} \#\rel(w) \times |D|^{|\{x_{p+2},\dots,x_n\}\setminus \var(f_\tau)|} \\
                                                    &= \nrel(v,d) \times \prod_{w \in f_\tau \setminus \{v\}} \#\rel(w) \times |D|^{|\{x_{p+2},\dots,x_n\}\setminus \var(f_\tau)|} \\
    &=\nrel(v,d) P
  \end{align*}
  where $P$ is a product containing at most $|X|$ precomputed integers since , all of them of size at most $|D|^{|X|}$ and hence, it can be evaluated in time $\bigo(|X|^2\log |X|)$. Indeed, every value $\#\rel(w)$ for $w \in f_\tau \setminus \{v\}$ have been precomputed and can be accessed in constant time. Moreover, $\var(f_\tau) = \bigcup_{v \in f_\tau} \var(v)$. Hence, since $\var(v)$ has been precomputed too, we can compute  $e := \size{\{x_{p+2},\dots,x_n\}\setminus\var(f_\tau)}$ in $\bigo(|X|)$ and then $|D|^e$ in $\bigo(|X|^2\log |X|)$.

  Now observe that the smallest value $d$ such that $\#\sigma_{x_{p+1} \leq d}(\sigma_\tau(\rel(C))) \geq N$ has to be the label of one ingoing edge of $v$. Indeed, if $d$ does not label any ingoing edge of $v$, then $\nrel(v,d) = \nrel(v,d_0)$ where $d_0 < d$ is the greatest value smaller than $d$, which labels one ingoing edge of $v$ (or $0$ if no such $d_0$ exists). Hence the smallest value $d$ we are looking for has to be the label of one ingoing edge of $v$ and is the smallest one among them such that $\nrel(v,d) \geq {N \over P}$. Since $P$ has been evaluated already, we can evaluate $N \over P$ in time $\bigo(|X|^2\log|X|)$.

  To find the smallest value $d$ where $\nrel(v,d)$ is greater than $N \over P$, it remains to do a binary search among the ingoing edges of $v$, which have already been sorted by increasing label. Comparing $\nrel(v,d)$ to $N \over P$ can be done in $\bigo(|X|)$ for any $d$ since the values do not exceed $|D|^{|X|}$. There are at most $|D|$ ingoing edge, meaning this binary search can be performed in time $\bigo(|X|\log|D|)$. Hence finding the value $d$ we are interested in has a total complexity of $\bigo(|X|^2\log |X| + |X|\log |D|)$.

  Now that we know the value $d \in D$ we are looking for, it remains to compute $\#\sigma_{x_{p+1} < d}(\sigma_\tau(\rel(C)))$. This value is equal to $\#\sigma_{x_{p+1} \leqslant d'}(\sigma_\tau(\rel(C)))$ where $d'$ is the value labeling the ingoing edge of $v$ just before $d$. If no such value exist, then  $\#\sigma_{x_{p+1} < d}(\sigma_\tau(\rel(C)))$ is $0$. Otherwise, we can compute it as before in time $\bigo(|X|^2\log|X|)$.

  In the second case where there is no decision-gate labeled by $x_{p+1}$ in $f_\tau$, we have that $x_{p+1} \notin \var(f_\tau)$ since the circuit is ordered. Hence, we apply a similar reasoning to obtain:
 
  \begin{align*}
    \#\sigma_{x_{p+1}  \leqslant d }(\sigma_{\tau}(\rel(C))) & = \#\sigma_{x_{p+1}
      \leq d}(D^{\{x_{p+1}\}}) \cdot
    \size{D}^{\size{\{x_{p+2},\dots,x_n\}\setminus\var(f_\tau)}} \cdot
                                                               \prod_{w\in f_\tau}\#\rel(w) \\
    & = \mathsf{rank}(d) \times P
  \end{align*}

  where $\mathsf{rank}(d)$  is the number of elements smaller than $d$ in $D$ and $P$ a product whose elements have been precomputed and which can hence be evaluated in time $\bigo(|X|^2\log |X|)$. Hence, the smallest value $d \in D$ such that $\#\sigma_{x_{p+1}\leq d}(D^{\{x_{p+1}\}}) \geq N$ is the one whose rank is $\lceil {N \over P} \rceil$. Moreover,  $\#\sigma_{x_{p+1}  < d }(\sigma_{\tau}(\rel(C))) = \max(0, (\lceil {N \over P} \rceil - 1)P$).
\end{proof}

In short, we can follow the edges in the circuit by choosing the correct
edge from the precomputed values in \nrel. A short visual example of the
followed paths for different direct access tasks over the same annotated
circuit is presented in \cref{fig:da}. Notice how in the case where $k =
13$, the fact that we meet a $\times$-gate implies that we follow both
paths at once. At one point of the algorithm, a frontier containing both
the gates for $x_2$ and $x_3$ exists. The values shown at the right of the
reached decision gates show how the index of the searched tuples evolves
during a run of the search algorithm.

The proof of \cref{thm:rankedcircuit} is now an easy corollary of \cref{lem:compl-rk,lem:oracle-complexity}. Indeed, after having precomputed \nrel{} and \var{} using \cref{lem:precomputation-complexity}, we can answer direct access tasks using the oracle based algorithm from \cref{lem:compl-rk}. \cref{lem:oracle-complexity} shows that these oracle accesses are in fact tractable in ordered circuits.

We conclude this section with a high level description of the direct access algorithm using pseudo code. The first procedure we need is the procedure {\sc Find-Frontier} from \cref{alg:countext}. Given a circuit $C$ and a prefix assignment $\tau$ of $\{x_1,\dots,x_p\}$, this procedure returns the frontier of $\tau$ by following every decision-gates whose variable is assigned by $\tau$ and sinking every $\times$-gate, implementing the procedure described in~\cref{lem:building-ftau}. Procedure {\sc Count-Extensions} then computes the number of tuples in $\rel(C)$ compatible with $\tau$ and such that $x_{p+1} \leq d$ for some domain value $d \in D$. It uses the frontier of $\tau$ and the equality described in \cref{lem:oracle-complexity}. Observe that one needs access to $\nrel(v,d)$, $\var(v)$ and $|\rel(v)|$ for this, which can be accessed from the annotations of the circuit described in the previous section. 

The last function is the direct access function. The {\sc Next-Value} function takes a circuit, a prefix assignment $\tau$ of $x_1,\dots,x_{p-1}$, an index $k$ and finds the value on $x_p$ of the $k^{\mathsf{th}}$ tuple of $\sigma_\tau(\rel(C))$. It does so by using calls to the {\sc Count-Extensions} function and does a binary search over domain value. We then implements the approach described in \ref{lem:compl-rk} in the {\sc Direct-Access} function by iteratively constructing $\tau$ using {\sc Next-Value} and also updating the index to take into account discarded tuples from $\rel(C)$.

\begin{algorithm}[htp]
  \caption[Count extension]{Given an \ocircuit{} $C$ on domain $D$, a prefix assignment $\tau$ assigning $\{x_1,\dots, x_p\}$ and a domain value $d$, {\sc Count-Extensions} returns the number of tuples in $\rel(C)$, compatible with $\tau$ such that $x_{p+1} \leq d$.}
  \label{alg:countext}
  \begin{algorithmic}[1]
    \Procedure{Find-Frontier}{$C$, $\tau$}
    \State $tmp \gets \{\out(C)\}$\;
    \State $F \gets \emptyset$\;
    \While{$tmp \neq \emptyset$}
    \For{$v \in tmp$}
    \eIf{$v$ is a $\bot$-gate}{\Return $\emptyset$}
    \eIf{$v$ is a $\times$-gate}{$tmp \gets (tmp \setminus \{v\}) \cup \mathsf{input}(v)$;}
    \If{$v$ is a decision-gate on $x$ and $\tau(x)$ is defined}
      \eIf{$v$ has no incoming edge labeled by $\tau(x)$}{\Return $\emptyset$}
      \State $(w,v) \gets$ the incoming edge of $v$ labeled by $\tau(x)$\;
      \State $tmp \gets (tmp \setminus \{v\}) \cup \{w\}$\;
      \Else
      \State{$F \gets F \cup \{v\}$}\;
      \State{$tmp \gets tmp \setminus \{v\}$}\;
    \EndIf
    \EndFor
    \EndWhile
    \State{\Return $F$}
    \EndProcedure
    \Procedure{Count-Extensions}{$C$, $\tau$}
    \State $F \gets $ {\sc Find-Frontier}($C$,$\tau$)\;
    \State $Y \gets \{x_{p+2},\dots,x_n\} \setminus \bigcup_{v \in F} \var(v)$
    \If{$F$ contains a decision-gate $w$ on variable $x_{p+1}$}
    \State \Return $|D|^{|Y|} \times \prod_{w \in F \setminus \{v\}} |\rel(w)| \times \nrel(v,d)$
    \Else
    \State \Return $|D|^{|Y|} \times \prod_{w \in F} |\rel(w)| \times {\mathsf{rank}(d)}$
    
    \EndIf
    \EndProcedure
  \end{algorithmic}
\end{algorithm}

\begin{algorithm}[htp]
  \caption[Direct Access]{Given a \ocircuit{} $C$ on variable $(x_1,\dots, x_n)$, domain $D = (d_1,\dots,d_{|D|})$ and an index $k$, return the $k^{\mathsf{th}}$ tuple in $\rel(C)$}
  \label{alg:da}
    \begin{algorithmic}[1]
      \Procedure{Next-Value}{$C$, $\tau$, $k$, $m$, $M$}
      \If{$m = M$}
      \eIf{{\sc Count-Extensions}($C$,$\tau$,$d_m$)$\geq k$}{\Return $d_m$ {\bf Else} \Return $\bot$}
      \Else
      \State $mi = \lfloor {m + M \over 2} \rfloor$\;
      \eIf{{\sc Count-Extensions}$(C,\tau,d_{mi})\geq k$}{\Return {\sc Next-Value}$(C,\tau,k,m,mi)$}
      \State {\bf else} \Return {\sc Next-Value}$(C, \tau, k, mi, M)$      
      \EndIf
      \EndProcedure
  \end{algorithmic}

  \begin{algorithmic}[1]
    \Procedure{Direct-Access}{$C$, $k$}
    \State $p \gets 1$\;
    \State $\tau \gets \langle \rangle$\;
    \While{$p \leq n$}    
    \State $d \gets $ {\sc Next-Value}$(C,\tau,k,1,|D|)$\;
    \eIf{$d = \bot$}{\Return $\bot$}
    \State $\tau \gets \tau \times [x_p \gets d]$\;
    \State $p \gets p+1$\;
    \State $i \gets \rank(d)$ (that is, $d=d_i$)\;
    \eIf{$i > 1$}{$k \gets k-${\sc Count-Extensions}$(C,\tau,d_{i-1})$}
    \EndWhile
    \EndProcedure
  \end{algorithmic}
\end{algorithm}

\begin{rem}
  Observe that in {\sc Next-Value} and \cref{lem:oracle-complexity}, we implemented the oracle from \cref{lem:compl-rk} using a binary search on the ingoing edges of decision gates, which is the origin of the $\bigo(\log |D|)$ factor in the access time. Actually, the oracle call that we use boils down to solving a problem known as $\mathsf{FindNext}$ in priority queues: given values $v_1 < \dots < v_p$ and $N$, find the smallest $j$ such that $v_j \geq N$. Some data structures such as van Emde Boas trees~\cite{van1975preserving} support this operation in time $\log\ \log\ p$ which would allow us to improve the direct access to $\bigo(\poly(|Q|)\log\ \log\ |D|)$ time for a small sacrifice in the preprocessing (inserting in van Emde Boas trees takes time $\bigo(\log\ \log\ |D|)$) if we store values $\nrel(v,d)$ in such data structures instead of ordered lists.
\end{rem}

\begin{rem}
  We observe that our use of the {\sc Find-Frontier} function is not optimal. Indeed, each call to {\sc Find-Frontier} are dealt with independently while they are actually only updating a given frontier by iteratively adding new values in $\tau$. Hence, in practice, we could get better performances by updating the frontier at each iteration of the for-loop in {\sc Direct-Access} instead of recomputing it from scratch. 
\end{rem}

% \begin{thm}
%   Let $(X,\prec)$ be a finite ordered set and let $C$ be a
%   $\prec$-\ocircuit on variables $X$ and domain $D$. We can answer direct
%   access tasks on $\rel_X(C)$ with preprecossing time $\bigo(|C| \cdot
%   \poly n \cdot \polylog |D|)$ and access time $\bigo(\poly n \cdot
%   \polylog |D|)$ where $n=|X|$.
% \end{thm}

%%% Local Variables:
%%% mode: latex
%%% TeX-master: "main"
%%% End:

\section{From join queries to \ocircuit{s}}
\label{sec:cqtocircuits}

In this section, we present a simple top-down algorithm, that can be seen as an
adaptation of the exhaustive DPLL algorithm from~\cite{sang04}, such that given a signed join query $Q$, $\prec$ an order on its variables and $\db$ a database, it returns a $\succ$-\ocircuit{} $C$ such that $\rel(C) =
Q(\db)$. Exhaustive DPLL is an algorithm that was originally devised to solve the \#SAT problem. It was observed by Huang and Darwiche~\cite{huang_dpll_2005} that the trace of this algorithm implicitly builds a Boolean circuit, corresponding to the \dcircuit{s} on domain $\{0,1\}$, enjoying interesting tractability properties. We show how to adapt it to the framework of signed join queries. The algorithm itself is presented in \cref{sec:exha-dpll-sign}. We study the complexity of this algorithm in \cref{sec:compl-exha-dpll} depending on the structure of $Q$ and $\prec$, using hypergraph structural parameters introduced in \cref{sec:hyperorder-width}.

\subsection{Exhaustive DPLL for signed join queries}
\label{sec:exha-dpll-sign}

The main idea of DPLL for signed join queries is the following: given an order $\prec$ on the variables of a join query $Q$ and a database $\db$, we construct a $\succ$-\ocircuit{} (where $x \succ y$ is defined as a shorthand for $y \prec x$)\footnote{It may look strange at this point that the order given in the input differs from the order used by the output circuit. We will see later that $\prec$ corresponds to a structural parameters, generalizing existing results~\cref{sec:hyperorder-width} but we have to follow it in reverse to build the circuit.} computing $\ans$ by successively testing the variables of $Q$ with decision gates, from the highest to the lowest wrt $\prec$. At its simplest form, the algorithm picks the highest variable $x$ of $Q$ wrt $\prec$, creates a new decision gate $v$ on $x$ and then, for every value $d \in D$, sets $x$ to $d$ and recursively computes a gate $v_d$ computing the subset of $\ans$ where $x=d$. We then add $v_d$ as an input of $v$. This approach is however not enough to get interesting tractability results. We hence add the following optimizations. First, we keep a cache of already computed queries so that if we recursively call the algorithm twice on the same input, we can directly return the previously constructed gate. Moreover, if we detect that the answers of $Q$ are the Cartesian product of two or more subqueries $Q_1, \dots, Q_k$, then we create a new $\times$-gate $v$, recursively call the algorithm on each component $Q_i$ to construct a gate $w_i$ and plug each $w_i$ to $v$. Detecting such cases is mainly done syntactically, by checking whether the query can be partitioned into subqueries having disjoint variables. However, this approach would fail to give good complexity bounds in the presence of negative atoms. To achieve the best complexity, we also remove from $Q$ every negative atom as soon as it is satisfied by the current partial assignment. This allows us to discover more cases in which the query has more than one connected component.

The complexity of the previously described algorithm may however vary if one is
not careful in the way the recursive calls are actually made. %
We hence give a more formal presentation of the algorithm, whose pseudocode is
given in \cref{alg:dpll}. %
We prove upper bounds on the runtime of \cref{alg:dpll} parametrised by the
structure of the query in \cref{sec:compl-exha-dpll}. %
Since we are not insterested in complexity analysis yet, we deliberately let the
underlying data structures for encoding relations unspecified and delay this
discussion to \cref{sec:compl-exha-dpll}.

\begin{algorithm}[htp]
  \caption[Compilation algorithm]{An algorithm to compute a $\succ$-ordered \dcircuit{} computing $\ans$}
  \label{alg:dpll}
  \begin{algorithmic}[1]
    \Procedure{$\DPLL$}{$Q,\tau,\db,\prec$} %\Comment{$v$ is a gate}
    \eIf{$(Q, \tau)$ is in $\cache$}{\Return $\cache(Q, \tau)$}
    \eIf{$Q$ is inconsistent with $\tau$}{\Return $\bot$-gate}
    \eIf{$\tau$ assigns every variable in $Q$}{\Return $\top$-gate}

    \State $x \gets \max_{\prec}  \var(Q)$
    \For{$d \in D$} \label{line:dpllfor} % TODO: beware, that may be too much and we may want to filter out a bit around here; check leapfrog.
    \State $\tau' \gets \tau \times [x \gets d]$ \label{line:buildtau}
    \If{$Q$ is inconsistent with $\tau'$} \label{line:inconsistency}
     {$v_d \gets $ $\bot$-gate}
    \Else
    \State Let $Q_1, \dots, Q_k$ be the $\tau'$-connected components of $\simp[\tau']$
    \For{$i=1\ to\ k$}
    \State $w_i \gets \DPLL(Q_i, \tau_i,  \db, \prec)$ where $\tau_i=\tau'|_{\var(Q_i)}$ \label{line:dpllreccall}
    \EndFor
    \If{$k=1$}
    \State $v_d \gets w_1$
    \Else
    \State $v_d \gets \New $ $\times$-gate with inputs $w_1, \dots, w_k$
    \EndIf
    \EndIf

    \EndFor
    \State $v \gets \New$ $\dec$-gate connected to $v_d$ by a $d$-labelled edge for every $d \in D$ \label{line:dplldec}
    \State $\cache(Q,\tau) \gets v$
    \State \Return $v$

    \EndProcedure
  \end{algorithmic}
\end{algorithm}

A few notations are used in \cref{alg:dpll}. Given a database $\db$ on domain $D$ and a tuple $\tau \in D^Y$ (where $Y$ may contain variables not present in $Q$), we denote by $\ans[Q][\tau]$ the set of tuples $\sigma \in D^{\var(Q) \setminus Y}$ that are answers of $Q$ when extended with $\tau$. More formally, $\sigma \in \ans[Q][\tau]$ if and only if  $(\sigma \times \tau)|_{\var(Q)} \in \ans$. Given an atom $R(\tup)$, a database $\db$ and a tuple $\tau \in D^Y$, we say that \emph{$R(\tup)$ is inconsistent with $\tau$ with respect to $\db$} (or simply inconsistent with $\tau$ when $\db$ is clear from context) if there is no $\sigma \in R^\db$ such that  $\tau \simeq \sigma$. Observe that if $Q$ contains a positive atom $R(\tup)$ that is inconsistent with $\tau$ then $\ans[Q][\tau] = \emptyset$. Similarly, if $Q$ contains a negative atom $\neg R(\tup)$ such that $\tau$ assigns every variable of $\tup$ to a domain value and $\tau(\tup) \in R$, then $\ans[Q][\tau]=\emptyset$. If one of these cases arises, we say that $Q$ is \emph{inconsistent with $\tau$}. Now observe that if $\neg R(\tup)$ is a negative atom of $Q$ such that $R(\tup)$ is inconsistent with $\tau$, then $\ans[Q][\tau] = \ans[Q'][\tau] \times D^W$ where $Q' = Q \setminus \{ \neg R(\tup)\}$ and $W = \var(Q) \setminus (\var(Q') \cup \var(\tau))$  (some variables of $Q$ may only appear in the atom $\neg R(\tup)$). This motivates the following definition: the \emph{simplification of $Q$ with respect to $\tau$ and $\db$}, denoted by $\simpd$ or simply by $\simp$ when $\db$ is clear from context, is defined to be the subquery of $Q$ obtained by removing from $Q$ every negative atom $\neg R(\tup)$ of $Q$ such that $R(\tup)$ is inconsistent with $\tau$. From what precedes, we clearly have $\ans[Q][\tau] = \ans[Q'][\tau] \times D^W$ where $Q'=\simpd$ and $W = \var(Q) \setminus (\var(Q') \cup \var(\tau))$.

For a tuple $\tau \in D^Y$ assigning a subset $Y$ of variables of $Q$, the \emph{$\tau$-intersection graph $\igraph[\tau][Q]$ of $Q$} is the graph whose vertices are the atoms of $Q$ having at least one variable not in $Y$, and there is an edge between two atoms $a,b$ of $Q$ if $a$ and $b$ share a variable that is not in $Y$. Observe that $\igraph$ does not depend on the values of $\tau$ but only on the variables it sets. Hence it can be computed in polynomial time in the size of $Q$ only. A connected component $C$ of $\igraph$ naturally induces a subquery $Q_C$ of $Q$ and is called a \emph{$\tau$-connected component}. $Q$ is partitioned into its $\tau$-connected components and the set of atoms whose variables are completely set by $\tau$. More precisely, $Q = \bigcup_{C \in \mathcal{CC}} Q_C \cup Q'$ where $\mathcal{CC}$ are the connected components of $\igraph$ and $Q'$ contains every atom $a$ of $Q$ on variables set $\tup$ such that $\tup$ only has variables in $Y$. Observe that if $\tau$ is an answer of $Q'$, then $\ans[Q][\tau] = \bigtimes_{C \in \mathcal{CC}} \ans[Q_C][\tau_C]$ where $\tau_C = \tau|_{\var(Q_C)}$ since if $C_1$ and $C_2$ are two distinct $\tau$-connected components of $\igraph$, then $\var(Q_{C_1}) \cap \var(Q_{C_2}) \subseteq Y$.

% \begin{exa}
%   \label{ex:dpllex}
% We illustrate the previous definitions on the signed join query $Q(x_1,\dots,x_5)$ defined as $\neg R(x_1,\dots,x_5), S(x_1,x_2,x_3), T(x_1,x_4,x_5)$ and database $\db$ on domain $\{0,1\}$ with $R^\db = \{(1,1,1,1,1)\}$. Let $\tau = [x_1 \gets 0]$. The $\tau$-intersection graph of $Q$ is a path where $\neg R(x_1,\dots,x_5)$ is connected to $S(x_1,x_2, x_3)$ and $T(x_1, x_4, x_5)$. There is no edge between $S(x_1,x_2, x_3)$ and $T(x_1, x_4, x_5)$ since $x_1$ is their only common variable and it is assigned by $\tau$. Hence, $Q$ has one $\tau$-connected component containing every atom of $Q$. Now, $\simp = S(x_1,x_2,x_3), T(x_1,x_4,x_5)$ since $R(0,x_2,\dots,x_5)$ is inconsistent with $\tau$ wrt $\db$. The $\tau$-intersection graph of $\simp$ consequently consists in two isolated vertices $S(x_1,x_2,x_3)$ and $T(x_1, x_4, x_5)$. Thus $\simp$ has two $\tau$-connected components. This example also illustrates the role of simplification for discovering Cartesian products.
% \end{exa}

\begin{exa}
	\label{ex:dpllex}
  We illustrate the previous definitions and a run of \cref{alg:dpll} on the following query: $Q(x_1,x_2,x_3,x_4) = \neg S(x_1,x_2,x_3,x_4) \wedge T(x_1,x_3) \wedge R(x_2,x_4)$ and relations given in \cref{fig:dpllex:tables} on domain $\{0,1\}$.\footnote{An interactive version of this example and others can be found at \url{https://florent.capelli.me/algorithms/dpll}.} The circuit built by \cref{alg:dpll} is depicted in \cref{fig:dpllex:circuit}. Observe that the first recursive call happens on assignment $\tau_0 = [x_1 \gets 0]$ which is still consistent with $S$. Hence, the $\tau_0$-intersection graph of $Q$ is a path connecting $T, S$ and $R$ with $S$ between $T$ and $R$. It only has one connected component, hence no Cartesian product gate is produced and a decision gate on $x_2$ is created. The next recursive call happens on assignment $\tau_1 = [x_1 \gets 0, x_2 \gets 0]$. The $\tau_1$-intersection graph of $Q$ is the same as before and has only one connected component. A decision gate on $x_3$ is then created and a recursive call happens with $\tau_2 = [x_1 \gets 0,x_2 \gets 0,x_3 \gets 0]$. Now, $\neg S$ and $\tau_2$ are inconsistent, hence a $\bot$-gate is created and the algorithm backtracks and sets $x_3$ to $1$ to build the partial assignment $\tau' = [x_1 \gets 0,x_2 \gets 0,x_3 \gets 1]$ on Line~\ref{line:buildtau}. Now observe that $\simp[\tau']$ only contains atom $T(x_2,x_4)$ because $S(x_1,x_2,x_3,x_4)$ is inconsistent with $\tau'$ and hence atom $\neg S$ is removed. Moreover, all variables of $R(x_1,x_3)$ are set by $\tau'$ and it is also removed in $\simp[\tau']$. The $\tau'$-intersection graph of $Q$ hence has only one vertex $T(x_2,x_4)$. A recursive call is then issued with input $(T(x_2,x_4), [x_2 \gets 0])$ since $\tau'|_{x_2,x_4} = [x_2 \gets 0]$. A decision gate is created for $x_4$ and both values $0,1$ give answers of $Q$ which creates two $\top$-gates.

  Now, the algorithm backtracks to the decision gate labelled by $x_2$ and deals with this part of the recursive call. Later,  the algorithm will eventually backtrack to the first call in the stack and set $\tau' = [x_1 \gets 1]$ on Line~\ref{line:buildtau}. As before, $\neg S$ is simplified in $\simp[\tau']$. Hence the $\tau'$-intersection graph of $\simp[\tau']$ has only one vertex for $T$ and one vertex of $R$. But both atoms do not share variables, hence a Cartesian product gate is created and two recursive calls happen. The first one has parameter $(R(x_1,x_3), [x_1 \gets 1])$ and the other one $(T(x_2,x_4), [])$.  We focus on this second recursive call. The first call it makes is on $(T(x_2,x_4), [x_2 \gets 0])$. But now, recall that this call has already been made (see the end of previous paragraph). Hence the gate computing it is stored in the cache and a cache hit occurs, leading to sharing in the circuit.

  %NEEDFIX Vertical Label alignment required
  % Fixed Table to be aligned with eachother
  \begin{figure}
    \centering
    \begin{subfigure}[b]{0.48\linewidth}
    \centering
      \begin{tabular}{cccccc}
        $S$ &  & $x_1$ & $x_2$ & $x_3$ & $x_4$ \\ \midrule
            && 0 & 0 & 0 & 0 \\
            \\

        $R$& & $x_2$ & $x_4$ \\ \midrule
          && 0 & 0 \\
          && 0 & 1 \\
          && 1 & 1 \\
          && 1 & 0 \\
          \\
        $T$&  & $x_1$ & $x_3$ \\ \midrule
          && 0 & 1 \\
          && 1 & 1 \\
      \end{tabular}

        \caption{Database on domain $\{0,1\}$.}
        \label{fig:dpllex:tables}
      \end{subfigure}
      \hspace{1mm}
      \begin{subfigure}[b]{0.48\linewidth}
        \includegraphics{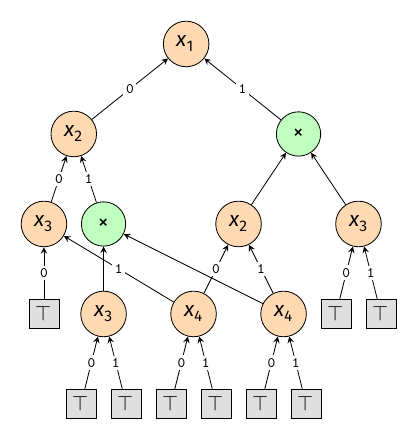}
        \caption{Produced circuit.}
        \label{fig:dpllex:circuit}
      \end{subfigure}
  
  \caption{A run of \cref{alg:dpll} with $Q(x_1,x_2,x_3,x_4) = \neg S(x_1,x_2,x_3,x_4) \wedge R(x_2,x_4) \wedge T(x_1,x_3)$. Details are given in \cref{ex:dpllex}}

  \end{figure}
\end{exa}

\cref{alg:dpll} uses the previous observations to produce a $\succ$-\ocircuit{}. More precisely:
\begin{thm}
  Let $Q$ be a signed join query, $\db$ a database and  $\prec$ an order on $\var(Q)$, then $\DPLL(Q,\emptytup,\db,\prec)$ constructs a $\succ$-\ocircuit{} $C$ and returns a gate $v$ of $C$ such that $\rel_{\var(Q)}(v) = \ans$.
\end{thm}
\begin{proof}
  The proof is by induction on the number of variables of $Q$ that are not assigned by $\tau$. We claim that $\DPLL(Q,\tau,\db,\prec)$ returns a gate computing $\ans[Q][\tau]$ which is stored into $cache(Q,\tau)$. If every variable is assigned, then $\DPLL(Q,\tau,\db,\prec)$ returns either a $\top$-gate or a $\bot$-gate depending on whether $\tau$ is inconsistent with $Q$ or not, which clearly is $\ans[Q][\tau]$. Otherwise, it returns and adds in the cache a decision gate $v$ connected to a gate $v_d$ by a $d$-labelled edge for each $d \in D$. We claim that $v_d$ computes $\ans[Q][\tau\times[x \gets d]]$. It is enough since in this case, by definition of the relation computed by a decision-gate, $v$ computes $\bigcup_{d \in D} \ans[Q][\tau\times[x \gets d]] \times [x \gets d] = \ans[Q][\tau]$.

  To prove that $v_d$ computes $\ans[Q][\tau']$ where $\tau'=\tau\times[x \gets d]$, we separate two cases: if $\tau'$ is inconsistent with $Q$ then $\ans[Q][\tau']$ is empty and $v_d$ is a $\bot$-gate, which is what is expected. Otherwise, let $Q_1,\dots,Q_k$ be the $\tau'$-connected components of $\simp[\tau']$ and let $\tau_i = \tau'|_{\var(Q_i)}$. From what precedes, we have $\ans[Q][\tau'] = \bigtimes_{i=1}^k \ans[Q_i][\tau_i]$. The algorithm uses a gate $w_i$ from Line~\ref{line:dpllreccall}, obtained from a recursive call to $\DPLL(Q_i, \tau_i, \db, \prec)$ where the number of variables not assigned by $\tau_i$ in $Q_i$ is less than the number of variables unassigned by $\tau$ in $Q$. Hence, by induction, $w_i$ computes $\ans[Q_i][\tau_i]$ and since $v_d$ is a $\times$-gate connected to each $w_i$, we indeed have $\rel(v_d) = \bigtimes_{i=1}^k \ans[Q_i][\tau_i]$.
\end{proof}

The worst case complexity of $\DPLL$ may be high when no cache hit occurs which would result in at least $|\ans|$ recursive calls. However, when $\prec$ has good properties with respect to $Q$, we can prove better bounds. \cref{sec:compl-exha-dpll} gives upper bounds on the complexity of $\DPLL$ depending on $\prec$, using measures  defined in \cref{sec:hyperorder-width}.

\subsection{Complexity of exhaustive DPLL}
\label{sec:compl-exha-dpll}

The complexity of DPLL on a conjunctive query $Q$ and order $\prec$ can be bounded in terms of the hyperorder width of $H(Q)$ wrt $\prec$:

\begin{thm}
  \label{thm:compilation}
  Let $Q$ be a signed join query, $\db$ a database over domain $D$ and $\prec$ an order on $\var(Q)$. Then $\DPLL(Q,\emptytup,\db,\prec)$ produces a $\succ$-\ocircuit{} $C$ of size $\bigo((\poly_k |Q|) \cdot |\db|^{k} \cdot |D|)$ such that $\rel(C) = \ans$ and:
  \begin{itemize}
  \item $k=\fhow{H(Q)}[\prec]$ if $Q$ is positive,
  \item $k=\show{H(Q)}[\prec]$ otherwise.
  \end{itemize}
  Moreover, the runtime of $\DPLL(Q,\emptytup,\db,\prec)$ is at most $\bigot(\poly_k(|Q|) \cdot |\db|^{k} \cdot |D|)$.
\end{thm}

We now proceed to prove Theorem~\ref{thm:compilation}. %
In this section, we fix a signed join query $Q$ that has exactly one
$\emptytup$-component. %
This can be done without loss of generality  since the case where $Q$ has many $\emptytup$-components
can be easily dealt with by constructing the Cartesian product of each
$\emptytup$-component of $Q$. %
We also fix a database $\db$ and an order $\prec$ on $\var(Q)=\{x_1,\dots,x_n\}$
where $x_1 \prec \dots \prec x_n$. %
We let $D$ be the domain of $\db$, $n$ be the number of variables of $Q$ and $m$
be the number of atoms of $Q$. %
To ease notation, we will write $X$ instead of $\var(Q)$. For $i \leqslant n$,
we denote $\{x_1,\dots, x_i\}$ by $X_{\preceq x_i}$. %
Similarly, $X_{\prec x_i}=X_{\preceq x_i} \setminus \{x_i\}$, $X_{\succ
  x_i}=\var(Q) \setminus X_{\preceq x_i}$ and $X_{\succeq x_i}=\var(Q) \setminus
X_{\prec x_i}$. %S: changed i to x_i as in the sequel, variables are used and not
                % indices.

%
Finally, we let $\rec$ be the set of $(K,\sigma)$ such that $\DPLL(Q,\emptytup,\db,\prec)$ makes at least one recursive call to $\DPLL(K,\sigma,\db,\prec)$ and such that $K$ is consistent with $\sigma$. We start by bounding the size of the circuit and the runtime in terms of the number of recursive calls:

\begin{lem}
  \label{lem:sizefromrec}  $\DPLL(Q,\emptytup,\db,\prec)$ produces a circuit of size at most $\bigo(|\rec| \cdot |D| \cdot \poly(|Q|))$ in time $\bigot(|\rec| \cdot |D|\cdot\poly(|Q|))$.
\end{lem}

\begin{proof}
Given $(K,\sigma) \in \rec$, we bound the number of edges created in the circuit during the first recursive call with these parameters. There are at most $m+1$ such edges for each $d \in D$. Indeed, for a value $d \in D$, there are at most $m$ $\sigma'$-connected components for $\sigma'=\sigma \cup [x\gets d]$ hence the first recursive call creates at most $m$ edges between $v_d$ and $w_i$ and one edge between $v$ and $v_d$. Observe that any other recursive call with these parameters will not add any extra edges in the circuit since it will result in a cache hit. Moreover, any recursive call of the form $(K,\sigma)$ with $K$ inconsistent with $\sigma$ will return a $\bot$-gate without creating any new edge. Hence, the size of the circuit produced in the end is at most $|\rec| \cdot |D| \cdot (m+1) = \bigo(|\rec| \cdot |D| \cdot \poly(Q))$.

Moreover, each operation in \cref{alg:dpll} can be done in time polynomial in $|Q|$ if one stores the relation using a well-chosen data structure. Indeed, if one sees a relation $R$ on variables $x_1 \prec \dots \prec x_n$ as a set of words on an alphabet $D$ whose first letter is $x_n$ and last is $x_1$, one can store it as a trie of size $\bigot(|R|)$ and project $R$ on $x_1, \dots, x_{n-1}$ in $\bigot(1)$ (this is the encoding used in the Triejoin algorithm from~\cite{triejoin}). Hence, we can test for inconsistency in time $\bigot(|Q|)$ after having fixed the highest variables in $Q$ to a value $d \in D$ by going over every atom of $Q$. During a recursive call $(K,\sigma)$ where $K$ is inconsistent with $\sigma$, we have to check for this inconsistency before returning $\bot$ and cannot really do it for free. To simplify the analysis, we assume that this check is performed before calling the function and hence we can assume every recursive call $(K,\sigma)$ where $K$ is inconsistent with $\sigma$ is done for free. In other words, we incur the cost of this call to the calling function. 

Moreover computing the $\sigma'$-connected components can be done in polynomial time in $|Q|$ since it boils down to finding the connected components of a graph having at most $m$ nodes. Such a graph can be constructed in polynomial time in $|Q|$ by testing intersections of variables in atoms. Finally, from the previous discussion, a recursive call to \cref{alg:dpll} creates at most $m+1$ edges for each $d \in D$. Moreover, reading and writing values in the cache can be done in time $\bigot(\poly(|Q|))$ by using a hash table. Indeed, the cached values are subqueries of $Q$ together with partial variable assignments, hence they can be encoded with $\bigo(|Q|\log |D|)$ bits.  If we account for the cost of reading the cache in a recursive call and the cost of checking inconsistency (as explained before) directly on Line~\ref{line:dpllreccall}, the time for each $(K,\sigma) \in \rec$ is $\bigot(|D| \cdot \poly(|Q|))$ and the total time is $\bigot(|\rec| \cdot |D|\cdot\poly(|Q|))$.
\end{proof}

It remains to bound the size of $\rec$ which is done in two steps. %
\cref{lem:reccall} characterises the structure of the elements of $\rec$ and
\cref{lem:rechyp} shows connections with the structure of the hypergraph of
$Q$. %
We need a few notations.
Let $Q' \subseteq Q$ be a subquery of $Q$ and $x,y$ two variables of $Q'$ such
that $y \prec x$. An \emph{$x$-path to $y$ in $Q'$} is a list $x_0, a_0, x_1,
a_1,\dots, a_{p-1}, x_p$ where $a_i$ is an atom of $Q'$ on variables
$\tup[x_i]$, $x_i$ and \(x_{i+1}\) are variables of $\tup[x_i]$, $x_0=x$, $x_p =
y$ and $x_i \preceq x$ for every $i \leqslant
p$. % S: added x_{i+1} is in \tup[x_i], otherwise it is not
    % a path
Thus, \(x\)-paths to \(y\) in the hypergraph of $Q'$ are simply paths that start
from $x$, end in \(y\) and are only allowed to go through variables smaller than
$x$. % S: I changed the
% sentence, but now it is just a rephrasal of
% the formal explanation and it seems useless.
By extension, given an atom \(a\) of \(Q'\), an \emph{\(x\)-path to \(a\) in
  \(Q'\)} is an \(x\)-path to some variable \(y\) in \(a\). %
The \emph{$x$-component of $Q'$} is the set of atoms $a$ of $Q'$ such that there
exists an $x$-path to $a$ in $Q'$. %
An \emph{\(x\)-access to \(y\) in \(Q'\)}, is an \(x\)-path to some atom \(a\)
such that is a variable of \(a\). %
Notice that there is an \(x\)-access to \(y\) in \(Q'\) iff \(y\) occurs in the
\(x\)-component of \(Q'\). %
Notice that when there is an \(x\)-access to \(y\) in \(Q'\), it may be the case
that \(x \prec y\).%

% Given a conjunctive query $Q$, we define $\qabs[Q]$ to be the conjunctive query whose atoms are $\atomp[Q] \cup \{R(\tup) \mid \neg R(\tup) \in \atomn\}$, that is, $\qabs[Q]$ is the positive conjunctive query obtained by changing every negative atom of $Q$ into a positive one.

It turns out that the recursive calls performed by $\DPLL$ are in correspondence with the $x$-components of some  $x \in X$ and $Q' \subseteq Q$ where $Q'$ is obtained from $Q$ by removing negative atoms. Intuitively, these removed atoms are the ones that cannot be satisfied anymore by the current assignment of variables.

\begin{figure}
  \centering
  \includegraphics[scale=1.25]{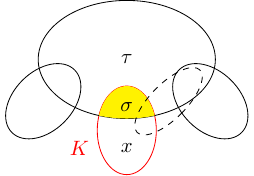}
  \caption{Illustration of \cref{lem:reccall}: $\tau$ is the assignments of variables so far when the first recursive call to $(K,\sigma)$ occurs and it separates $Q$ into disjoint $\tau$-components. Some  negative atoms of $Q$ (represented with dashed edges) have been simplified away in $\simp$. $K$, in red, is a $\tau$-connected component of $\simp$: it is connected only through the atoms not simplified in $\simp$.  $\sigma$ is the part of $\tau$ that sets variables in $K$, and $x$ is the largest variable of $K$ not assigned by $\sigma$. \cref{lem:reccall} proves that $K$ is exactly the $x$-component of $\simp$, that is, every atom that can be reached from $x$ using atoms in $\simp$ and variables smaller than $x$.}
    \label{fig:reccall}
\end{figure}

This observation is stated formally in \cref{lem:reccall} whose proof is quite technical, due in part to heavy notations and concepts that are necessary to truly formalise what is happening. A picture illustrating the proof of \cref{lem:reccall} can be found on \cref{fig:reccall}.

\begin{lem}
  \label{lem:reccall}
  Let $(K,\sigma) \in \rec$ and let $x$ be the biggest variable of $K$ not assigned by $\sigma$. There exists $\tau$ a partial assignment of $X_{\succ x}$ such that: $\sigma = \tau|_{\var(K)}$ and $K$ is the $x$-component of $\simp$.
\end{lem}
\begin{proof}
  The proof is by induction on the order of recursive calls. We start by the first call, $(Q, \emptytup)$. Since $Q$ has one $\emptytup$-component, the $x_n$-component of $Q$ is $Q$ itself. Moreover, since $\simp[\emptytup] = Q$, we have that $Q$ is the $x_n$-component of $\simp[\emptytup]$. Now let $(K,\sigma) \in \rec$. By definition, the recursive call is made during the execution of another recursive call with parameters $(K',\sigma')$.  Assume by induction that for $(K', \sigma')$, the statement of Lemma~\ref{lem:reccall} holds. In other words, let $x'$ be the biggest variable of $K'$. Then $K'$ is the $x'$-component of $\simp[\tau']$ for some partial assignment $\tau'$ of $X_{\succ x'}$ such that $\tau'|_{\var(K')} = \sigma'$.  The recursive call then made to $(K,\sigma)$ has the following form: $\sigma = \sigma''|_{\var(K)}$ where $\sigma'' = \sigma' \cup [x'\gets d]$ for some $d \in D$ and $K$ is a $\sigma''$-component of $\simp[\sigma''][K']$.

  We claim that $K$ is the $x$-component of $\simp[\tau]$, where $\tau = \tau'
  \cup [x' \gets d]$. %
  First observe that every atom $a$ of $K$ is in $\simp[\tau]$. Indeed, if $a$
  is positive, then $a$ is also in $\simp[\tau]$ by definition. %
  Now if $a = \neg R(\tup)$ is negative, we claim that $a$ is not inconsistent
  with $\tau$. %
  Indeed, by induction, $a$ is in $\simp[\tau']$, hence it is not inconsistent
  with $\tau'$. %
  Now since $a$ is also in $K$ and $K$ is a subset of $\simp[\sigma''][K']$ and
  $\sigma''(x')=d$, we know that $a$ is not inconsistent with $\tau' \cup [x
  \gets d]$ which is $\tau$ by definition. %
  Hence $a$ is in $\simp[\tau]$.

  Now let $a$ be an atom of $K$. %
  % Since $x$ is a variable of $K$, there is an atom $a_0$ in $K$ that contains
  % $x$. % Sylvain : this is not useful to mention the bag containing x.
  $K$ is a $\sigma''$-connected component of $\simp[\sigma''][K']$. %
  Hence, by definition, we have a path in $K$ from $x$ % (starting with atom
  % $a_0$) Sylvain : here again this does not bring anything
  to some variable $y$ in $a$ that does not use any variable assigned by
  $\sigma''$, which is equivalent to saying that it does not use any variable
  assigned by $\sigma$ since, by definition, $\sigma = \sigma''|_{\var(K)}$. %
  As $x$ is defined to be the biggest variable of $K$ that is not assigned by
  $\sigma$, the path from $x$ to $y$ only uses variables smaller than
  $x$. %
  In other words, there is an $x$-path to $y$ in $K$. That is, every atom $a$ of
  $K$ is in the $x$-component of $\simp[\tau]$.

  Now let $a$ be an atom that is in the $x$-component of $\simp[\tau]$. %
  % We first show that $a$ is in $K'$. %
  % First of all, observe that $a$ is in $\simp[\tau']$ since
  % $\simp[\tau]\subseteq \simp[\tau']$. %
  Let \(y\) be a variable in \(a\) such that there is an \(x\)-path to $y$ in
  $\simp[\tau]$. % that uses only variables smaller than $x$. %
  %Let $a_0$ be the atom of $K$ containing $x$ used in the path from \(x\) to
  %\(y\). %
  We call this path \(p\). %
  We first show that \(p\) is also in \(K'\). %
  Since $\simp[\tau] \subseteq \simp[\tau']$, \(p\) is also an \(x\)-path to
  \(y\) of $\simp[\tau']$. %
  % Hence, there is a path from \(x\) to \(y\) using only variables smaller than
  % $x$, hence also smaller than $x'$. %
  As \(x\) is occurring in \(K\) and \(K\subseteq K'\), \(x\) occurs in some
  atom of \(K'\). %
  Moreover, as \(K'\) is the \(x'\)-component of \(\simp[\tau']\), there is an
  \(x'\)-path to \(x\), call it \(p'\), in \(K'\). %
  Since \(K' \subseteq \simp[\tau']\), \(p'\) is also an \(x'\)-path to \(x\) in
  \(\simp[\tau']\). %
  The path \(p'p\) obtained by concatenating the \(p'\) and \(p\) in
  \(\simp[\tau']\) is an \(x'\)-path to \(y\) in \(\simp[\tau']\). %
  Since \(K'\) is the \(x'\)-component of \(\simp[\tau']\), \(p'p\) is in \(K'\)
  and hence \(p\) is in \(K'\). %
  We now show that \(p\) is in \(K\). %
  By definition, \(K\) is the \(\sigma''\)-connected component of
  $\simp[\sigma''][K']$ that contains \(x\). %
  In particular, it contains every path from \(x\) to other variables in \(K'\)
  that does not pass through \(x'\). %
  As \(p\) is an \(x\)-path to \(y\), it cannot pass through \(x'\) (since
  \(x\prec x'\))  and is thus
  in \(K'\). %
  This finally shows that \(a\) is in \(K\) and concludes the proof. %

  % Now, since $a$ is in $\simp[\tau]$ and since $a \in K'$ from what precedes,
  % $\tau \supseteq \sigma''$ and $K' \subseteq Q$, we have that $a$ is an atom of
  % $\simp[\sigma''][K']$. In particular, $a$ is in the $\sigma''$-component of
  % $\simp[\sigma''][K']$ that contains $x$, that is, it is in $K$, which
  % concludes the proof.
\end{proof}

\begin{figure}
  \centering
  \includegraphics{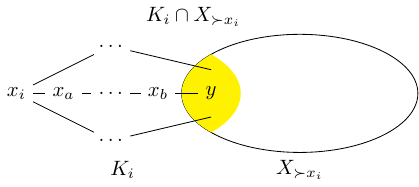}
    \includegraphics[page=2]{images/neighbor.pdf}
  \caption{Illustration of \cref{lem:rechyp}: in the first picture, $K_i$ is pictured on the left and the yellow area represents the variables in $K_i$ that are greater than $x_i$, that is, that can be reached by following paths of variables smaller than $x_i$. When removing variables smaller than $x_i$, these paths are progressively merged into the neighborhood of $x_i$ resulting in the second picture: in $H_i$, the neighborhood of $x_i$ is exactly the yellow area from the first picture, that is, the vertices of $K_i$ greater than $x_i$. Every other variable from $K_i$ have been removed. The proof of \cref{lem:rechyp} consists in proving by induction on paths length that both yellow areas are the same.} 
    \label{fig:neighbors}
\end{figure}

The following lemma establishes a connection between $x$-components and the structure of the underlying hypergraph. In essence, it allows us to bound the number of atoms needed to cover $X_{\succ x}$ in an $x$-component using the signed hyperorder width. \cref{fig:neighbors} illustrates the lemma and give an idea of the proof.

\begin{lem}
  \label{lem:rechyp}
  Let $Q$ be a signed join query on variables set $X=\{x_1,\dots,x_n\}$. Let $1 \leq i \leq n$ and $K_i$ the $x_i$-component of $Q$. We let $H$ be the hypergraph of $Q$, $H_1 = H$ and $H_{j+1} = H_j / x_j$. We have $\neigh[x_i][H_i] = \var(K_i) \cap X_{\succeq x_i}$.
\end{lem}
\begin{proof}
  The proof is by induction on $i$. We start by proving the equality for $i=1$.  Since there are no variables of $K_1$ smaller than $x_1$, it is clear that $K_1 = \{a \in \atoms(Q) \mid x_1 \in \var(a)\}$. Hence $\var(K_1) \cap X_{\succeq x_1} = \var(K_1)$. Moreover, $\neigh[x_1][H]$ is exactly the set of variables of atoms of $Q$ containing $x_1$ since there is one hyperedge in $H$ per atom of $Q$. Hence, $\var(K_1)\cap X_{\succeq x_1} = \neigh[x_1][H] = \neigh[x_1][H_1]$, the last equality being a consequence of $H_1 = H$.

  Now assume that the equality has been established up to $x_{i-1}$. %
  We start by proving that $\var(K_i) \cap X_{\succeq x_i} \subseteq
  \neigh[x_i][H_i]$. %
  Let $w \in \var(K_i) \cap X_{\succeq x_i}$. %
  First assume $w \in \neigh[x_i][H]$. %
  That is, there is an edge $e$ in $H$ such that both $w$ and $x_i$ are in
  $e$. %
  Moreover, it is easy to see that $e \setminus \{x_1,\dots,x_{i-1}\}$ is an
  edge of $H_i$. %
  Hence it is clear that $w \in \neigh[x_i][H_i]$. %
  Otherwise, if $w \notin \neigh[x_i][H]$, as \(w\in \var(K_i)\), by definition
  of \(K_i\), there is an \(x_i\)-access of length greater than \(1\) to \(w\)
  in \(K_i\). %
  Let $x_j$ be the biggest node on that path that is different from \(x_i\). %
  By definition, an $x_i$-access is an \(x_i\)-path and $j<i$. %
  Moreover, splitting that of \(x_i\)-access at \(x_j\) gives an
  \(x_j\)-access to \(x_i\) and an \(x_j\)-access to \(w\) in \(K_i\). %
  Thus, \(x_i\) and \(w\) occur in the \(x_j\)-component of \(K_i\). %
  Since \(K_i\subseteq H\), they also occur in \(K_j\), the \(x_j\)-component of
  \(H\). %
  By induction, we thus have $w \in \neigh[x_j][H_j]$ and $x_i \in
  \neigh[x_j][H_j]$. %
  Hence $w$ and $x_i$ are neighbours in $H_{j+1}$ since the edge
  $\neigh[x_j][H_j]\setminus \{x_j\}$ has been added in $H_{j+1}$. %
  In particular, since $i > j$, it means that $H_i$ contains the edge
  $\neigh[x_j][H_j] \setminus \{x_j, \dots, x_{i-1}\}$, which contains both $w$
  and $x_i$, hence $w \in \neigh[x_i][H_i]$. %
  This establishes the first part of the proof, that is, $\var(K_i) \cap
  X_{\succeq x_i} \subseteq \neigh[x_i][H_i]$.

  We now prove the converse inclusion. Let $w \in \neigh[x_{i}][H_{i}]$. %
  By definition, $w \in X_{\succeq x_i}$ since $x_1, \dots, x_{i-1}$ have been
  removed in $H_{i}$. %
  It remains to prove that $w$ is in an atom $a$ such that there is an
  $x_i$-path to $a$. %
  If $x_i$ and $w$ are neighbours in $H$, then it means that they appear
  together in an atom of $Q$ and it is clear that $w \in \var(K_i)$. %
  Otherwise, let $j$ be the smallest value for which $w \in \neigh[x_i][H_{j}]$
  which exists since $w \in \neigh[x_i][H_i]$. %
  We must have $j>1$ since $x_i$ and $w$ are not neighbours in $H=H_1$. %
  The minimality of $j$ implies that $x_i$ and $w$ are not neighbours in
  $H_{j-1}$. %
  Since the only edge added in $H_j$ is $\neigh[x_{j-1}][H_{j-1}] \setminus
  \{x_{j-1}\}$, it means that both $x_i$ and $w$ are neighbours of $x_{j-1}$ in
  $H_{j-1}$, that is, $x_i \in \neigh[x_{j-1}][H_{j-1}]$ and $w \in
  \neigh[x_{j-1}][H_{j-1}]$. %
  By induction, both $x_i$ and $w$ are variables of $K_{j-1}$. %
  In other words, there is an \(x_{j-1}\)-access to \(x_i\) and an
  \(x_{j-1}\)-access to \(w\) in \(K_{j-1}\). %
  Since \(j-1<i\), the paths that constitute these \(x_{j-1}\)-accesses only
  pass through variables \(y\) so that \(y\prec x_{j-1}\prec x_i\). %
  This shows that that there is an \(x_i\)-access to \(w\) in \(K_{j-1}\). %
  So \(w\) is in the \(x_i\)-component of \(K_{j-1}\) and thus in the
  \(x_i\)-component of \(H\). %
  Hence $w \in \var(K_i)$. %
\end{proof}

We are now ready to prove the upper bound on $|\rec|$ depending on the width of $\prec$. 

\begin{lem}
  \label{lem:boundR}
  Let $m$ be the number of atoms of $Q$ and $n$ the number of variables. We have:
  \begin{itemize}
  \item if $Q$ is a positive join query, $|\rec| \leqslant n|\db|^k$ where $k=\fhow{H(Q)}[\prec]$.
  \item otherwise $|\rec| \leqslant nm^{k+1}|\db|^k$ where $k=\show{H(Q)}[\prec]$.
  \end{itemize}
\end{lem}
\begin{proof}

  We start with the case where $Q$ is a positive join query. Let $(K,\sigma) \in \rec$. In this case, we know by \cref{lem:reccall} that $K$ is the $x_i$-component of $\simp$ for some $\tau \supseteq \sigma$. Now, since $Q$ does not have negative atoms, $Q = \simp$ since $\simp$ is obtained from $Q$ by removing negative atoms only. In other words, $K$ is the $x_i$-component of $Q$ and $\sigma$ assigns the variables of $K$ that are greater than $x_i$. We also know that $\sigma$ is not inconsistent with the atoms of $K$ by definition of $\rec$. Hence, $\sigma$ satisfies every atom of $K$ when projected on $X_{\succ x_i}$. By \cref{lem:rechyp}, $\var(K) \cap X_{\succ x_i} = N_{x_i}(H_i)$ where $H_i$ is defined as in \cref{lem:rechyp}. Thus, by definition, there exits a fractional cover of $\neigh[x_i][H_i]$ using the atoms of $Q$ with value at most $k = \fhow{H(Q)}[\prec]$. In other words, $\sigma$ can be seen as the projection on $\var(K) \cap X_{\succ x_i}$ of an answer of the join of the atoms involved in the fractional cover. By \cref{thm:agm}, this join query has at most $|\db|^k$ answers. Hence, there are at most $n|\db|^k$ possible elements in $\rec$: there are at most $n$ $x_i$-components (one for each $i \leq n$), and at most $|\db|^k$ associated assignments $\sigma$ to each component.

  The case of signed queries is similar but requires more attention. Again, for $(K,\sigma) \in \rec$, we know that $K$ is the $x_i$-component of $\simp$ for some $\tau \supseteq \sigma$ and, as before, $\tau$ is consistent with every positive atom in $\simp$. Moreoever, if $\neg R(\tup)$ is an atom of $\simp$, then $\tau$ is consistent with $R(\tup)$, since otherwise  $\neg R(\tup)$ would not be in $\simp$. Now, let $H'$ be the hypergraph of $\simp$. By definition, it  is a subhypergraph of $H(Q)$, where only negative edges have been removed. Hence, by \cref{lem:rechyp}, $\var(K) \cap X_{\succ x_i} = \neigh[x_i][H'_i]$ is covered by at most $\show{H'}[\prec] \leqslant \show{H(Q)}[\prec] = k$ edges. Hence, $\sigma$, which corresponds to $\tau$ restricted to $\var(K) \cap X_{\succ x_i}$, can be seen as the projection to $\var(K) \cap X_{\prec x_i}$ of an answer of a positive join query having at most $k$ atoms of $Q$. Indeed, even if an edge of $H(Q)$ used to cover $\var(K) \cap X_{\succ x_i}$ corresponds to a negative atom $\neg R$ of $Q$, we know that $\tau$ is consistent with $R$, otherwise, the atom $\neg R$ would have been simplified by $\tau$ in $\simp$.

  Consider $I_k(Q)$ to be the following set: $I_k(Q)$ contains every pair $(L,\alpha)$ such that $L$ is a subset of at most $k$ atoms  $P_1,\dots,P_{k_1}, \neg N_1, \dots, \neg N_{k_2}$ of $Q$ (that is $k_1+k_2 \leq k$) and $\alpha$ is a solution of the join query with atoms $P_1,\dots,P_{k_1}, N_1, \dots, N_{k_2}$ on $\db$. By definition, it is easy to see that $|I_k(Q)| \leq m^{k+1}|\db|^k$. Indeed, there are at most  $\sum_{j=1}^k {m \choose j} \leq m^{k+1}$ choices for $L$ and since $\alpha$ is an answer of a positive join query having at most $k$ relations, there are at most $|\db|^k$ such answers.
  
  From what precedes, we know that for every $(K,\sigma) \in \rec$, there exists some $\alpha$ such that $(K,\alpha) \in I_k(Q)$ and $\sigma = \alpha|_{\var(K) \cap X_{\succ x_i}}$ for some $x_i \in X$. Hence there are at most $nm^{k+1}|\db|^k$ elements in $\rec$, the extra $n$-factor originating from the choice of $x_i$.
\end{proof}

Now \cref{thm:compilation} is a direct corollary of \cref{lem:sizefromrec,lem:boundR}. If $Q$ is not $\emptytup$-connected, then the first recursive call of $\DPLL$ will simply break $Q$ into at most $\bigo(m)$ connected components and recursively call itself on each now $\emptytup$-component of $Q$.

Observe that our result is based on signed hyperorder width, which is not the
fractional version we had in \cref{sec:lowerbound}. %
This is because with our current understanding of the algorithm, we are not able
to prove a polynomial combined complexity bound on the runtime of DPLL when only
the signed fractional hyperorder width is bounded. %
Indeed, the proof of \cref{lem:boundR} breaks in this case. %
The proof would be analoguous up to the definition of $I_k(Q)$. %
Indeed, to account for fractional cover, we should not consider $I_k(Q)$ anymore
but $J_k(Q)$ defined as follows: $J_k(Q)$ contains every pair $(L,\alpha)$ such
that $L$ is a subset of atoms $P_1,\dots,P_{k_1}, \neg N_1, \dots, \neg N_{k_2}$
of $Q$ \textbf{having a fractional cover of at most $k$} and $\alpha$ is a
solution of the join query with atoms $P_1,\dots,P_{k_1}, N_1, \dots, N_{k_2}$
on $\db$. %
Now, bounding the number of $\alpha$ once $L$ is fixed by $|\db|^k$ can still be
done by \cref{thm:agm} but we cannot bound the number of distinct $L$ by
$m^{k+1}$ anymore. %
It is not clear to us whether this number can be bounded by a polynomial (where
$k$ is considered a constant) in $m$. %
The best that we can do is to bound this number by $2^m$, that is, just saying
that $L$ is one subset of atoms of $Q$. %
By doing this, we can still bound $|\rec|$ by $2^mn|\db|^k$ which gives:

\begin{thm}
  \label{thm:compilation-fractional}
  Let $Q$ be a signed join query with $n$ variables and $m$ atoms, $\db$ a database over domain $D$ and $\prec$ an order on $\var(Q)$. Then $\DPLL(Q,\emptytup,\db,\prec)$ produces a $\succ$-\ocircuit{} $C$ of size $\bigo((2^{m} \poly(|Q|) \cdot |\db|^{k} \cdot |D|)$ such that $\rel(C) = \ans$ where $k =\sfhow{H(Q)}[\prec]$.
  Moreover, the runtime of $\DPLL(Q,\emptytup,\db,\prec)$ is at most $\bigot(2^m\poly(|Q|) \cdot |\db|^{k} \cdot |D|)$.
\end{thm}

%Another improvement that could be made in Theorem~\ref{thm:compilation} is to have a dependency of $|\db|^k$ instead of $|\db|^{k+1}$. The extra $|\db|$ comes from the for-loop on Line~\ref{line:dpllfor} that explores every element of the domain. One could improve the complexity here by exploring only the values $d \in D$ such that setting $x$ to $d$ does not make $Q$ inconsistent. One could use the Leapfrog join proposed in the Triejoin algorithm~\cite[Section 3.1 and 3.2]{triejoin} to explore these candidates and we believe it would shave the extra $|\db|$ factor. However, the complexity analysis is already complicated enough and we decided to leave this analysis for future investigations. 

%%% Local Variables:
%%% mode: latex
%%% TeX-master: "main"
%%% End:

\subsection{Binarisation}
\label{sec:binarisation}

In this section, we explain how one can remove a $|D|$ factor in the complexity of DPLL. This is desirable since this extra factor shows a gap between our approach and the optimal bound. We start by observing that this factor is not an over-estimation in the analysis of the algorithm but that it can effectively appear on a concrete example.

\begin{exa}
  \label{ex:nonlinear}
Consider the query $Q := A(x_1) \wedge B(x_2) \wedge \neg R(x_1,x_2)$ (which has already been considered in~\cite{ZhaoFOK24} for similar reasons). Consider order $x_1,x_2$ and take $A(x_1) = [d]$, $B(x_2) = [d]$ and $R(x_1,x_2) = \{(v,v) \mid v \in [d]\}$ (that is $\neg R(x_1,x_2)$ enforces $x_1 \neq x_2$). The size of the database is $\bigo(d)$, and the signed hyperorder width of $Q$ is $1$, but DPLL will produce a quadratic size circuit. Indeed, $\neg R(x_1,x_2)$ prevents DPLL from creating any Cartesian product. Moreover, for $d' \in [d]$, $\simp[[x_1 \gets d']]$ is $\big(B(x_2) \wedge x_1 \neq x_2\big)$. Hence, for $d' \neq d''$, $(\simp[[x_1\gets d']], [x_1 \gets d'])$ and $(\simp[[x_1\gets d'']], [x_1 \gets d''])$ are syntactically distinct since they do not assign the same value to $x_1$. It means that  recursive calls on $x_1 \gets d'$ for $d' \in [d]$ never hit the cache. Hence, the resulting circuit will have size $\bigo(d^2)$ since the returned circuit is essentially a tree with one path for each solution of the form $\{x_1 \gets d_1, x_2 \gets d_2\}$ and $d_1 \neq d_2$. 
\end{exa}

To overcome this inefficiency, we rely on a simple trick that was already used
in the design of worst-case optimal join algorithms~\cite{CapelliIS24} which
consists in encoding the domain over the Boolean domain and transforming the
query and the database accordingly. %
It turns out that this transformation can be done in a way that preserves the
structure of the query. %
Moreover, the isomorphism between the answers of the original query and the
answers of the transformed one preserves the order. %
This allows us to get direct access to the answers of the original query from
the answers of the transformed one.

In this section we fix a query $Q$ on variables set $X$, a database $\db$ on domain $D = [d]$ and let $b = \lceil \log(d) \rceil$. For $v \in [d]$, we denote by $\bin[v]$ the binary encoding of $v$ over $b$ bits and let $\bin[v][b][i]$ be the $i^{\mathsf{th}}$ bit of $\bin[v]$ for every $1 \leq i \leq b$. For a variable $x \in X$, we introduce fresh variables $x^1, \dots, x^b$ and let $\bin[X] = \{ x^i \mid x \in X, 1 \leq i \leq b\}$. For a tuple $\tau \in D^Y$, we let $\bin[\tau]$ be the tuple in $D^{\bin[Y]}$ such that for every $y \in Y$ and $1 \leq i \leq b$, $\bin[\tau](y^i) = \bin[\tau(y)][b][i]$.  For a relation $R \subseteq D^{Y}$, we let $\bin[R] = \{\bin[\tau] \mid \tau \in R \} \subseteq D^{\bin[Y]}$. For an atom $A$ (positive or negative) on variables set $Y$, we let $\bin[A]$  be the corresponding atom on variables set $\bin[Y]$.

We let $\bin[Q]$ be the query on variables set $\bin[X]$ whose atoms are $\bin[A]$ for each atom $A$ of $Q$. Given a database $\db$ for $Q$, we let $\bin[\db]$ be a database for $\bin[Q]$ where every relation symbol $R$ appearing in $Q$ and interpreted by $R^\db$ in $\db$ is now interpreted by $\bin[R^\db]$ in $\bin[\db]$.

\begin{exa}
  Let $Q := A(x_1), B(x_2), \neg R(x_1,x_2)$ and $D = [3]$. We have $\bin[Q][2]$ is \[\bin[A][2](x_1^1, x_1^2), \bin[B][2](x_2^1, x_2^2), \neg \bin[R][2](x_1^1, x_1^2, x_2^1, x_2^2).\]

  The binarised version $\bin[\db]$ of the database $\db$ for $Q$ given in \cref{fig:ex-bin:normal} is given in \cref{fig:ex-bin:binarised}.

  %NEEDFIX, Inspect Figure and Change to booktabs if required
  \begin{figure}
    \centering
    \begin{subfigure}[b]{0.35\linewidth}
      \begin{tabular}[t]{c|c}
        $A$ & $x_1$ \\ \midrule
          & 0 \\
          & 1 \\
          & 2 
      \end{tabular}
      \begin{tabular}[t]{c|c}
        $B$ & $x_2$ \\ \midrule
          & 1 \\
          & 2 \\
          & 3 
      \end{tabular}
      \begin{tabular}[t]{c|c c}
        $R$ & $x_1$ & $x_2$ \\ \midrule
          & 0  & 0 \\
          & 1  & 1 \\
          & 2  & 2 \\
          & 3  & 3 
      \end{tabular}

      \caption{Original database}
      \label{fig:ex-bin:normal}
    \end{subfigure}
%    \hfill
    \quad
    \begin{subfigure}[b]{0.6\linewidth}
      \centering

      \begin{tabular}[t]{c | c c}
        $\bin[A][2]$ & $x_1^1$ & $x_1^2$ \\ \midrule
          & 0 & 0 \\
          & 1 & 0 \\
          & 0 & 1 
      \end{tabular}
      \begin{tabular}[t]{c | c c}
        $\bin[B][2]$ & $x_2^1$ & $x_2^2$ \\ \midrule
          & 1 & 0 \\
          & 0 & 1 \\
          & 1 & 1  
      \end{tabular}
      \begin{tabular}[t]{c | c c c c}
        $\bin[R][2]$ & $x_1^1$ & $x_1^2$ & $x_2^1$ & $x_2^2$ \\ \midrule
          & 0 & 0 & 0 & 0 \\
          & 1 & 0 & 1 & 0 \\
          & 0 & 1 & 0 & 1 \\
          & 1 & 1 & 1 & 1 \\
      \end{tabular}

      \caption{Binarised database with $b=2$}
      \label{fig:ex-bin:binarised}
    \end{subfigure}

    \caption{Binarisation of a database}
  \end{figure}
\end{exa}

\noindent
Binarisation is interesting since  there is a natural isomorphism between $\ans$ and $\ans[\bin[Q]][][\bin[\db]]$. Indeed, given $\tau \in \ans$, it is easy to see that $\bin[\tau]$ is an answer of $\bin[Q]$ on $\bin[\db]$. Similarly given $\tau \in \{0,1\}^{\tilde{X}}$, we let  $\debin[\tau] \in D^X$ be the tuple defined as for every $x \in X$, $\debin[\tau] = \sum_{i=1}^b 2^{i-1}\tau(x^i)$. It is clear that for every $\tau \in \ans[\bin[Q]][][\bin[\db]]$, $\debin[\tau] \in \ans[Q]$. Moreover, for an order $\prec$ on $X$, we denote by $\prec^b$ the order on $\bin[X]$ defined as $x^i \prec^b y^j$ if and only if $x \prec y$ or $x = y$ and $i>j$. In other words, if $x_1,\dots,x_n$ is the order on $X$, the order $\prec^b$ on $\bin[X]$ is $x_1^b, \dots, x_1^1, \dots, x_n^b, \dots, x_n^1$.  The order $\prec^b$ naturally induces an order $\lexprec^b$ on $\{0,1\}^{\bin[X]}$. For $\tau,\tau' \in [d]^{X}$, we have that $\tau \lexprec \tau'$ if and only if $\bin[\tau] \lexprec^b \bin[\tau']$. This is because the natural ordering on $[d]$ can be seen as the lexicographical order on its bit representation, starting from the most significant bit of $\tau(x)$ which is $\bin[\tau](x^b)$ to the least significant bit of $\tau(x)$ which is $\bin[\tau](x^1)$. Hence, we have the following:

\begin{prop}
  \label{thm:order-preserving-isomorphism}
  Let $Q$ be a join query on variables set $X$, $\prec$ an order on $X$, and $\db$ a database for $Q$ on domain $[d]$ for some $d \in \N$.  Let $b = \lceil \log(d) \rceil$. The function $\debin[\cdot]$ is an isomorphism between $\ans[\bin[Q]][][\bin[\db]]$ and $\ans$ that can be computed in time $\bigo(nb)$ such that if $\tau$ is the $k^{th}$ answer of $\bin[Q]$ on $\bin[\db]$ for the order $\lexprec^b$ then $\debin[\tau]$ is the $k^{th}$ solution of $Q$ for the order $\lexprec$.
\end{prop}

\noindent
A consequence of \cref{thm:order-preserving-isomorphism} is that if one has direct access for the answers of $\bin[Q]$ on database $\bin[\db]$ for the order $\prec_b$ with preprocessing time $P$ and access time $A$, then we directly get direct access for $Q$ on database $\db$ with preprocessing time $P$ and access time $A + nb$. Indeed, we can perform the preprocessing as for $\bin[Q]$. To get the $k^{th}$ answer of $Q$, one computes the $k^{th}$ answer $\tau$ of $\bin[Q]$ and returns $\debin[\tau]$.

This is interesting since the analysis of DPLL made in \cref{thm:compilation}
has a linear dependency in the size of the domain, which is bounded by $2$ when
considering $\bin[\db]$ instead of $\db$. Since the number of atoms does not
change and the number of variables only increases by a logarithmic factor
$\log |D|$, running DPLL on $\bin[Q]$ and $\bin[\db]$ would offer an improvement
on the complexity, as long as the hyperorder width of $Q$ for $\prec$ and of
$\bin[Q]$ for $\prec^b$ are the same, without negatively affecting our ability to solve direct access tasks. This is indeed true and we now show:

% In other words, from \cref{thm:compilation} and \cref{thm:rankedcircuit}, we can get direct access to $\bin[Q]$ on database $\bin[\db]$ with preprecossing time $\bigo(\poly_k(Q) |\bin[\db]|^k)$ for $k = \show{H(\bin[Q])}[\succ_b]$ (see \cref{sec:ra-scq} for details). Now, it turns out that $\bin[Q]$ does not increase the width of the query:

\begin{thm}
  \label{thm:bin-preserves-width} For every query $Q$ and $b \in \N$, $\show{H(\bin[Q])}[\prec^b] = \show{H(Q)}[\prec]$ and $\sfhow{H(\bin[Q])}[\prec^b] = \sfhow{H(Q)}[\prec]$.
\end{thm}

\noindent
The rest of this section is dedicated to the proof of \cref{thm:bin-preserves-width}. It is based on the following notion: given a signed hypergraph $H=(V,E_+,E_-)$ and a vertex $u \in V$, we define the signed hypergraph $\clone[H][u]$ to be the hypergraph obtained by \emph{cloning $u$ in $H$}, that is, by adding a new vertex $u'$ in every edge where $u$ appears. More formally, the vertices of $\clone$ are $V \cup \{u'\}$ where $u'$ is a fresh vertex not in $V$ and the positive (resp. negative) edges of $\clone$ are $\{e \cup \{u'\} \mid e \in E_+, u \in e\} \cup \{e \in E_+ \mid u \notin e\}$ (resp. $\{e \cup \{u'\} \mid e \in E_-, u \in e\} \cup \{e \in E_- \mid u \notin e\}$). We will use $\clone[e][u]$ as a shorthand for $e \cup \{u'\}$ if $u \in e$ and for $e$ otherwise. 

It is easy to see that $H(\bin[Q])$ can be obtained from $H(Q)$ by iteratively
cloning $b-1$ times each variable $x$ of $Q$. %
Proving \cref{thm:bin-preserves-width} boils down to proving that cloning a vertex
$u$ does not change the width of a hypergraph as long as the copy $u'$ of $u$ is
removed right after $u$. %
In other words, we will show that $\show{H}[\prec] = \show{H'}[\prec_u]$ and
$\sfhow{H}[\prec] = \sfhow{H'}[\prec_u]$ where $H' = \clone$ and $\prec_u$ is
the order obtained from $\prec$ by inserting $u'$ after $u$. %
% On an intuitive level, this is easy to see but it is tedious to fully
% formalize. % S: this sentence does not bring any information
The first thing needed for the proof is to observe that cloning and removing
vertices commute in the following way:

\begin{lem}
  \label{lem:cloning-permutes} Let $H=(V,E)$ be a hypergraph and $u \in V$. For every $v \neq u$, $\clone[H/v] = \clone/v$.
\end{lem}
\begin{proof}
  Since $u \neq v$, $\clone[H/v]$ and $\clone/v$ have the same vertex set. %
  It is then enough to show that they have the same edges. %
  Let $f$ be an edge of $\clone[H/v]$. %
  By definition, either $f = \clone[e \setminus \{v\}]$ for $e$ an edge of $H$
  or $f = \clone[\oneigh[v]]$. %
  In the first case, $f = \clone[e] \setminus \{v\}$ since $u \neq v$ and so it
  is an edge of $\clone[H]/v$. %
  In the second case, we claim that $f = \oneigh[v][\clone]$, and hence $f$ is
  also an edge of $\clone/v$.

  Indeed, we first show $f \subseteq \oneigh[v][\clone]$. %
  Let $w \in f$ and \(u'\) be the vertex added in \(\clone\). %
  If $w \neq u'$, it means that $w \in \oneigh[v]$. %
  Hence there is an edge $e' \in E$ such that $\{w,v\} \subseteq e'$. %
  In this case, $w$ is in $\clone[e']$. %
  In particular, $w \in \oneigh[v][\clone]$. %
  Now if $w = u'$, then it can only happen if $u \in f$, that is, $u \in
  \oneigh[v]$. %
  In this case, there is $e' \in E$ such that $\{u,v\} \subseteq e'$ and in
  particular $\{v,u'\} \subseteq \clone[e']$. %
  It means that $u' \in \oneigh[v][\clone]$.

  We now show $\oneigh[v][\clone] \subseteq f$. %
  Let $w \in \oneigh[v][\clone]$ and \(u'\) be the vertex added to \(\clone\). %
  As before, if $w \neq u'$, then $w \in \oneigh[v]$ and $w \in
  \clone[\oneigh[v]]$. %
  If $w = u'$, then $u$ must be in $\oneigh[v][\clone]$, but then it means it is
  in $\oneigh[v]$ too. Hence $u' \in \clone[\oneigh[v]]$.

  It remains to show that every edge $f$  of $\clone/v$ is also an edge of $\clone[H/v]$. This is completely symmetrical to the proof above. Indeed, either $f = \clone[e] \setminus \{v\}$ for some $e$ and since $u \neq v$, we have $f = \clone[e \setminus \{v\}]$ and it proves that $f$ is an edge of $\clone[H/v]$, or we have $f = \oneigh[v][\clone]$ but we just proved that $f = \clone[\oneigh[v]]$ which is an edge of $\clone[H/v]$.
\end{proof}

We now address the missing case from \cref{lem:cloning-permutes} where $v=u$:

\begin{lem}
  \label{lem:cloning-rm-u}
  Let $H=(V,E)$ be a hypergraph and $u \in V$. Let $H' = \clone / u$.  Then $H/u = H' \setminus \{u'\}$ and $\oneigh[u'][H'] = \oneigh[u][H]$.
\end{lem}
\begin{proof}
  Let $f$ be an edge of $H'$. We show that $f \setminus \{u'\}$ is an edge of $H/u$. By definition, $f$ is either of the form $\clone[e] \setminus \{u\}$ for $e$ an edge of $H$ or $\oneigh[u][\clone]$. In the first case, $f \setminus \{u'\} = e \setminus \{u\}$  which is an edge of $H/u$. In the second case, we clearly have $f = \oneigh[u][\clone] = \oneigh[u] \cup \{u'\}$. Hence $f \setminus \{u'\} = \oneigh[u]$ is an edge of $H/u$.

  Similarly, let $f$ be an edge of $H/u$. We show that either $f$ or $f \cup \{u'\}$ is an edge of $H'$. By definition, $f$ is either equal to $e \setminus \{u\}$ for some edge $e$ of $H$ or $f = \oneigh[u]$. In the first case, there are two subcases depending on whether $u \notin e$ or $u \in e$.
  Assume $u \notin e$ first. Then $f=e$ is also an edge of $\clone$, and hence of $\clone/u = H'$. Now if $u \in e$, then $\clone[e] = e \cup \{u'\}$ is an edge of $\clone$ and then $(e \setminus \{u\}) \cup \{u'\} = f \cup \{u'\}$  is an edge of $H'$.

  In the second case, we have $f =\oneigh[u]$. But then, as before $\oneigh[u][\clone] = f \cup \{u'\}$. Hence $f \cup \{u'\}$ is an edge of $\clone/u = H'$.

  From this analysis, observe that $u'$ in $H'$ only appears in $\oneigh[u][\clone]$ or in edge $f$ of the form $(e \setminus \{u\}) \cup \{u'\}$ where $e$ is an edge of $H$ such that $u \in e$. Hence, $\oneigh[u'][H'] = \oneigh[u][H]$. 
\end{proof}

We are now ready to prove the following, which  implies \cref{thm:bin-preserves-width}:
\begin{lem}
  \label{lem:cloning-preserves-width} Let $H=(V,E)$ be a hypergraph, $\prec$ an order on $V$ and $u \in V$. Let $u'$ be the clone of $u$ in $H' = \clone$ and $\prec_u$ be the order on $V \cup \{u'\}$ where $u'$ is put right after $u$. Then  $\how{H}[\prec] = \how{H'}[\prec_u]$  and $\fhow{H}[\prec] = \fhow{H'}[\prec_u]$.
\end{lem}
\begin{proof}
  We write $V$ as $u_1,\dots,u_n$, with $u_1 \prec \dots \prec u_n$. Assume $u = u_j$ and let $H_0=H$ and $H_{i} = H/u_1/\dots/u_i$. Moreover, let $H'_0 = H'$ and for $i<j$, let $H'_i = H'/u_1/\dots/u_i$. For $i \geq j$, let $H'_i = H'/u_1/\dots/u_j/u'/u_{j+1}/\dots/u_i$. By \cref{lem:cloning-permutes}, for $i <j$, $H'_i=\clone[H_i][u_j]$. Hence, it directly follows that $\neigh[u_{i+1}][H_i]  = \neigh[u_{i+1}][H'_i] \setminus \{u'\}$ and that $\rho(\neigh[u_{i+1}][H_i], H) = \rho(\neigh[u_{i+1}][H'_i], H')$ and  $\rho^*(\neigh[u_{i+1}][H_i], H) = \rho^*(\neigh[u_{i+1}][H'_i], H')$.

  Up to $i=j$, removing $u_i$ from $H_i$ or $H_i'$ results in the same effect on the width. Now, consider the case where $u'$ is removed from $H'_j$. By \cref{lem:cloning-rm-u}, the neighbourhood of $u'$ in $H'_j$ is the same as the neighbourhood of $u_j$ in $H_{j-1}$. Thus it can be covered with the same (fractional) number of edges and hence removing $u'$ from $H'_j$ does not increase the width of the order.

  Finally, we observe that by \cref{lem:cloning-rm-u} again,  $H'_j/u' = H_j$. Indeed, since $\oneigh[u'][H'_j] =\oneigh[u_j][H_{j-1}]$, $H'_j/u'$ does not introduce any new edge and then $H'_j/u' = H'_j \setminus \{u'\}$ which is equal to $H_{j-1}/u_j = H_j$ by \cref{lem:cloning-rm-u}. Hence by a direct induction, $H'_i = H_i$ for $i>j$. Hence both elimination orders have the same width.
\end{proof}

It only remains to generalise \cref{lem:cloning-preserves-width} to signed hyperorder width:
\begin{lem}
  \label{lem:cloning-preserves-signed-width} Let $H=(V,E_+ \cup E_-)$ be a signed hypergraph, $\prec$ an order on $V$ and $u \in V$. Let $u'$ be the clone of $u$ in $H' = \clone$ and $\prec_u$ be the order on $V \cup \{u'\}$ where $u'$ is put right after $u$. Then  $\show{H}[\prec] = \show{H'}[\prec_u]$  and $\sfhow{H}[\prec] = \sfhow{H'}[\prec_u]$.
\end{lem}
\begin{proof}
Let $H_0 \subseteq^- H$ be a negative subhypergraph of $H$ and let $H_0' = \clone[H_0]$. It is easy to see that $H_0'$ is a negative subhypergraph of $H'$. By \cref{lem:cloning-preserves-width}, $\how{H_0}[\prec] = \how{H'_0}[\prec_u]$ and $\fhow{H_0}[\prec] = \fhow{H'_0}[\prec_u]$. Hence $\show{H}[\prec] \leq \show{H'}[\prec_u]$ and $\sfhow{H}[\prec] \leq \sfhow{H'}[\prec_u]$. Similarly, for a negative subhypergraph $H_0'$ of $H'$, there is a negative subhypergraph $H_0$ of $H$ such that $H_0' = \clone[H_0]$. We hence similarly get $\show{H}[\prec] \geq \show{H'}[\prec_u]$ and $\sfhow{H}[\prec] \geq \sfhow{H'}[\prec_u]$. 
\end{proof}

\cref{thm:bin-preserves-width} is a direct consequence of \cref{lem:cloning-preserves-signed-width} since $H(\bin[Q])$ is obtained by cloning each vertex of $H(Q)$ $b$ times and the copies $x^i$ of vertex $x$ are consecutive in $\prec^b$. 

\begin{rem}
  Consider again the instance $Q=A(x_1) \wedge B(x_2) \wedge (x_1 \neq x_2)$ from \cref{ex:nonlinear} where we proved that caching would never occur with this instance, resulting in a circuit of size at least $\bigot(d^2)$. However, when running DPLL on $\bin[Q]$, we can see that caching will now often happen. Indeed, assume that the algorithm is in a state where every copy $x_1^b,\dots,x_1^1$ of $x_1$ was set already to values $v_b,\dots,v_1$ and the algorithm is now setting variables $x_2^b,\dots,x_2^1$. Observe that as soon as some copy $x_2^i$ is set to the value $1-v_i$, then $\neg \bin[R]$ is satisfied since the value assigned to $x_1$ is necessarely different from the value assigned to $x_2$. Hence, the atom $\neg \bin[R]$ is now simplified away and caching can occur. Each subset of $\bin[B]$ where some prefix of $x_2^b,\dots,x_2^1$ will be cached, which boils down to $d \cdot \log d$ values, less than the $\bigot(d^2)$ from \cref{ex:nonlinear}.
\end{rem}

We conclude this section by stating how one can use binarisation as a means of improving the construction of \ocircuit{s} representing answers of signed conjunctive queries. 

\begin{thm}
  \label{thm:bin-compilation}
  Let $Q$ be a signed join query, $\db$ a database over domain $D$ and $\prec$ an order on $\var(Q)$. Then $\DPLL(\bin[Q],\emptytup,\bin[\db],\prec_b)$ produces a $\succ_b$-\ocircuit{} $C$ of size $\bigot((\poly_k |Q|) \cdot |\db|^{k})$ on domain $\{0,1\}$ such that $\rel(C) = \ans[\bin[Q]]$ and:
  \begin{itemize}
  \item $k=\fhow{H(Q)}[\prec]$ if $Q$ is positive,
  \item $k=\show{H(Q)}[\prec]$ otherwise.
  \end{itemize}
  Moreover, the runtime of this construction is at most $\bigot(\poly_k(|Q|) \cdot |\db|^{k})$.
\end{thm}
\begin{proof}
  This is simply \cref{thm:compilation} applied to $\bin[Q]$ with $|D|=2$ together with the fact that $\sfhow{H(\bin[Q])}[\prec_b] = \sfhow{H(Q)}[\prec]$ and $\fhow{H(\bin[Q])}[\prec_b] = \fhow{H(Q)}[\prec]$ from \cref{thm:bin-preserves-width}.
\end{proof}
%%% Local Variables:
%%% mode: latex
%%% TeX-master: "main"
%%% TeX-engine: luatex
%%% End:

%%% Local Variables:
%%% mode: latex
%%% TeX-master: "main"
%%% End:

\section{Tractability results for join and conjunctive queries}
\label{sec:ra-scq}

In this section, we connect the tractability result on direct access on ordered circuit of \cref{sec:ocircuits} with the algorithm presented in \cref{sec:cqtocircuits} to obtain tractability results concerning the complexity of direct access on signed join queries. Moreover, we show how our approach can be used to get direct access for signed conjunctive queries, that is, signed join queries with projected variables.

\begin{thm}
  \label{thm:ra-sjq} Given a signed join query $Q$,  an order $\prec$ on $\var(Q)$ and a database $\db$ on domain $D$, we can solve the direct access problem for $\lexprec$ with preprocessing time $\bigot(|\db|^{k}\poly_k(|Q|))$ and access time $\bigo(\poly(|Q|) \cdot (\log |D|)^3(\log \log |D|))$ and:
  \begin{itemize}
  \item if $Q$ does not contain any negative atom, then $k=\fhtw{H(Q)}[\succ]$,
  \item otherwise $k=\show{H(Q)}[\succ]$.
  \end{itemize}
\end{thm}
\begin{proof}

  We construct a $\prec_b$ ordered-circuit computing $\ans[\bin[Q]]$ using \cref{thm:bin-compilation} for $b = \lceil \log |D| \rceil$ on domain of size $2$ and variables set $\bin[X] = \var(\bin[Q])$. We preprocess this circuit as in \cref{thm:rankedcircuit}. Constructing the circuit and preprocessing it constitutes the preprocessing phase of our direct access algorithm and can be executed in $\bigot(|\db|^{k}\poly_k(|Q|))$ by \cref{thm:bin-compilation}. 
  
  Now, to find the $i^{th}$ solution in $\ans$, we simply find the $i^{th}$ solution of $\rel(C)$ using the algorithm of \cref{sec:ra-ocircuits}, which is the $i^{th}$ solution of $\ans[\bin[Q]][][\bin[\db]]$ and by \cref{thm:order-preserving-isomorphism}, we can reconstruct from it the $i^{th}$ solution of $\ans$ in time $\bigo(\log |D|)$. By \cref{thm:rankedcircuit}, the access time is hence $\bigo(\poly(|Q|) \cdot (\log |D|)^3(\log \log |D|))$ because the number of variables of $C$ is $\bigo(|\var(Q)| \log |D|)$. 
\end{proof}

The main difference between \cref{thm:ra-sjq} and \cref{thm:re-scq-optimal} is that the complexity given in \cref{thm:ra-sjq} is polynomial in $|Q|$ for bounded signed hyperorder width or for bounded fractional hypertree width, which matches \cref{thm:complexity-posqueries} in this case. If we are now interested in signed \emph{fractional} hyperorder width, then we are not able to prove that DPLL produces a circuit that is polynomial in $|Q|$. However, by \cref{thm:compilation-fractional}, we know that DPLL on the binarisation of $Q$ produces a circuit of size $\bigot(2^mn|\db|^k)$, which then matches the complexity given in \cref{thm:complexity-posqueries} for preprocessing but has a better access time since the degree of $\log |D|$ does not depend on $|Q|$ anymore. Formally, we get this improved version of the upper bound of \cref{thm:re-scq-optimal}. In particular, observe that even if preprocessing is exponential in the query size, this is not the case for the access time:

\begin{thm}
  \label{thm:ra-sjq-fractional} Given a signed join query $Q$,  an order $\prec$ on $\var(Q)$ and a database $\db$ on domain $D$, we can solve the direct access problem for $\lexprec$ with preprocessing time $\bigot(|\db|^{\sfhow{H(Q)}[\succ]}2^{\size{\atoms(Q)}} \poly(|Q|))$ and access time $\bigo(\poly(|Q|) \cdot (\log |D|)^3(\log \log |D|))$.
\end{thm}

\paragraph*{Direct access for conjunctive queries.} As mentioned, \cref{thm:ra-sjq} allows to recover the tractability of direct access for positive join queries with bounded fractional hypertree width proven in~\cite{Carmeli2023,bringmann2022tight} and restated here in \cref{thm:complexity-posqueries}. However, \cite{Carmeli2023} (see also~\cite[Theorem 39]{bringmann2022arxiv} which is the arXiv version of \cite{bringmann2022tight}) also generalises the algorithm to conjunctive queries, that is, join queries with projections. Here, we demonstrate the versatility of the circuit-based approach by showing how one can also handle quantifiers directly on the circuit:
\begin{thm}
  \label{thm:existentialcircuit}
  Let $C$ be a  $\prec$-\ocircuit{} on domain $D$, variables set $X=\{x_1,\dots,x_n\}$ such that $x_1 \prec \dots \prec x_n$ and $j \leqslant n$. One can compute in time $\bigo(|C| \cdot \poly(n)\cdot\polylog(|D|))$ a $\prec$-\ocircuit{} $C'$ of size at most $|C|$ such that $\rel(C') = \rel(C)|_{\{x_1,\dots,x_{j}\}}$.
\end{thm}
\begin{proof}
Let $v$ be a decision gate on variable $x_k$ with $k > j$. By definition, every decision-gate in the circuit rooted at $v$ tests a variable $y \in \{x_{k+1},\dots, x_n\}$. Hence $\rel(v) \subseteq D^Y$ with $Y \subseteq \{x_k,\dots,x_n\}$. Moreover, one can compute $|\rel(v)|$ for every gate $v$ of $C$ in time $\bigo(|C|\cdot \poly(n) \cdot\polylog(|D|))$ as explained in \cref{lem:precomputation-complexity}. Hence, after this preprocessing, we can decide whether $\rel(v)$ is the empty relation in time $\bigo(1)$ by simply checking whether $|\rel(v)| = \nrel(v,d_0) \neq 0$  where $d_0$ is the largest element of $D$. We construct $C'$ by replacing every decision-gate $v$ on a variable $x_k$ with $k > j$ by a constant gate $\top$ if $\rel(v) \neq \emptyset$ and $\bot$ otherwise. We clearly have that $|C'| \leqslant |C|$ and from what precedes, we can compute $C'$ in $\bigo(|C|\cdot\poly(n)\cdot\polylog(|D|))$. Moreover, it is straightforward to show by induction that every gate $v'$ of $C'$ which corresponds to a gate $v$ of $C$ computes $\rel(C)|_{\{x_1,\dots,x_j\}}$, which concludes the proof.
\end{proof}

Now we can use \cref{thm:existentialcircuit} to handle conjunctive queries by first using \cref{thm:compilation} on the underlying join query to obtain a $\prec$-circuit and then by projecting the variables directly in the circuit. This approach works only when the largest variables in the circuits are the quantified variables. It motivates the following definition: given a hypergraph $H=(V,E)$, an elimination order $(v_1,\dots,v_n)$ of $V$ is $S$-connex if and only if there exists $i$ such that $\{v_i,\dots,v_n\}=S$. In other words, the elimination order starts by eliminating $V\setminus S$ and then proceeds to $S$. Given a conjunctive query $Q$ and an elimination order $\prec$ on $\var(Q)$, we say that the elimination is free-connex if it is a $\free(Q)$-connex elimination order of $H(Q)$ where $\free(Q)$ are the free variables of $Q$.\footnote{The notion of $S$-connexity already exists for tree decompositions. We use the same name here as the existence of an $S$-connex tree decomposition of (fractional) hypertree width $k$ is equivalent to the existence of an $S$-connex elimination order of (fractional) hyperorder width $k$.}
We directly have the following: 

\begin{thm}
  \label{thm:ra-scq} Given a signed conjunctive query $Q(Y)$, a free-connex order $\prec$ on $\var(Q)$ and a database $\db$ on domain $D$, we can solve the direct access problem for $\lexprec$ with preprocessing time $\bigot(|\db|^{k}\poly_k(|Q|))$ and access time $\bigo(\poly(|Q|) \cdot (\log |D|)^3 \log \log |D|)$ and:
  \begin{itemize}
  \item if $Q$ does not contain any negative atom, then $k=\fhtw{H(Q)}[\succ]$,
  \item Otherwise $k=\show{H(Q)}[\succ]$.
  \end{itemize}
\end{thm}
\begin{proof}
  By running $\DPLL(\bin[Q], \emptytup, \bin[\db], \prec^b)$ for $b = \lceil \log |D| \rceil$, one obtains a $\succ^b$-ordered circuit computing $\ans[\bin[Q]][][\bin[\db]]$. The size of the circuit is $\bigot(|\db|^{k}\poly_k(|Q|))$ by \cref{thm:compilation}. Now, $\prec^b$ is free-connex, hence $\succ^b$ is of the form $z_1^b \succ \dots \succ z_1^1 \succ \dots \succ z^b_n \succ \dots \succ z_n^1$ and there exists $i$ such that $\{z_1,\dots,z_j\}=\free(Q)$.  Hence by Theorem~\ref{thm:existentialcircuit}, we can construct a $\succ^b$-\ocircuit{} of size at most $\bigo(|\db|^{k}\poly_k(|Q|))$ computing $\ans[\bin[Q]][][\bin[\db]]|_{\bin[\free(Q)]}$, which concludes the proof using \cref{thm:rankedcircuit}. 
\end{proof}
We observe that our notion of free-connex elimination order for $Q$ is akin to \cite[Definition 38]{bringmann2022arxiv} with one main difference:  in \cite{bringmann2022arxiv}, it is allowed to only specify a preorder on $\free(Q)$ and the complexity of the algorithm is then stated with the best possible compatible ordering, which would be possible in our framework too. Now, Theorem~\ref{thm:ra-scq} constructs a direct access for $\lexprec$ when $\prec$ is free-connex, so Theorem~\ref{thm:ra-scq} proves the same tractability result as \cite[Theorem 39]{bringmann2022arxiv} with the same complexity.

Observe also that there is another way of recovering \cref{thm:ra-scq}. We can modify DPLL so that when only projected variables remain, instead of compiling, it calls an efficient join algorithm that can decide whether the answer set of the query is empty or not and returns either $\top$ or $\bot$ in the circuit depending on the answer of this join algorithm. This approach could be more efficient because it is able to exploit more complex measures such as the submodular width of the resulting hypergraph, using a join algorithm such as PANDA~\cite{abo2017shannon}.

%%% Local Variables:
%%% mode: latex
%%% TeX-master: "main"
%%% End:

\section{Negative join queries and SAT}
\label{sec:sat}

One particularly interesting application of our result is when it is applied to  negative join queries, that is, join queries where every atom is negated. Tractability results for negative join queries have been established previously in the literature. One of the first results in this direction was the study of $\beta$-acyclic negative join queries. It was shown in~\cite{Brault-Baron12} that the negative conjunctive queries where evaluation can be done in linear time in the database are exactly the $\beta$-acyclic conjunctive queries. A slight generalisation of this result to acyclic signed join queries can be found in Brault Baron's thesis~\cite{brault2013pertinence}. This result was generalised for counting in~\cite{BraultCM15} and recently to queries with more complex agregation in~\cite{ZhaoFOK24}. A generalisation of $\beta$-acyclic queries, namely negative join queries with bounded nest set width, was proposed by Lazinger in~\cite{lanzinger2023tractability} where evaluation of such queries was shown to be tractable. 

Another interesting application of our result is to offer new tractability results for aggregation problems in SAT solving. Indeed, the SAT problem inputs are CNF (Conjunctive Normal Form) formulas, which can be seen as a particular case of negative join queries. Indeed, a CNF formula $F$ with $m$ clauses can directly be transformed into a negative join query $Q_F$ with $m$ atoms having the same hypergraph, a database $\db_F$ on domain $\{0,1\}$ of size at most $m$ such that $\ans[Q_F][][\db_F]$ is the set of satisfying assignments of $F$: a clause can be seen as the negation of a relation having exactly one tuple. For example, $x \vee y \vee \neg z$ can be seen as $\neg R(x,y,z)$ where $R$ contains the tuple $(0,0,1)$. Hence, any polynomial time algorithm in \emph{combined complexity} on negative join queries directly transfers to CNF formulas, where many tractability results for SAT and \#SAT are known when the hypergraph of the input CNF is restricted~\cite{OrdyniakPS13,BovaCMS15,PaulusmaSS13,SlivovskyS13,SamerS10,SaetherTV14} (see~\cite[Chapter 2]{CapelliPhD} for a survey).

In this section, we show that \cref{thm:ra-sjq} generalises many of these results because most of the hypergraph families known to give tractability results for negative join queries can be shown to have bounded signed hyperorder width. We now study the notion of signed hyperorder width restricted to negative join queries and compare it to existing measures from literature.

\subsection{Hyperorder width for negative join queries}
\label{sec:betahow}

If a join query is negative, then we can consider its signed hypergraph to be simply the hypergraph induced by its negative atoms. Hence, the notion of signed hyperorder width can be simplified without separating between positive and negative edges. We hence use the following definitions: for a hypergraph $H=(V,E)$ and an order $\prec$ on $V$, the \emph{$\beta$-hyperorder width $\bhow{H}[\prec]$ of $\prec$ for $H$} is defined as $\max_{H' \subseteq H} \how{H'}[\prec]$. The \emph{$\beta$-hyperorder width $\bhow{H}$ of $H$} is defined as the width of the best possible elimination order, that is, $\bhow{H} = \min_\prec \bhow{H}[\prec]$. We define similarly the $\beta$-fractional hyperorder width of an order $\prec$ and of a hypergraph -- $\bfhow{H}[\prec]$ and $\bfhow{H}$ -- by replacing $\how{\cdot}$ by $\fhow{\cdot}$ in the definitions.
Observe that for every negative join query $Q$, $\iota^s(Q,\succ)$  corresponds to $\bfhow{H(Q)}[\prec]$ where $H(Q)$ is seen as a (non signed) hypergraph. 

We now turn to comparing the notion for $\beta$-hyperorder width with other measures from the literature. The definition of $\beta$-hyperorder width can be seen as a hereditary closure of generalised hypertree width. Indeed, it is a well known fact that hypertree width is not hereditary, that is, a subhypergraph may have a greater hypertree width than the hypergraph itself. Take for example the hypergraph consisting of a triangle on vertices $\{1,2,3\}$ together with the hyperedge $\{1,2,3\}$. This hypergraph is $\alpha$-acyclic, and hence has hypertree width $1$ while its subhypergraph consisting of the triangle has fractional hypertree width $3/2$ and generalised hypertree width $2$.

The fact that fractional hypertree width is not hereditary has traditionally been worked around by taking the largest width over every subhypergraph. This led to the definition of the \emph{$\beta$-fractional hypertree width $\bfhtw{H}$} of $H$, defined as $\bfhtw{H}=\max_{H' \subseteq H} \fhtw{H'}$~\cite{gottlob2004hypergraphs}. The $\beta$-hypertree width $\bhtw{H}$ is defined similarly by replacing $\fhtw{\cdot}$ by $\htw{\cdot}$. If one plugs the ordered characterisation of $\fhtw{H'}$ in this definition, we have the following equivalent way of defining fractional $\beta$-hypertree width: $\bfhtw{H}=\max_{H' \subseteq H} \min_{\prec} \fhow{H'}[\prec]$. Now recall that $\bfhow{H} =  \min_{\prec} \max_{H' \subseteq H} \fhow{H'}[\prec]$. Hence, the difference between $\bfhtw{H}$ and $\bfhow{H}$ boils down to inverting the $\min$ and the $\max$ in the definition. It is easy to see then that $\bfhtw{H}\leqslant\bfhow{H}$ and  $\bhtw{H} \leqslant \bhow{H}$ for every $H$ (see \cref{thm:bfhow-vs-rest} for details). The main advantage of the $\beta$-fractional hyperorder width is that it comes with a natural notion of decomposition --- the best elimination order $\prec$ --- that can be used algorithmically. No such decomposition for $\bfhtw{\cdot}$ that can be used algorithmically has yet been found.

The case where $\bfhtw{H}=1$, known as $\beta$-acyclicity, is the only one for which tractability results are known, for example SAT~\cite{OrdyniakPS13}, \#SAT or \#CQ for $\beta$-acyclic instances~\cite{Capelli17,BraultCM15}. This is due to the fact that in this case,  an order-based characterisation is known. The elimination order is based on the notion of nest points. In a hypergraph $H=(V,E)$, a \emph{nest point} is a vertex $v \in V$ such that $E(v)$ can be ordered by inclusion, that is, $E(v) = \{e_1,\dots,e_p\}$ with $e_1 \subseteq \dots \subseteq e_p$. A \emph{$\beta$-elimination order $(v_1,\dots,v_n)$ for $H$} is an ordering of $V$ such that for every $i\leqslant n$, $v_i$ is a nest point of $H \setminus \{v_1,\dots,v_{i-1}\}$. We show in \cref{thm:bfhow-vs-rest} that $\beta$-elimination orders correspond exactly to elimination orders having $\beta$-hyperorder width $1$, hence proving that our width notion generalises $\beta$-acyclicity.

We actually prove a more general result: the notion of $\beta$-acyclicity has been recently generalised by Lanzinger in~\cite{lanzinger2023tractability} using a notion called nest sets. A set of vertices $S \subseteq V$ is a \emph{nest set of $H$} if $\{e \setminus S \mid e \in E, e \cap S \neq \emptyset\}$ can be ordered by inclusion. A \emph{nest set elimination order} is a list $\Pi=(S_1, \dots, S_p)$ such that:
\begin{itemize}
\item $\bigcup_{i=1}^p S_i=V$,
\item $S_i \cap S_j = \emptyset$ and
\item $S_i$ is a nest set of $H \setminus \bigcup_{j<i} S_j$.
\end{itemize}
The width of a nest set elimination order is defined as $\nsw{H}[\Pi]=\max_{i} |S_i|$ and the \emph{nest set width $\nsw{H}$ of $H$} is defined to be the smallest possible width of a nest set elimination order of $H$. It turns out that our notion of width generalises the notion of nest set width; that is, we have $\bhow{H} \leqslant \nsw{H}$. More particularly, any order $\prec$ obtained from a nest set elimination order $\Pi=(S_1, \dots, S_p)$ by ordering each $S_i$ arbitrarily verifies $\nsw{H}[\Pi] \geqslant \bhow{H}[\prec]$.

We summarise and give formal proofs of the above discussion in the following theorem:
\begin{thm}
  \label{thm:bfhow-vs-rest}
  For every hypergraph $H=(V,E)$, we have:  $\bhtw{H} \leqslant \bhow{H} \leqslant \nsw{H}$. In particular, $H$ is $\beta$-acyclic if and only if $\bhow{H}=1$.
\end{thm}

The proof mainly follows from the following lemma. Intuitively, it says that if $S$ is a nest set of $H$ of size $k$, then if we iteratively remove the vertices of $S$ in $H$, then the edges it introduces can always be covered by $k$ edges from the original hypergraph. This is because they are made of some vertices from $S$ (at most $k$ of them) and some other vertices that are all covered by the maximal edge covering $\{e \setminus S \mid e \cap S \neq \emptyset\}$. 
\begin{lem}
  \label{lem:nestsetelim}
  Let $H=(V,E)$ be a hypergraph and $S$ be a nest set of $H$ of size $k$. We let $f$ be the maximal element (for inclusion) of $\{e \setminus S \mid e \in E, e \cap S \neq \emptyset\}$, which exists by definition and $(s_1, \dots, s_k)$ an ordering of $S$. For every $i \leqslant k$ and edge $e$ of $H/s_1/\dots/s_i$,  either $e \cap S \neq \emptyset$ and $e \subseteq f \cup S$ or $e \cap S = \emptyset$ and $e$ is an edge of $H \setminus \{s_1, \dots, s_i\}$.
\end{lem}
\begin{proof}
  We prove this lemma by induction on $i$. For $i=0$, it is clear since if $e \cap S \neq \emptyset$, then $e \setminus S \subseteq f$ by definition of $f$. Hence $e \subseteq f \cup S$. Now, assuming the hypothesis holds for some $i$, let $H_i = H/s_1/\dots/s_{i}$ and $H_{i+1}=H_i/s_{i+1}$. By definition, the edges of $H_{i+1}$ are (i) the edges of $H_i$ without the vertex $s_{i+1}$ or (ii) the additional edge $\oneigh[s_{i+1}][H_i]$. Let $e$ be an edge of $H_{i+1}$ that is not $\oneigh[s_{i+1}][H_i]$. We have two cases: either $e$ is already in $H_i$, in which case the induction hypothesis still holds. Or $e$ is not in $H_i$, which means that $e = e' \setminus \{s_{i+1}\}$ for some edge $e'$ of $H_i$ containing $s_{i+1}$. In particular, $s_{i+1} \in e'\cap S$ and then $e' \cap S \neq \emptyset$. By induction, we have $e' \subseteq f \cup S$ and then $e = e'\setminus \{s_{i+1}\} \subseteq f \cup S$ and the induction hypothesis follows. Finally, assume $e = \oneigh[s_{i+1}][H_i]$, that is, $e$ is obtained by taking the union of every $e'$ in $H_i$ where $s_{i+1}$ is in $e'$ and removing $s_{i+1}$. But then $e' \cap s_{i+1} \neq \emptyset$.  By induction again, $e' \subseteq f \cup S$ hence $\oneigh[s_{i+1}][H_i] \subseteq f \cup S$ and the induction hypothesis follows.
\end{proof}

\begin{proof}[Proof of \cref{thm:bfhow-vs-rest}]
  The inequality $\bhtw{H} \leqslant \bhow{H}$ is straightforward using:
  \begin{itemize}
  \item $\bhtw{H} = \max_{H' \subseteq H} \min_{\prec} \how{H'}[\prec]$
  \item $\bhow{H} = \min_{\prec} \max_{H' \subseteq H}  \how{H'}[\prec]$
  \end{itemize}
  
  \noindent
  Indeed, let $\prec_0$ be an elimination order that is minimal for $\bhow{H}[\prec]$. By definition, for $H' \subseteq H$, $\how{H'}[\prec_0] \geqslant \min_{\prec} \how{H'}[\prec]$. Hence \[
    \bhow{H} = \max_{H' \subseteq H} \how{H'}[\prec_0] \geqslant \max_{H' \subseteq H} \min_{\prec} \how{H'}[\prec] = \bhtw{H}.
  \]
  We now prove $\bhow{H} \leqslant \nsw{H}$. Let $k=\nsw{H}$ and $\Pi=(S_1, \dots, S_p)$ a nest set elimination of $H$ of width $k$, that is, for every $i$, $|S_i|\leqslant k$. Let $\prec$ be an order on $V=(v_1,\dots,v_n)$ with $v_1 \prec \dots \prec v_n$, obtained from $\Pi$ by ordering each $S_i$ arbitrarily, that is, if $x \in S_i$ and $y \in S_j$ with $i<j$, we require that $x \prec y$. We claim that $\bhow{H}[\prec]\leqslant \nsw{H}[\Pi]$. First of all, we observe that if $(S_1,\dots,S_p)$ is a nest set elimination order for $H$, then it is also a nest set elimination order for every $H' \subseteq H$, which is formally proven in~\cite[Lemma 4]{lanzinger2023tractability}\footnote{Lemma 4 of~\cite{lanzinger2023tractability} establishes the result for a connected subhypergraph of $H$ but the same proof works for non-connected subhypergraphs.}. Consequently, it is enough to prove that $\how{H}[\prec] \leqslant k$. Indeed, if we prove it, then we know that for every hypergraph $H' \subseteq H$, $\prec$ is also a nest set elimination order for $H'$, and hence, $\how{H'}[\prec] \leqslant k$ too.

  This follows from \cref{lem:nestsetelim}. Indeed, let $(v_1,\dots,v_t)$ be the prefix of $(v_1,\dots,v_n)$ such that $S_1=\{v_1,\dots,v_t\}$. By \cref{lem:nestsetelim}, when $v_{i+1}$ is removed from $H_i^\prec = H/v_1/\dots/v_i$, then $N_{i+1}=\neigh[v_{i+1}][H_i]$ is included in $f \cup S_1$ since $N_{i+1} \cap S_1 \neq \emptyset$ (both sets contain $v_{i+1}$). Hence $N_{i+1}$ is covered by at most $t$ edges: $f$ -- which contains at least one element of $S_1$ -- plus at most one edge for each remaining element of $S_1$. Hence, up to the removing of $v_t$, the hyperorder width of $\prec$ is at most $t \leqslant k$. Now, when removing $(v_1,\dots,v_t)$ from $H$, by \cref{lem:nestsetelim} again, $H_t^\prec = H \setminus \{v_1,\dots,v_t\}$  since no edge of $H_t^\prec$ has a non-empty intersection with $S_1$. It follows that $S_2$ is a nest set of $H_t^\prec$ and we can remove it in a similar way to $S_1$ and so on. Hence $\bhow{H}[\prec] \leqslant k = \nsw{H}[\Pi]$ which settles the inequality stated in the theorem.

  It directly implies that $H$ is $\beta$-acyclic if and only if $\bhow{H}=1$. Indeed, if $H$ is $\beta$-acyclic, then $\nsw{H}=1$ (the definition of nest set width elimination order of width $1$ directly corresponds to the definition of $\beta$-elimination orders) and  $\bhow{H} \leqslant \nsw{H}=1$ by the previously established bound.
\end{proof}

The goal of this paper is not to give a thorough analysis of $\beta$-fractional hyperorder width so we leave for future research several questions related to it. We observe that we do not know the exact complexity of computing or approximating the $\beta$-fractional hyperorder width of an input hypergraph $H$. It is very likely hard to compute exactly since it is not too difficult to observe that when $H$ is a graph, $\bfhow{H}$ is sandwiched between the half of the treewidth of $H$  and the treewidth of $H$ itself and it is known that treewidth is $\NP$-hard to compute~\cite{Bodlaender93a}. This does not rule out the possibility that deciding whether $\bhow{H} \leq k$ is tractable for every $k$ as it is the case for treewidth~\cite{Bodlaender93a}, but we observe that deciding whether $\fhow{H} \leq 2$ is known to be $\NP$-hard~\cite{fischl2018general}.  We also leave open many questions concerning how $\beta$-fractional hyperorder width compares with other widths such as (incidence) treewidth, (incidence) cliquewidth or MIM-width. For these measures of width, \#\SAT, a problem close to computing the number of answers in signed join queries, is known to be tractable (see~\cite[Chapter 2]{CapelliPhD} for a survey). % This may be shorten for the conference version
We leave open the most fundamental question of comparing the respective powers of $\bfhtw{\cdot}$ and $\bfhow{\cdot}$:
\begin{openquestion}
  Does there exist a family $(H_n)_{n \in \N}$ of hypergraphs such that there exists $k \in \N$ such that for every $n$, $\bfhtw{H_n} \leq k$ while $\bfhow{H_n}$ is unbounded?
\end{openquestion}

% This may be shorten for the conference version
One may wonder why the definition of $\beta$-hyperorder width has not appeared earlier in the literature, as it just boils down to swapping a $\min$ and a $\max$ in the definition of $\beta$-hypertree width while enabling an easier algorithmic treatment. We argue that the expression of hypertree width in terms of elimination orders -- which is not the widespread way of working with this width in previous literature -- is necessary to make this definition interesting. Indeed, if one swaps the $\min$ and $\max$ in the traditional definition of $\beta$-hypertree width, we get the following definition: $\bphtw{H}[T]=\min_{T} \max_{H' \subseteq H} \htw{H'}[T]$ where $T$ runs over every tree decomposition of $H$ and hence is valid for every $H' \subseteq H$ since, as every edge of $H$ is covered by $T$, so are the edges of $H'$. This definition, while being obtained in the same way as $\bhow{\cdot}$, is not really interesting however because it does not generalise the notion of $\beta$-acyclicity:

\begin{prop}
  \label{lem:bpfhtw} There exists a family of $\beta$-acyclic hypergraphs $(H_n)$ such that for every $n \in \N$, $\bphtw{H_n}=n$.
\end{prop}
\begin{proof}
  Consider the hypergraph $H_n$ whose vertex set is $[n]$ and edges are $\{0,i\}$ for $i>0$ and $[n]$. That is $H_n$ is a star centered in $0$ and additionnally has an edge containing every vertex. $H_n$ is clearly $\beta$-acyclic (any elimination order that ends with $0$ is a $\beta$-elimination order) but we claim that $\bphtw{H_n}=n$. Indeed, let $T$ be a tree decomposition for $H_n$. By definition, it contains a bag that covers $[n]$. Now consider the subhypergraph $H'_n$ of $H_n$ obtained by removing the edge $[n]$. The hypertree width of $T$ with respect to $H'_n$ is $n$ since one needs the edge $\{i,0\}$ for every $i$ to cover vertex $i$ in the bag $[n]$ since $i$ appears only in this edge.
\end{proof}

\subsection{Applications}

In the previous section, we have shown that $\beta$-hyperorder width generalises $\beta$-acyclicity and nest set width. Hence, \cref{thm:ra-sjq} can be used to show that direct access is tractable for the class of queries with bounded nest set width. In particular, counting the number of answers is tractable for this class, a question left open in~\cite{lanzinger2023tractability}:

\begin{thm}
  Let $Q$ be a negative join query of bounded nest set width and $\db$ be a database. Then we can compute $\size{\ans}$ in polynomial time. 
\end{thm}
\begin{proof}
  Let $k$ be the nest set width of $Q$. It is proven in~\cite[Theorem 15]{lanzinger2023tractability} that a nest set elimination order for $Q$ can be found in time $2^{\bigo(k^2)}\poly(|H(Q)|)$ which induces an elimination order $\prec$ for $H(Q)$ of $\beta$-hyperoder width $k$. Now, using DPLL on this order and on the binarised form of $Q$, we can construct a circuit computing $\ans$ in time $\bigot((\poly(H(Q))|\db|)^k)$ and extract $|\ans|$ from it using \cref{lem:precomputation-complexity} in time $\bigot((\poly(H(Q))|\db|)^k)$.
\end{proof}

Similarly, our result can directly be applied to \#SAT, the problem of counting the number of satisfying assignments of a CNF formula. Hence, \cref{thm:ra-sjq} generalises both \cite{Capelli17} and \cite{BraultCM15} by providing a compilation algorithm for $\beta$-acyclic queries to any domain size and to the more general measure of $\beta$-hyperorder width. It also shows that not only counting is tractable but also the more general direct access problem.

\begin{figure}
  \centering
  \includegraphics[scale=0.6]{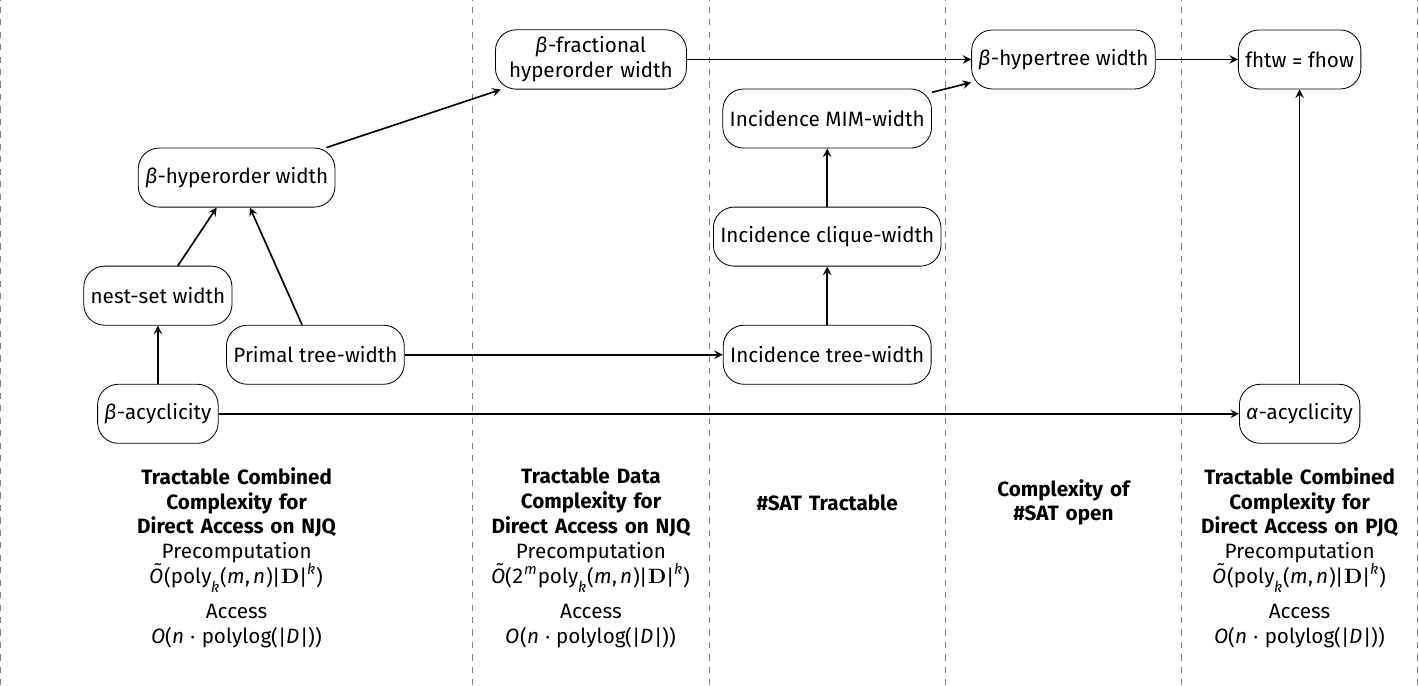}
    \caption[Landscape of hypergraph measures]{Landscape of hypergraph
      measures and known inclusions with tractability results for
      direct access on negative join queries (NJQ), direct access on
      positive join queries (PJQ), and \#SAT on CNF formulas. Here, $n$ is the
      number of variables, $m$ the number of atoms, $\db$ the
      database, $D$ the domain and $k$ the width measure ($k=1$ for
      $\alpha$- and $\beta$-acyclicity). In the case of CNF formulas, $m$ stands for the
      number of clauses, the size of the database is at most $m$ and
      the domain is $\{0,1\}$. An arrow between two classes indicates inclusion for a fixed $k$.

      All tractability results for direct access are consequences of \cref{thm:ra-scq} applied to either fractional hyperorder width in the positive case or $\beta$-hyperorder width in the negative case. The $\beta$-fractional hyperorder width is a consequence of \cref{thm:scq-optimal}. Comparison between $\beta$-hyperorder width, nest-set width and $\beta$-hypertree width is a consequence of \cref{thm:bfhow-vs-rest}. Tractability of \#SAT for MIM-width is from~\cite{SaetherTV14}, where most inclusions below can be found. The inclusion between MIM-width and $\beta$-hypertree width can be found in~\cite[Theorem 2.33]{CapelliPhD}.
    } 
  \label{fig:tractabilityDA}
\end{figure}

\cref{fig:tractabilityDA} summarises our contributions for join queries with negations and summarises how our contribution is located in the landscape of known tractability results. The two left-most columns of the figure are contributions of this paper (Theorem~\ref{thm:ra-sjq}), the right-most column is known from \cite{bringmann2022tight} but can be recovered in our framework (see discussion below). A complete presentation of the results stated in this figure can be found in \cite[Chapter 2]{CapelliPhD}.

%%% Local Variables:
%%% mode: latex
%%% TeX-master: "main"
%%% End:

\section{Conclusion and Future Work}
\label{sec:conclusion}

In this paper, we have proven new tractability results concerning the direct access of the answers of signed conjunctive queries. In particular, we have introduced a framework unifying the positive and the signed case using a factorised representation of the answer set of the query. Our complexity bounds, when restricted to the positive case, match the existing optimal one and we have proven that our algorithm remains optimal for signed join queries without self-join (in terms of data complexity), closing the two main open questions of the conference version of this paper~\cite{confversion}.  Our approach opens many new avenues of research. In particular, we believe that the circuit representation that we use is promising for answering different kinds of aggregation tasks and hence generalising existing results on conjunctive queries to the case of signed conjunctive queries. For example, we believe that FAQ and AJAR queries~\cite{abo2016faq, ajar} could be solved using this data structure. Indeed, it looks possible to annotate the circuit with semi-ring elements and to project them out in a similar fashion as \cref{thm:existentialcircuit}. Similarly, we believe that the framework of~\cite{eldar2023direct} for solving direct access tasks on conjunctive queries with aggregation operators may be generalised to the class of \ocircuit{s}. Finally, contrary to the positive case, we do not yet know what is the complexity of solving direct access tasks on signed join queries with self-joins. We know that self-joins can only make things easier and have exhibited an example at the end of \cref{sec:optimally-solving-sjq} where having a self-join between the positive and the negative part leads to drastic improvement in the preprocessing complexity. We do not know however whether there are other more subtle cases where self-joins help, as it was shown for enumeration~\cite{carmeli2023conjunctive}.

% Actually, our algorithm to answer direct access tasks on the circuit is completely independent from the way the circuit has been constructed, as long as it has the necessary syntactic properties. Hence, we have hope larger class blabla.

%%% Local Variables:
%%% mode: latex
%%% TeX-master: "main"
%%% End:

\newpage

\bibliographystyle{alphaurl}
\bibliography{biblio}

\newpage
\appendix 
\section{Proof of \cref{lem:assignment-of-tau-x}}
\label{app:assignment}

\assignmentTauX*

\begin{proof}
  We now formally prove this claim. %
  Let $A = \{d \mid \#\sigma_{x \leqslant d}(R) \geqslant k\}$. %
  We start by showing that $\tau(x) \in A$, meaning
  $\#\sigma_{x \leqslant \tau(x)} \geqslant k$. %
  Let $\alpha \lexpreceq \tau$. %
  Since $x$ is the smallest variable, it follows that
  $\alpha \in \sigma_{x \leqslant \tau(x)}(R)$ as
  $\alpha(x) \leqslant \tau(x)$. %
  Since there exists exactly $k$ such assignments $\alpha$ (by definition of
  $\tau$ which is the \kt{} of $R$), we have
  $\#\sigma_{x \leqslant \tau(x)}(R) \geqslant k$. %

  We now show that, given a value $d' < \tau(x)$, we have that $d' \notin A$ and
  therefore $\tau(x)$ is indeed the smallest value in $A$. %
  Let $\alpha \in \sigma_{x \leqslant d'}$. %
  It follows that $\alpha(x) \leqslant \tau(x)$, and therefore that
  $\alpha < \tau$. %
  We therefore have that
  $\sigma_{x \leqslant d'}(R) \subset \{\alpha \mid \alpha \lexprec
  \tau\}$. %
  By definition of $\tau$ as the \kt{}, the latter set has less than $k - 1$
  elements. %
  Hence $d' \notin A$. %
  % This implies that for any $d \in A, \tau(x) \leqslant d$. %
  This shows that $\tau(x)$ is indeed the smallest value $d$ such that there
  exists at least $k$ tuples $\alpha$ where $\alpha(x) \leqslant d$. %

  The second part of the lemma follows from the following observation: when
  assigning a value $d$ to the variable $x$, one actually \emph{eliminates} a
  certain number of tuples from the initial set. Specifically, the tuples that
  assign a different value to $x$.

  By definition, $k$ is the cardinal of the set $\{\tau' \mid \tau' \lexpreceq
  \tau\}$. This set can be written as the disjoint union of the set of tuples
  where $\tau'(x) < d$ (which are all smaller than $\tau$) and the set of tuples
  smaller than $\tau$ where $\tau'(x) = d$. We therefore have $k = \#\{\tau
  \mid \tau(x) < d\} + \#\{\tau' \mid \tau' \lexprec \tau, \tau'(x) = d\}$. By
  definition, the first set is $\sigma_{x < d}(R)$. The second part of the sum
  is exactly the index of the tuple in the subset of $R$ where $\tau(x) = d$. We
  can rewrite the sum as $k = \#\sigma_{x < d}(R) + k'$, implying $k' = k -
  \#\sigma_{x < d}(R)$. 
\end{proof}

%%% Local Variables:
%%% mode: LaTeX
%%% TeX-master: "main"
%%% End:

\end{document}